%% file: JNN-RNN-RC-paper.tex
\documentclass[a4paper,fleqn]{cas-sc}

\input{./utils/utils-packages.tex}

\input{./utils/utils-commands.tex}

\input{./utils/utils-colors.tex}

\usepackage[authoryear]{natbib}

\begin{document}
\let\WriteBookmarks\relax
\def\floatpagepagefraction{1}
\def\textpagefraction{.001}
\shorttitle{Forecasting of Complex Dynamics with RNNs}
\shortauthors{PR Vlachas et~al.}


\title [mode = title]{
Backpropagation Algorithms and Reservoir Computing in Recurrent Neural Networks for the Forecasting of Complex Spatiotemporal Dynamics
}

\author[1]{P. R. Vlachas}[type=editor, orcid=0000-0002-3311-2100]
\ead{vlachas@collegium.ethz.ch}
\credit{Conceptualization, Methodology, Software, Data Curation, Visualization, Writing - Original Draft}

\author[2,3]{J. Pathak}
\ead{jpathak@umd.edu}
\credit{Conceptualization, Methodology, Software, Data Curation, Writing - Review \& Editing}

\author[4,5]{B. R. Hunt}
\ead{bhunt@umd.edu}
\credit{Conceptualization, Supervision, Writing - Review \& Editing}

\author[6]{T. P. Sapsis}
\ead{sapsis@mit.edu}
\credit{Conceptualization, Supervision, Writing - Review \& Editing}

\author[2,3,4]{M. Girvan}
\ead{girvan@umd.edu}
\credit{Conceptualization, Supervision}

\author[2,3,7]{E. Ott}
\credit{Conceptualization, Supervision, Writing - Review \& Editing}
\ead{edott@umd.edu}

\author[1]{P. Koumoutsakos}
\credit{Conceptualization, Methodology, Supervision, Writing - Review \& Editing}
\cormark[1]
\ead{petros@ethz.ch}
\cortext[cor1]{Corresponding author}

\address[1]{Computational Science and Engineering Laboratory, ETH Z\"urich, Clausiusstrasse 33, Z\"urich CH-8092, Switzerland}
\address[2]{Institute for Research in Electronics and Applied Physics, University of Maryland, College Park, Maryland 20742, USA}
\address[3]{Department of Physics, University of Maryland, College Park, Maryland 20742, USA}
\address[4]{Institute for Physical Science and Technology, University of Maryland, College Park, Maryland 20742, USA}
\address[5]{Department of Mathematics, University of Maryland, College Park, Maryland 20742, USA}
\address[6]{Department of Mechanical Engineering, Massachusetts Institute of Technology, 77 Massachusetts Avenue, Cambridge, Massachusetts 02139-4307, USA}
\address[7]{Department of Electrical and Computer Engineering, University of Maryland, Maryland 20742, USA}

\begin{abstract}
We examine the efficiency of Recurrent Neural Networks in forecasting the spatiotemporal dynamics of high dimensional and reduced order complex systems using Reservoir Computing (RC) and  Backpropagation through time (BPTT) for gated network architectures.
We highlight advantages and limitations of each method and discuss their implementation for parallel computing architectures. 
We  quantify the relative prediction accuracy of these algorithms for the long-term forecasting of chaotic systems using as benchmarks the Lorenz-96 and the Kuramoto-Sivashinsky (KS) equations.
We find that, when the full state dynamics are available for training, RC outperforms BPTT approaches in terms of predictive performance and in capturing of the long-term statistics, while at the same time requiring much less training time.
However, in the case of reduced order data, large scale RC models can be unstable and more likely than the BPTT algorithms to diverge.
In contrast, RNNs trained via BPTT show superior forecasting abilities and capture well the dynamics of reduced order systems.
Furthermore, the present study quantifies for the first time the Lyapunov Spectrum of the KS equation with BPTT, achieving similar accuracy as RC.
This study establishes that RNNs are a potent computational framework for the learning and forecasting of complex spatiotemporal systems.
\end{abstract}




\maketitle

\input{./sections/sections-1-introduction.tex}

\input{./sections/sections-2-sequence-modelling.tex}

\input{./sections/sections-3-comparison-metrics.tex}

\input{./sections/sections-4-lorenz96.tex}

\input{./sections/sections-5-parallel.tex}

\input{./sections/sections-6-lyapunov-spectrum.tex}
\input{./sections/sections-7-conclusion.tex}

\input{./sections/sections-8-misc.tex}

\printcredits

\bibliographystyle{cas-model2-names}

\bibliography{JNN-RNN-RC-paper}


\bio{}
\endbio

\newpage
\appendix

\input{./appendix/appendix-0-rc-memory.tex}

\input{./appendix/appendix-1-rc-noise-reg.tex}

\input{./appendix/appendix-2-dim-red.tex}

\input{./appendix/appendix-3-lyapunov-spectrum.tex}

\input{./appendix/appendix-4-hyperparameters.tex}

\input{./appendix/appendix-5-lorenz96-f8-additional.tex}

\input{./appendix/appendix-6-lorenz96-f10-additional.tex}

\input{./appendix/appendix-7-BBTT.tex}

\input{./appendix/appendix-8-parallel-overfitting.tex}

\end{document}

%% file: utils/utils-packages.tex
\usepackage{hyperref}       
\usepackage{url}            
\usepackage{booktabs}       
\usepackage{amsfonts}       
\usepackage{nicefrac}       
\usepackage{microtype}      
\usepackage{algpseudocode}
\usepackage{textcomp}
\usepackage{amssymb}
\usepackage{amsmath}
\usepackage{bm}
\usepackage{caption, subcaption}
\usepackage{anyfontsize}
\usepackage{etoolbox}
\usepackage{tikz}
\usetikzlibrary{tikzmark}
\usetikzlibrary{calc}

\usepackage{cleveref}
\usepackage{dirtytalk}

\usepackage{algorithm}
\usepackage{algpseudocode}

\usepackage{etoolbox}
\usepackage{tikz}
\usetikzlibrary{tikzmark}
\usetikzlibrary{calc}

\errorcontextlines\maxdimen

\newcommand{\ALGtikzmarkcolor}{black}
\newcommand{\ALGtikzmarkextraindent}{4pt}
\newcommand{\ALGtikzmarkverticaloffsetstart}{-.5ex}
\newcommand{\ALGtikzmarkverticaloffsetend}{-.5ex}
\makeatletter
\newcounter{ALG@tikzmark@tempcnta}

\newcommand\ALG@tikzmark@start{%
    \global\let\ALG@tikzmark@last\ALG@tikzmark@starttext%
    \expandafter\edef\csname ALG@tikzmark@\theALG@nested\endcsname{\theALG@tikzmark@tempcnta}%
    \tikzmark{ALG@tikzmark@start@\csname ALG@tikzmark@\theALG@nested\endcsname}%
    \addtocounter{ALG@tikzmark@tempcnta}{1}%
}

\def\ALG@tikzmark@starttext{start}
\newcommand\ALG@tikzmark@end{%
    \ifx\ALG@tikzmark@last\ALG@tikzmark@starttext
    \else
        \tikzmark{ALG@tikzmark@end@\csname ALG@tikzmark@\theALG@nested\endcsname}%
        \tikz[overlay,remember picture] \draw[\ALGtikzmarkcolor] let \p{S}=($(pic cs:ALG@tikzmark@start@\csname ALG@tikzmark@\theALG@nested\endcsname)+(\ALGtikzmarkextraindent,\ALGtikzmarkverticaloffsetstart)$), \p{E}=($(pic cs:ALG@tikzmark@end@\csname ALG@tikzmark@\theALG@nested\endcsname)+(\ALGtikzmarkextraindent,\ALGtikzmarkverticaloffsetend)$) in (\x{S},\y{S})--(\x{S},\y{E});%
    \fi
    \gdef\ALG@tikzmark@last{end}%
}

\apptocmd{\ALG@beginblock}{\ALG@tikzmark@start}{}{\errmessage{failed to patch}}
\pretocmd{\ALG@endblock}{\ALG@tikzmark@end}{}{\errmessage{failed to patch}}
\makeatother

\usepackage{standalone}

\usepackage{tikz}
\usepackage{bm}
\tikzset{every path/.append style={line width=0.8pt}}
\usetikzlibrary{arrows}
\usetikzlibrary{arrows.meta}
\usepackage{scalefnt}
\usetikzlibrary{shapes}

%% file: utils/utils-commands.tex
\newcommand{\blueRectangle}{\raisebox{1pt}{\tikz{\node[rectangle,draw=blue, scale=0.9] at (0.25,-0.05) {};
}}}
\newcommand{\greenCross}{\raisebox{1pt}{\tikz{
\draw[mark=x,scale=2,line width=1.8pt,green] plot coordinates {(0.125,-0.05)};
 }}}
\newcommand{\redCircle}{\raisebox{1pt}{\tikz{\node[circle,draw=red, scale=0.65] at (0.25,-0.05) {};
}}}
\newcommand{\orangeDiamond}{\raisebox{1pt}{\tikz{\node[diamond,draw=orange, scale=0.6]  at (0.25,0) {};
}}}

\newcommand{\bluelineRectangle}{\raisebox{1pt}{\tikz{\node[rectangle,blue,fill=blue, scale=0.95] at (0.25,0) {};
\draw[-,blue,solid,line width = 2](0,0) -- (0.5,0);
}}}

\newcommand{\greenlineX}{\raisebox{1pt}{\tikz{
\draw[mark=x,scale=2,line width=1.8pt,green] plot coordinates {(0.125,0)};
\draw[-,green,solid,line width=2pt](0,0) -- (0.5,0);
 }}}

\newcommand{\redlineCircle}{\raisebox{1pt}{\tikz{\node[circle,red,fill=red, scale=0.7] at (0.25,0) {};
\draw[-,red,solid,line width = 2pt](0,0) -- (0.5,0);
}}}

\newcommand{\orangelineDiamond}{\raisebox{1pt}{\tikz{\node[diamond,orange,fill=orange, scale=0.6]  at (0.25,0) {};
\draw[-,orange,solid,line width = 2pt](0,0) -- (0.5,0);
}}}

\newcommand{\bluevioletlineStar}{\raisebox{1pt}{\tikz{\node[star,blueviolet,fill=blueviolet,scale=0.5]  at (0.27,0) {};
\draw[-,blueviolet,solid,line width = 2pt](0,0) -- (0.5,0);
}}}

\newcommand{\blacklineDashed}{\raisebox{3pt}{\tikz{\draw[-,black,dashed,line width = 2pt](0,0mm) -- (5mm,0mm);}}}

\newcommand{\x}{\bm{x}}
\newcommand{\h}{\bm{h}}
\newcommand{\bmo}{\bm{o}}
\newcommand{\bmr}{\bm{r}}
\newcommand{\bmb}{\bm{b}}
\newcommand{\bmg}{\bm{g}}
\newcommand{\bmz}{\bm{z}}
\newcommand{\bmc}{\bm{c}}
\newcommand{\W}{\bm{W}}
\newcommand{\R}{\mathbb{R}}

\newcommand{\violinBlueRectangle}{\raisebox{1pt}{\tikz{\node[rectangle,blue,fill=blue,minimum width=0.5cm,minimum height=0.25cm, scale=0.95, opacity=0.4] at (0.0,0) {};
}}}

\newcommand{\violinGreenRectangle}{\raisebox{1pt}{\tikz{\node[rectangle,green,fill=green,minimum width=0.5cm,minimum height=0.25cm, scale=0.95, opacity=0.4] at (0.0,0) {};
}}}

\newcommand{\violinRedRectangle}{\raisebox{1pt}{\tikz{\node[rectangle,red,fill=red,minimum width=0.5cm,minimum height=0.25cm, scale=0.95, opacity=0.4] at (0.0,0) {};
}}}

\newcommand{\violinOrangeRectangle}{\raisebox{1pt}{\tikz{\node[rectangle,orange,fill=orange,minimum width=0.5cm,minimum height=0.25cm, scale=0.95, opacity=0.4] at (0.0,0) {};
}}}

%% file: utils/utils-colors.tex
\definecolor{blue}{rgb}{0,0,1} 
\definecolor{green}{rgb}{0.0, 0.5020, 0.0} 
\definecolor{red}{rgb}{1,0,0} 
\definecolor{black}{rgb}{0,0,0} 
\definecolor{blueviolet}{rgb}{0.5412,0.1686,0.8863} 
\definecolor{orange}{rgb}{1,0.6471,0} 
\definecolor{cornflowerblue}{rgb}{0.3922,0.5843,0.9294}

%% file: sections/sections-1-introduction.tex
\section{Introduction}
\label{sec:intro}

In recent years we have observed significant advances in the field of machine learning (ML) that rely on potent algorithms and their deployment on powerful computing architectures. Some of these advances have been materialized by deploying ML algorithms on dynamic environments such as video games~\citep{ha2018world,schrittwieser2019mastering} and simplified physical systems (AI gym)~\citep{brockman2016openai,mnih2015human,silver2016mastering}.
Dynamic environments are often encountered across disciplines ranging from engineering and  physics to finance and social sciences. They can serve as bridge for scientists and engineers to advances in machine learning and at the same time they present a fertile ground for the development and testing of advanced ML algorithms~\citep{hassabis2017neuroscience}. The deployment of advanced machine learning algorithms to complex systems is in its infancy.
We believe that  it deserves  further exploration as it may have far-reaching implications for societal and scientific challenges ranging from weather and climate prediction~\citep{weyn2019can, gneiting2005weather}, to energy networks, medicine~\citep{esteva2017dermatologist,kurth2018exascale}, and the dynamics of ocean dynamics and turbulent flows~\citep{aksamit2019machine,sunderhauf2018limits,brunton2020machine}.

Complex systems  are characterized by multiple, interacting spatiotemporal scales that challenge classical numerical methods for their prediction and control. The  dynamics of such systems are typically chaotic and difficult to predict,  a critical issue in problems such as weather and climate prediction. Recurrent Neural Networks (RNNs), offer a potent method for addressing these challenges.
RNNs were developed for processing of sequential data, such as time-series~\citep{Hochreiter1997}, speech~\citep{graves2014towards}, and language~\citep{dong2015multi,Kyunghyun2014}.
Unlike classical numerical methods that aim at discretizing existing equations of complex systems, RNN models are data driven.
RNNs keep track of a hidden state, that encodes information about the history of the system dynamics. 
Such data-driven models are of great importance in applications to complex systems where equations based on first principles may not exist, or may be expensive to discretize and evaluate, let alone control, in real-time.

Early application of neural networks for modeling and prediction of dynamical systems can be traced to the  work of Lapedes et. al.~\citep{lapedes1987nonlinear}, where they demonstrated the efficiency of feedforward artificial neural networks (ANNs) to model deterministic chaos.  As an alternative to ANNs, wavelet networks were proposed in~\citep{Liangyue1995} for chaotic time-series prediction. 
However, these works have been  limited to intrinsically low-order systems, and they have been  often deployed in conjunction with dimensionality reduction tools.
As shown in this work, RNNs have the potential to overcome these scalability problems and be applied to high-dimensional spatio-temporal dynamics.
The  works of Takens~\citep{Takens1981} and  Sauer, Yorke and Casdagli~\citep{Sauer1991} showed that the dynamics on a D-dimensional attractor of a dynamical system can be unfolded in a time delayed embedding of dimension greater than 2D.
The identification of a useful embedding and the construction of a forecasting model have been the subject of life-long research efforts~\citep{Bradley2015}.
More recently, in~\citep{lusch2018deep}, a data-driven method based on the Koopman operator formalism~\citep{koopman1931hamiltonian} was proposed, using feed-forward ANNs to identify an embedding space with linear dynamics that is then amenable to theoretical analysis.  

There is limited work at the interface of RNNs and nonlinear dynamical systems~\citep{Vlachas2018,Zhong2018,Pathak2017,Pathak2018a,lu2018}. 
Here we examine and compare two of the most prominent nonlinear techniques in the forecasting of dynamical systems, namely RNNs trained with backpropagation and Reservoir Computing (RC).
We note that our RC implementation also uses a recurrent neural network, but according to the RC paradigm, it does not train the internal network parameters.
We consider the cases of fully observed systems as well as the case of partially observed systems such as reduced order models of real world problems, where typically we do not have access to all the degrees-of-freedom of the dynamical system.

Reservoir Computing (RC)  has shown significant success in modeling the full-order space dynamics of high dimensional chaotic systems.
This success  has sparked the interest of theoretical researchers that  proved universal approximation properties of these models~\citep{grigoryeva2018echo,gonon2019reservoir}. 
In~\citep{Pathak2017,Pathak2018b} RC is utilized to build surrogate models for chaotic systems and compute their Lyapunov exponents based solely on data. 
A scalable approach to high-dimensional systems with local interactions is proposed in~\citep{Pathak2018a}.
In this case, an ensemble of RC networks is used in parallel.
Each ensemble member is forecasting the evolution of a group of modes while all other modes interacting with this group is fed at the input of the network. The model takes advantage of the local interactions in the state-space to decouple the forecasting of each mode group and improve the scalability.

RNNs are  architectures designed to capture long-term dependencies in sequential data~\citep{Pascanu2013,Bengio1994,Hochreiter1998, Goodfellow2016}. The potential of RNNs for capturing temporal dynamics in physical systems was explored first using low dimensional RNNs~\citep{Elman90} without gates to predict unsteady boundary-layer development, separation, dynamic stall, and dynamic reattachment back in 1997~\citep{Faller1997}. The utility of RNNs was  limited  by the finding that  during the learning process the gradients may vanish or explode.  In turn, the recent  success of RNNs is largely attributed to a cell architecture termed Long Short-Term Memory (LSTM). LSTMs employ gates that effectively remember and forget information thus alleviating the problem of vanishing gradients~\citep{Hochreiter1998}.
In the recent years~\citep{Bianchi2017} RNN architectures have been bench-marked for short-term load forecasting of demand and consumption of resources in a supply network, while in~\citep{Laptev2017} they are utilized for extreme event detection in low dimensional time-series.
In~\citep{wan2018machine} LSTM networks are used as  surrogates to model the kinematics of spherical particles in fluid flows.
In~\citep{Vlachas2018} RNNs with LSTM cells were utilized in conjunction with a mean stochastic model to capture the temporal dependencies and long-term statistics in the reduced order space of a dynamical system and forecast its evolution.
The method demonstrated better accuracy and scaling to high-dimensions and longer sequences than Gaussian Processes (GPs).
In~\citep{Zhong2018} the LSTM is deployed to model the residual dynamics in an imperfect Galerkin-based reduced order model derived from the system equations.  RC and LSTM networks are applied in the long-term forecasting of partially observable chaotic chimera states in~\citep{Neofotistos2019}, where instead of a completely model-free approach, ground-truth measurements of currently observed states are helping to improve the long-term forecasting capability.
RNNs are practical and efficient data-driven approximators of chaotic dynamical systems, due to their \textbf{(1)} universal approximation ability~\citep{Schafer2006,Siegelmann1995} and \textbf{(2)} ability to capture temporal dependencies and implicitly identify the required embedding for forecasting.

Despite the rich literature on both methods there are limited comparative studies of the two frameworks. The present work aims to fill this gap by examining these two machine learning algorithms on challenging physical problems. We compare the accuracy, performance, and computational efficiency of the two methods on the full-order and reduced-order modeling of two prototype chaotic dynamical systems.
We  also examine the modeling capabilities of the two approaches for reproducing correct Lyapunov Exponents and frequency spectra.
Moreover, we include in the present work  some more recent RNN architectures, like Unitary~\citep{Arjovsky2016,Jing2017} and Gated Recurrent Units (GRUs)~\citep{Chung2014,Kyunghyun2014} that have  shown superior performance over LSTMs for a wide variety of language, speech signal and polyphonic music modeling tasks.

We are interested in model-agnostic treatment of chaotic dynamical systems, where the time evolution of the full state or some observable is available, but we do not possess any knowledge about the underlying equations.
In the latter case, we examine which method is more suitable for modeling temporal dependencies in the reduced order space (observable) of dynamical systems.
Furthermore, we evaluate the efficiency of an ensemble of RNNs in predicting the full state dynamics of a high-dimensional dynamical system in parallel and compare it with that of RC. 
Finally, we discuss the advantages, implementation aspects (such as RAM requirements and training time) and limitations of each model.
We remark that the comparison in terms of time and RAM memory consumption, does not aim to quantify advantages/drawback among models but rather provide information for the end users of the software.

We hope that the present  study may open to the ML community a new arena with highly structured and complex environments for developing and testing advanced new algorithms~\citep{hassabis2017neuroscience}.
At the same time it may offer a bridge to the physics community to appreciate and explore the importance of advanced ML algorithms for solving challenging physical problems~\citep{brunton2020machine}. 

The structure of the paper is as follows.
\Cref{sec:methods} provides an introduction to the tasks and an outline of the architectures and training methods used in this work.
\Cref{sec:comparison} introduces the measures used to compare the efficiency of the models.
In \Cref{sec:observable} the networks are compared in forecasting reduced order dynamics in the Lorenz-96 system.
In \Cref{sec:parallel}, a parallel architecture leveraging local interactions in the state space is introduced and utilized to forecast the dynamics of the Lorenz-96 system~\citep{Lorenz1995} and the Kuramoto-Sivashinsky equation~\citep{Kuramoto1978}.
In \Cref{sec:le} the GRU and RC networks are utilized to reproduce the Lyapunov spectrum of the Kuramoto-Sivashinsky equation, while \Cref{sec:conclusion} concludes the paper.

%% file: sections/sections-2-sequence-modelling.tex
\section{Methods - Sequence Modeling}
\label{sec:methods}

We  consider machine learning algorithms for time-series forecasting.
The models are trained on time-series of an observable $\bmo \in \R^{d_o}$ sampled at a fixed rate $1/\Delta t$, $\{ \bmo_1, \dots, \bmo_T \}$, where we eliminate $\Delta t$ from the notation for simplicity.
The models possess an internal high-dimensional hidden state denoted by $\h_t \in \R^{d_h}$ that enables the encoding of temporal dependencies on past state history.
Given the current observable $\bmo_t$, the output of each model is a forecast $\bm{\hat{o}}_{t+1}$ for the observable at the next time instant $\bm{o}_{t+1}$.
This forecast is a function of the hidden state.
As a consequence, the general functional form of the models is given by
\begin{gather}
\h_{t}=f_h^h(\bmo_t, \h_{t-1}), \quad
\bm{\hat{o}}_{t+1}=f_h^o(\h_{t}),
\end{gather}
where $f_h^h$ is the hidden-to-hidden mapping and $f_h^o$ is the hidden-to-output mapping.
All recurrent models analyzed in this work share this common architecture.
They differ in the realizations of $f_h^h$ and $f_h^o$ and in the way the parameters or \textbf{weights} of these functions are learned from data, i.e., \textbf{trained}, to forecast the dynamics.

\subsection{Long Short-Term Memory}
\label{sec:lstm}
\begin{figure}
\begin{subfigure}[t]{0.2\textwidth}
\centering
\includegraphics[height=3.5cm]{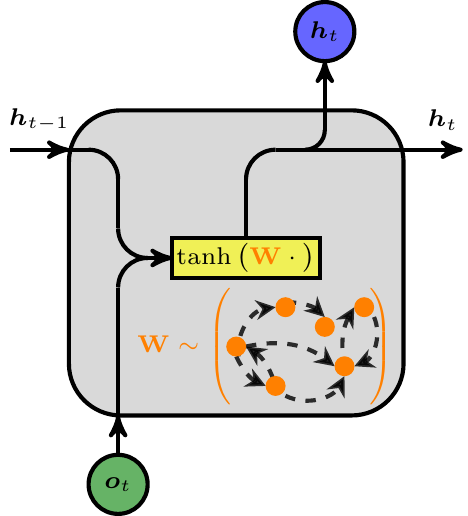}
\caption{RC Cell}
\label{fig:rc}
\end{subfigure}
\hfill
\begin{subfigure}[t]{0.2\textwidth}
\centering
\includegraphics[height=3.5cm]{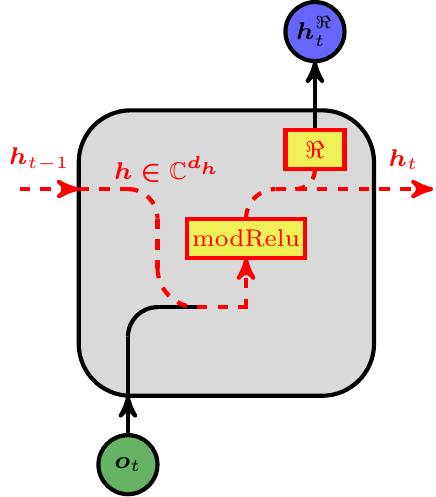}
\caption{Unit Cell}
\label{fig:unitary}
\end{subfigure}
\hfill
\begin{subfigure}[t]{0.25\textwidth}
\centering
\includegraphics[height=3.5cm]{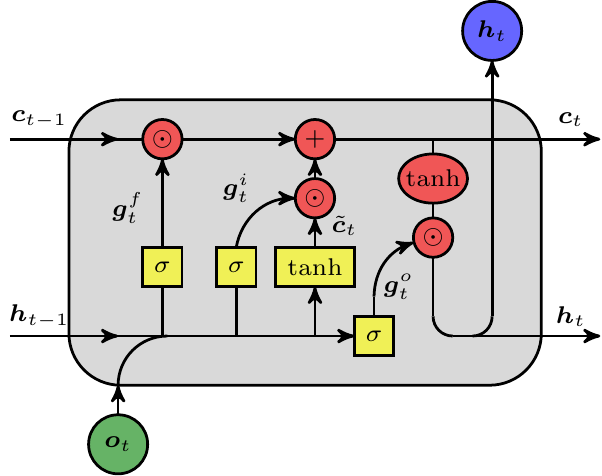}
\caption{LSTM Cell}
\label{fig:lstm}
\end{subfigure}
\hfill
\begin{subfigure}[t]{0.25\textwidth}
\centering
\includegraphics[height=3.5cm]{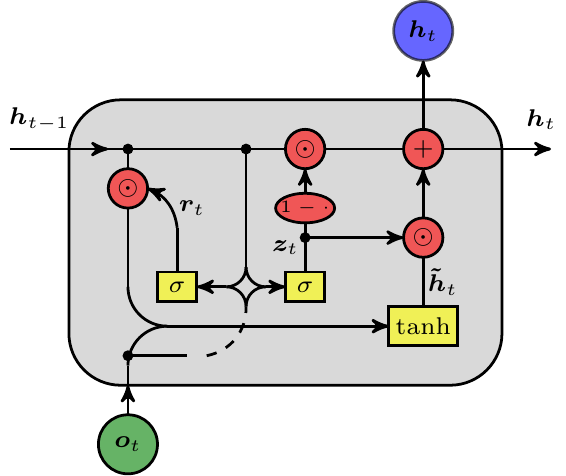}
\caption{GRU Cell}
\label{fig:gru}
\end{subfigure}
\caption{
The information flow for a Reservoir Computing (RC) cell, a complex Unitary cell (Unit), a Long Short-Term Memory (LSTM) cell and a Gated Recurrent Unit (GRU) cell.
The cells were conceptualized to tackle the vanishing gradients problem of Elman-RNNs.
The cell used in RC is the standard architecture of the Elman-RNN.
However, the weights of the recurrent connections are randomly picked to satisfy the echo state property and create a large reservoir of rich dynamics.
Only the output weights are trained (e.g., with ridge regression).
The Unitary RNN utilizes a complex unitary matrix to ensure that the gradients are not vanishing.
LSTM and GRU cells employ gating mechanisms that allow forgetting and storing of information in the processing of the hidden state.
Ellipses and circles denote entry-wise operations, while rectangles denote layer operations.
The information flow of the complex hidden state in the Unitary RNN is illustrated with dashed red color, while the untrained randomly picked weights of the RC with orange.
}
\end{figure}

In Elman RNNs~\citep{Elman90}, the vanishing or exploding gradients problem stems from the fact that the gradient is multiplied repeatedly during back-propagation through time~\citep{Werbos1988} with a recurrent weight matrix.
As a consequence, when the spectral radius of the weight matrix is positive (negative), the gradients are prone to explode (shrink).
The LSTM~\citep{Hochreiter1997} was introduced in order to alleviate the vanishing gradient problem of Elman RNNs~\citep{Hochreiter1998} by leveraging gating mechanisms that allow information to be forgotten.
The equations that implicitly define the recurrent mapping $f_h^h$ of the LSTM are given by
\begin{equation}
\begin{aligned}
\bmg^f_t &= \sigma_f \big(\W_f [\h_{t-1}, \bmo_t ] + \bmb_f\big) 
&& \bmg^{i}_t = \sigma_i \big( \W_i [\h_{t-1}, \bmo_t ] +\bmb_i \big) \\
\tilde{\bmc}_t &=\tanh \big( \W_c [\h_{t-1}, \bmo_t ] +\bmb_c \big) 
&& \bmc_t = \bmg^f_t \odot \bmc_{t-1} + \bmg^{i}_t \odot \tilde{\bmc}_t   \\
\bmg^o_t &= \sigma_h \big( \W_h [\h_{t-1}, \bmo_t ] + \bmb_h \big) 
&& \h_t =  \bmg^o_t \odot  \tanh(\bmc_t),
\end{aligned}
\label{eq:lstmequations}
\end{equation}
where $\bmg^f_t, \bmg^{i}_t, \bmg^{o}_t \in \R^{d_h}$,
are the gate vector signals (forget, input and output gates),
$\bmo_{t} \in \R^{d_o}$ is the observable input at time $t$,
$\h_{t} \in \R^{d_h}$ is the hidden state,
$\bmc_{t}\in \R^{d_h}$ is the cell state,
while $\W_f$, $\W_i$, $\W_c, \W_h$ $\in \R^{d_h \times (d_h+d_o)}$,
are weight matrices and $\bmb_f, \bmb_i, \bmb_c, \bmb_h \in \R^{d_h}$  biases.
The symbol $\odot$ denotes the element-wise product.
The activation functions $\sigma_f$, $\sigma_i$ and $\sigma_h$ are sigmoids.
For a more detailed explanation of the LSTM cell architecture refer to~\citep{Hochreiter1997}.
The dimension of the hidden state $d_{h}$ (number of hidden units) controls the capacity of the cell to encode history information.
The hidden-to-output functional form $f_h^o$ is given by a linear layer
\begin{gather}
\bm{\hat{o}}_{t+1} = \W_{o}\, \h_t,
\end{gather}
where $\W_{o} \in \mathbb{R}^{d_o \times d_h}$.
The forget gate bias is initialized to one according to~\citep{Jozefowicz2015} to accelerate training.
An illustration of the information flow in a LSTM cell is given in \Cref{fig:lstm}.

\subsection{Gated Recurrent Unit}
\label{sec:gru}

The Gated Recurrent Unit (GRU)~\citep{Kyunghyun2014} was proposed as a variation of LSTM utilizing a similar gating mechanism.
Even though GRU lacks an output gate and thus has fewer parameters, it achieves comparable performance with LSTM in polyphonic music and speech signal datasets~\citep{Chung2014}.
The GRU equations are given by
\begin{equation}
\begin{aligned}
&\bmz_t= \sigma_g \big(\W_z [\h_{t-1}, \bmo_t ] + \bmb_z\big) \\
& \bmr_t= \sigma_g \big(\W_r [\h_{t-1}, \bmo_t ] + \bmb_r\big) \\
 &\bm{\tilde{h}}_t = \tanh \big(\W_h \big[ \bmr_t \odot \h_{t-1}, \bmo_t \big] + \bmb_h\big) \\
 & \h_t = (1 - \bmz_t) \odot \h_{t-1} + \bmz_t \odot \bm{\tilde{h}}_t,
\label{eq:gruequations}
\end{aligned}
\end{equation}
where $\bmo_{t} \in \R^{d_o}$ is the observable at the input at time $t$,
$\bmz_t \in \R^{d_h}$ is the update gate vector,
$\bmr_t \in \R^{d_h}$ is the reset gate vector,
$\bm{\tilde{h}}_t \in \R^{d_h}$,
$\h_t \in \R^{d_h}$ is the hidden state,
$\W_z$, $\W_r$, $\W_h$ $\in \R^{d_h \times (d_h+d_o)}$
are weight matrices and $\bmb_z, \bmb_r, \bmb_h \in \R^{d_h}$ biases.
The gating activation $\sigma_g$ is a sigmoid.
The output $\bm{\hat{o}}_{t+1}$ is given by the linear layer:
\begin{gather}
\bm{\hat{o}}_{t+1} = \W_{o} \, \h_t,
\end{gather}
where $\W_{o} \in \mathbb{R}^{d_o \times d_h}$.
An illustration of the information flow in a GRU cell is given in \Cref{fig:gru}.

\subsection{Unitary Evolution}
\label{sec:unitary}

Unitary RNNs ~\citep{Arjovsky2016,Jing2017}, similar to LSTMs and GRUs, aim to alleviate the vanishing gradients problem of plain RNNs.
Here, instead of employing sophisticated gating mechanisms, the effort is focused on the identification of a re-parametrization of the recurrent weight matrix, such that its spectral radius is a-priori set to one.
This is achieved by optimizing the weights on the subspace of complex unitary matrices.
The architecture of the Unitary RNN is given by
\begin{equation}
\begin{aligned}
\h_{t}&=\operatorname{modReLU} \Big( \W_h \h_{t-1} + \, \W_o \bmo_t \Big) \\
\bm{\hat{o}}_{t+1}&=\W_o \, \Re \big( \h_{t} \big),
\end{aligned}
\end{equation}
where $\W_h \in \mathbb{C}^{d_h \times d_h}$ is the complex unitary \textbf{recurrent} weight matrix, $\W_o \in \mathbb{C}^{d_h \times d_o}$ is the complex \textbf{input} weight matrix,
$\h_t \in \mathbb{C}^{d_h}$ is the complex state vector,
$\Re(\cdot)$ denotes the real part of a complex number, $\W_o \in \mathbb{R}^{d_h \times d_h}$ is the real output matrix,
and the modified ReLU non-linearity $\operatorname{modReLU}$ is given by
\begin{equation}
\begin{aligned}
\Big( \operatorname{modReLU}(\bmz) \Big)_i = \frac{z_i}{|z_i|}
\odot \operatorname{ReLU} (|z_i| + b_i),
\end{aligned}
\end{equation}
where $|z_i|$ is the norm of the complex number $z_i$.
The complex unitary matrix $\W_h$ is parametrized as a product of a diagonal matrix and multiple rotational matrices.
The reparametrization used in this work is the one proposed in~\citep{Jing2017}.
The complex input weight matrix $\W_o \in \mathbb{C}^{d_h \times d_o}$ is initialized with $\W_o^{re}+j \, \W_o^{im}$, with real matrices $\W_o^{re}$, $\W_o^{im}$ $\in \mathbb{R}^{d_h \times d_o}$ whose values are drawn from a random uniform distribution $\mathcal{U}[-0.01, 0.01]$ according to~\citep{Jing2017}. 
An illustration of the information flow in a Unitary RNN cell is given in \Cref{fig:unitary}.

In the original paper of~\citep{Jing2017} the architecture was evaluated on a speech spectrum prediction task, a copying memory task and a pixel permuted MNIST task demonstrating superior performance to LSTM either in terms of final testing accuracy or wall-clock training speed.

\begin{figure*}
\centering
\includegraphics[width=\textwidth]{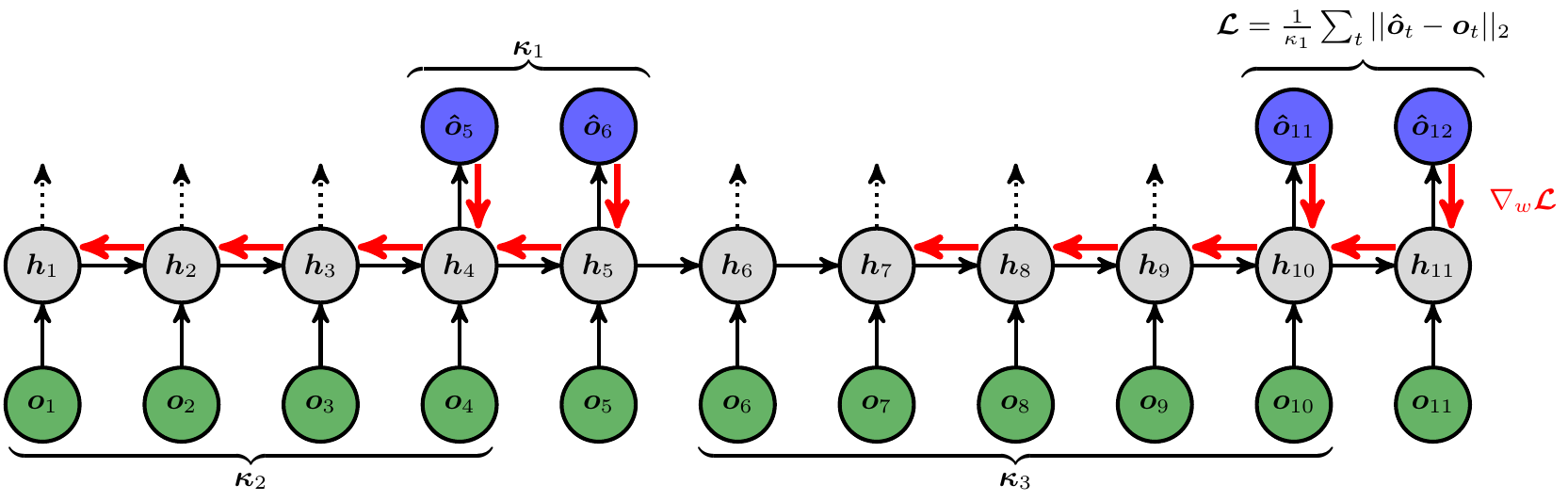}
\caption{
Illustration of an unfolded RNN.
Time-series data $\bm{o}$ are provided at the input of the RNN.
The RNN is forecasting the evolution of the observable at its outputs $\bm{\hat{o}}$.
The average difference (mean square error) between $\kappa_1$ iterative predictions (outputs) of the RNN  $\bm{\hat{o}}$ and the targets $\bm{o}$ from the time-series data is computed every $\kappa_3$ steps.
The gradient of this quantity, illustrated with red arrows, is back-propagated through time (BPTT) for $\kappa_2$ previous temporal time-steps, computing the gradients of the network parameters that are shared at each time layer.
The output of intermediate steps illustrated with dashed lines is ignored.
Stateless models initialize the hidden state before training at a specific fragment of the sequence of size $\kappa_2$ with zero (in this case $\bm{h}_6\hat{=}0$) and cannot capture dependencies longer than $\kappa_2$.
In this way, consecutive training batches (sequence fragments) do not have to be temporally adjacent. 
In stateful models, the hidden state is never set to zero and in order to train at a specific fragment of the sequence, the initial hidden state has to be computed from the previously processed fragment.
In order to eliminate the overlap between fragments, we teacher force the network with ground-truth data for $\kappa_3 \geq \kappa_2$ time-steps.
In our study we pick $\kappa_3=\kappa_2+\kappa_1-1$ as illustrated in the figure.
}
\label{fig:rnn_backprop}
\end{figure*}

\subsection{Back-Propagation Through Time}
\label{sec:training}

Backpropagation dates back to the works of~\citep{Dreyfus1962,Linnainmaa1976,Rumelhart1986}, while its extension to RNNs termed Backpropagation through time (BPTT) was presented in~\citep{Werbos1988,werbos1990}.
A forward pass of the network is required to compute its output and compare it against the label (or target) from the training data based on an error metric (e.g. mean squared loss).
Backpropagation amounts to the computation of the partial derivatives of this loss with respect to the network parameters by iteratively applying the chain rule, transversing backwards the network.
These derivatives are computed analytically with automatic differentiation.
Based on these partial derivatives the network parameters are updated using a first-order optimization method, e.g. stochastic gradient descent. 

The power of BBTT lies in the fact that it can be deployed to learn the partial derivatives of the weights of any network architecture with differentiable activation functions, utilizing state-of-the-art automatic differentiation software, while (as the data are processed in small fragments called batches) it scales to large datasets and networks, and can be accelerated by employing Graphics Processing Units (GPUs).
These factors made backpropagation the workhorse of state-of-the-art deep learning methods~\citep{Goodfellow2016}.

In our study, we utilize BBTT to train the LSTM (\Cref{sec:lstm}), GRU (\Cref{sec:gru}) and Unitary (\Cref{sec:unitary}) RNNs.
There are three key parameters of this training method that can be tuned.
The first hyperparameter $\kappa_1$ is the number of forward-pass timesteps performed to accumulate the error for back-propagation.
The second parameter is the number of previous time steps for the back-propagation of the gradient $\kappa_2$.
This is also denoted as truncation length, or sequence length.
This parameter has to be large enough to capture the temporal dependencies in the data.
However, as $\kappa_2$ becomes larger, training becomes much slower, and may lead to vanishing gradients. 
In the following, we characterize as stateless, models whose hidden state before $\kappa_2$ is hard-coded to zero, i.e.,  $\bm{h}_{-\kappa_2}=0$.
Stateless models cannot learn dependencies that expand in a time horizon larger that $\kappa_2$.
However, in many practical cases stateless models are widely employed assuming that only short-term temporal dependencies exist.
In contrast, stateful models propagate the hidden state $\bm{h}_{-\kappa_2} \neq 0$ between temporally consecutive batches.
In our study, we consider only \textbf{stateful} networks.

Training stateful networks is challenging because the hidden state $\bm{h}_{-\kappa_2}$ has to be available from a previous batch and the network has to be trained to learn temporal dependencies that may span many time-steps in the past.
In order to avoid overlap between two subsequent data fragments and compute $\bm{h}_{-\kappa_2}$ for the next batch update, the network is teacher-forced for $\kappa_3$ time-steps between two consecutive weight updates.
That implies providing ground-truth values at the input and performing forward passing without any back-propagation.
This parameter, has an influence on the training speed, as it determines how often the weights are updated.
We pick $\kappa_3=\kappa_2+\kappa_1 -1$ as illustrated in \Cref{fig:rnn_backprop}, and optimize $\kappa_1$ as a hyperparameter.

The weights of the networks are initialized using the method of Xavier proposed in~\citep{glorot2010understanding}.
We utilize a stochastic optimization method with adaptive learning rate called Adam~\citep{Kingma2015} to update the weights and biases.
We add Zoneout~\citep{Krueger2017} regularization in the recurrent weights and variational dropout~\citep{Gal2016} regularization at the output weights (with the same keep probability) to both GRU and LSTM networks to alleviate over-fitting.
Furthermore, following \citep{Vlachas2018} we add Gaussian noise sampled from $\mathcal{N}(0,\kappa_{n} \sigma)$ to the training data, where $\sigma$ is the standard deviation of the data.
The noise level $\kappa_n$ is tuned.
Moreover, we also vary the number of RNN layers by stacking \textbf{residual} layers~\citep{He2016DeepRL} on top of each other.
These deeper architectures may improve forecasting efficiency by learning more informative embedding at the cost of higher computational times.

In order to train the network on a data sequence of $T$ time-steps, we pass the whole dataset in pieces (batches) for many iterations (epochs).
An epoch is finished when the network has been trained on the whole dataset once.
At the beginning of every epoch we sample uniformly $B=32$ integers from the set $\mathcal{I}= \{1,\dots,T\}$, and remove them from it.
Starting from these indexes we iteratively pass the data through the network till we reach the last (maximum) index in $\mathcal{I}$, training it with BBTT.
Next, we remove all the intermediate indexes we trained on from $\mathcal{I}$.
We repeat this process, until $\mathcal{I}=\emptyset$, proclaiming the end of the epoch.
The batch-size is thus $B=32$.
We experimented with other batch-sizes $B\in \{8,16,64 \}$ without significant improvement in performance of the methods used in this work.

As an additional over-fitting counter-measure we use validation-based early stopping, where $90\%$ of the data is used for training and the rest $10\%$ for validation.
When the validation error stops decreasing for $N_{patience}=20$ consecutive epochs, the training round is over.
We train the network for $N_{rounds}=10$ rounds decreasing the learning rate geometrically by dividing with a factor of ten at each round to avoid tuning the learning rate of Adam.
When all rounds are finished, we pick the model with the lowest validation error among all epochs and rounds.

Preliminary work on tuning the hyperparameters of Adam optimizer apart from the learning rate ($\beta_1$ and $\beta_2$ in the original paper~\citep{Kingma2015}), did not lead to important differences on the results.
For this reason and due to our limited computational budget, we use the default values proposed in the paper~\citep{Kingma2015} ($\beta_1=0.9$ and $\beta_2=0.999$).

Due to the way we train the models (in multiple rounds by decreasing the learning rate when the validation loss saturates and by resetting Adam), we did not notice any important difference on the results.
The results were sensitive only when we had one single round and used  the learning rate as an additional hyperparameter.
Nevertheless, we agree that the hyperparameters of  Adam could be included in our studies but in our case this would have exceeded our  computing time allocations. We have added a remark on this issue on Section 2.4.

\subsection{Reservoir Computing}
\label{sec:rc}

Reservoir Computing (RC) aims to alleviate the difficulty in learning the recurrent connections of RNNs and reduce their training time ~\citep{Lukosevicius2009,Lukosevicius2012}.
RC relies on \textbf{randomly selecting} the recurrent weights such that the hidden state captures the history of the evolution of the observable $\bmo_t$ and train the hidden-to-output weights.
The evolution of the hidden state depends on the random initialization of the recurrent matrix and is driven by the input signal.
The hidden state is termed reservoir state to denote the fact that it captures temporal features of the observed state history.
This technique has been proposed in the context of Echo-State-Networks (ESNs)~\citep{Jaeger2004} and Liquid State Machines with spiking neurons (LSM)~\citep{Maass2002}.

In this work, we consider reservoir computers with $f_h^h$ given by the functional form
\begin{equation}
\h_{t}= \tanh \big(\W_{h,i} \bmo_t + \W_{h,h} \h_{t-1} \big),
\end{equation}
where $\W_{h,i} \in \R^{d_h \times d_o}$, and $\W_{h,h} \in \R^{d_h \times d_h}$.
Other choices of RC architectures are possible, including~\citep{Larger2012,Larger2017,Haynes2015,Antonik2017} 
Following \citep{Jaeger2004}, the entries of $\W_{h,i}$ are uniformly sampled from $[-\omega, \omega]$, where $\omega$ is a hyperparameter.
The reservoir matrix $\W_{h,h}$ has to be selected in a way such that the network satisfies the \say{echo state property}.
This property requires all of the conditional Lyapunov exponents of the evolution of $\h_t$ conditioned on the input (observations $\bmo_t$) to be negative so that, for large $t$, the reservoir state $\h_t$ does not depend on initial conditions.
For this purpose,  $\W_{h,h}$ is set to a large low-degree matrix, scaled appropriately to possess a spectral radius (absolute value of the largest eigenvalue) $\rho$ whose value is a hyperparameter adjusted so that the echo state property holds\footnote{Because of the nonlinearity of the tanh function, $\rho<1$ is not necessarily required for the echo state property to hold true.}.
The effect of the spectral radius on the predictive performance of RC is analyzed in~\citep{Jiang2019}.
Following~\citep{Pathak2018a} the output coupling $f_h^o$ is set to
\begin{equation}
\bm{\hat{o}}_{t+1}=\W_{o,h} \widetilde{\h}_t,
\end{equation}
where the augmented hidden state $\widetilde{\h}_t$ is a $d_h$ dimensional vector such that the $i$\textsuperscript{th} component of $\widetilde{\h}_t$ is $\widetilde{h}_t^i=h_t^i$ for half of the reservoir nodes and $\widetilde{h}_t^i=(h_t^i)^2$ for the other half, enriching the dynamics with the square of the hidden state in half of the nodes.
This was empirically shown to improve forecasting efficiency of RCs in the context of dynamical systems~\citep{Pathak2018a}.
The matrix $\W_{o,h} \in \R^{d_o \times d_h}$ is trained with regularized least-squares regression with Tikhonov regularization to alleviate overfitting~\citep{Tikhonov77, Yan2009} following the same recipe as in~\citep{Pathak2018a}.
The Tikhonov regularization $\eta$ is optimized as a hyperparameter.
Moreover, we further regularize the training procedure of RC by adding Gaussian noise in the training data.
This was shown to be beneficial for both short-term performance and stabilizing the RC in long-term forecasting.
For this reason, we add noise sampled from $\mathcal{N}(0,\kappa_{n} \sigma)$ to the training data, where $\sigma$ is the standard deviation of the data and the noise level $\kappa_n$ a tuned hyperparameter.


%% file: sections/sections-3-comparison-metrics.tex
\section{Comparison Metrics}
\label{sec:comparison}

The predictive performance of the models depends on the selection of model hyperparameters.
For each model we perform an extensive grid search of optimal hyperparameters, reported in the Appendix.
All model evaluations are mapped to a single Nvidia Tesla P100 GPU and are executed on the XC50 compute nodes of the Piz Daint supercomputer at the Swiss national supercomputing centre (CSCS).
In the following we quantify the prediction accuracy of the methods in terms of the normalized root mean square error, given by
\begin{equation}
\text{NRMSE}(\bm{\hat{o}}) = 
\sqrt{
\Big \langle
\frac{(\bm{\hat{o}} - \bm{o})^2}
{ \bm{\sigma}^2}
\Big \rangle
},
\label{eq:nrmse}
\end{equation}
where $\bm{\hat{o}} \in \mathbb{R}^{d_o}$ is the forecast at a single time-step, $\bm{o}  \in \mathbb{R}^{d_o}$ is the target value, and $\bm{\sigma}  \in \mathbb{R}^{d_o}$ is the standard deviation in time of each state component.
In \Cref{eq:nrmse}, the notation $\langle \cdot \rangle$ denotes the state space average (average of all elements of a vector).
To alleviate the dependency on the initial condition, we report the evolution of the NRMSE over time averaged over $100$ initial conditions randomly sampled from the attractor.

Perhaps the most basic characterization of chaotic motion is through the concept of Lyapunov exponents~\citep{ott2002}:
Considering two infinitesimally close initial conditions $U(t=0)$ and $U(t=0)+\delta U(t=0)$, their separation $| \delta U(t) |$ on average diverges exponentially in time, $| \delta U(t) | / | \delta U(t=0) | \sim \exp(\Lambda t)$, as $t \to \infty$.
Note that the dimensionality of the vector displacement $\delta U(t)$ is that of the state space.
In general, the Lyapunov exponent $\Lambda$ depends on the orientation ($\delta U(t) / |\delta U(t)|$) of the vector displacement $\delta U(t)$.
In the $t \to \infty$ limit, the number of possible values of $\Lambda$ is typically equal to the state space dimensionality.
We denote these values $\Lambda_1 \geq \Lambda_2 \geq \Lambda_3 \geq \dots$ and collectively call them the Lyapunov exponent spectrum (LS) of the particular chaotic system.
The Lyapunov exponent spectrum will be evaluated in \Cref{sec:le}.

However, we note that a special role is played by $\Lambda_1$, and only $\Lambda_1$, the largest Lyapunov exponent.
We refer to the largest Lyapunov exponent as the Maximal Lyapunov exponent (MLE).
Chaotic motion of a bounded trajectory is defined by the condition $\Lambda_1 > 0 $.
Importantly, if the orientation of $\delta U(t=0)$ is chosen randomly, the exponential rate at which the orbits separate is $\Lambda_1$ with probability one.
This is because in order for any of the other exponents ($\Lambda_2, \Lambda_3, \dots $) to be realized, $\delta U(t=0)$ must be chosen to lie on a subspace of lower dimensionality than that of the state space;
i.e., the orientation of $\delta U(t=0)$ must be chosen in an absolutely precise way never realized by random choice.
Hence, the rate at which typical pairs of nearby orbits separate is $\Lambda_1$, and $T^{\Lambda_1} = \Lambda_1^{-1}$, the \say{Lyapunov time}, provides a characteristic time scale for judging the quality of predictions based on the observed prediction error growth.
 
In order to obtain a single metric of the predictive performance of the models we compute the valid prediction time (VPT) in terms of the MLE of the system $\Lambda_1$ as
\begin{equation}
\text{VPT} = 
\frac{1}{\Lambda_1}
\underset{t_{f}}{\operatorname{argmax}} \:
\{
t_{f} \: | \:
\text{NRMSE}(\bm{o}_t)< \epsilon, \forall t \leq t_{f}
\}
\label{eq:validpredictiontime}
\end{equation}
which is the largest time $t_f$ the model forecasts the dynamics with a NRMSE error smaller than $\epsilon$ normalized with respect to $\Lambda_1$.
In the following, we set $\epsilon=0.5$.

In order to evaluate the efficiency of the methods in capturing the long-term statistics of the dynamical system, we evaluate the mean power spectral density (power spectrum) of the state evolution over all $i \in \{1, \dots, d_o \}$ elements $\bm{o}^i_t$ of the state (since the state $\bm{o}_t$ is a vector).
The power spectrum of the evolution of $\bm{o}^i_t$ is given by $PSD(f)=20 \log_{10} \Big(2\, | U(f) | \Big)$ dB, where $U(f)=FFT(\bm{o}^i_t)$ is the complex Fourier spectrum of the state evolution.

%% file: sections/sections-4-lorenz96.tex
\section{Forecasting Reduced Order Observable Dynamics in the Lorenz-96}
\label{sec:observable}

The accurate long-term forecasting of the state of a deterministic chaotic dynamical system is challenging as even a minor initial error can be  propagated exponentially in time due to the system dynamics even if the model predictions are perfect.
A characteristic time-scale of this propagation is the Maximal Lyapunov Exponent (MLE) of the system as elaborated in \Cref{sec:comparison}.
In practice, we are often interested in forecasting the evolution of an observable (that we can measure and obtain data from), which does not contain the full state information of the system.
The observable dynamics are more irregular and challenging to model and forecast because of the additional loss of information.

Classical approaches to forecast the observable dynamics based on Takens seminal work~\citep{Takens1981}, rely on reconstructing the full dynamics in a high-dimensional \textbf{phase space}.
The state of the phase space is constructed by stacking delayed versions of the observed state.
Assume that the state of the dynamical system is $\x_t$, but we only have access to the less informative observable $\bmo_t$.
The phase space state, i.e., the \textbf{embedding state}, is given by $\bmz_t=[\bmo_t, \bmo_{t-\tau}, \dots, \bmo_{t-(d-1)\tau}]$, where the time-lag $\tau$ and the embedding dimension $d$ are the embedding parameters.
For $d$ large enough, and in the case of deterministic nonlinear dynamical chaotic systems, there is generally a one-to-one mapping between a point in the phase space and the full state of the system and vice versa.
This implies that the dynamics of the system are deterministically reconstructed in the phase space~\citep{Kantz1997} and that there exists a phase space forecasting rule $\bmz_{t+1}=\mathcal{F}^{\bmz}(\bmz_t)$, and thus an observable forecasting rule $\bm{\hat{o}}_{t+1}=\mathcal{F}^{\bmo}(\bmo_t, \bmo_{t-\tau}, \dots, \bmo_{t-(d-1)\tau} )$.

The recurrent architectures presented in \Cref{sec:methods} fit to this framework, as the embedding state information can be captured in the high-dimensional hidden state $\h_t$ of the networks by processing the observable time series $\bmo_t$, without having to tune the embedding parameters $\tau$ and $d$.

In the following, we introduce a high-dimensional dynamical system, the  Lorenz-96 model and evaluate the efficiency of the methods to forecast the evolution of a reduced order observable of the state of this system.
Here the observable is not the full state of the system, and the networks need to capture temporal dependencies to efficiently forecast the dynamics.

\subsection{Lorenz-96 Model}
\label{sec:observable:lorenz96}

The Lorenz-96 model was introduced by Edward Lorenz~\citep{Lorenz1995} to model the large-scale behavior of the mid-latitude atmosphere. 
The model describes the time evolution of an atmospheric variable that is discretized spatially over a single latitude circle modelled in the high-dimensional state $\x=[\x_0,\dots,\x_{J-1}]$, and is defined by the equations
\begin{equation}
\frac{d \x_j}{dt} = (\x_{j+1}-\x_{j-2})\x_{j-1}-\x_j+F,
\label{eq:Lorenz96}
\end{equation}
for $j \in \{0,1,\dots,J-1 \}$, where we assume periodic boundary conditions $\x_{-1} = \x_{J-1}$, $\x_{-2} = \x_{J-2}$.
In the following we consider a grid-size $J = 40$ and two different forcing regimes, $F=8$ and $F=10$.

We solve \Cref{eq:Lorenz96} starting from a random initial condition with a Fourth Order Runge-Kutta scheme and a time-step of $\delta t=0.01$.
We run the solver up to $T=2000$ after ensuring that transient effects are discarded ($T_{trans}=1000$).
The first half $10^5$ samples are used for training and the rest for testing.
For the forecasting test in the reduced order space, we construct observables of dimension $\bm{d}_o \in \{35,40\}$ by performing Singular Value Decomposition (SVD) and keeping the most energetic $d_o$ components.
The complete procedure is described in the Appendix.
The $35$ most energetic modes taken into account in the reduced order observable, explain approximately $98 \%$ of the total energy of the system in both $F \in \{8,10 \}$.

As a reference timescale that characterizes the chaoticity of the system we use the Lyapunov time, which is the inverse of the MLE, i.e., $T^{\Lambda_1}=1/\Lambda_1$.
The Lyapunov spectrum of the Lorenz-96 system is calculated using a standard technique based on QR decomposition~\citep{Abarbanel2012}.
This leads to $\Lambda_1 \approx 1.68$ for $F=8$ and $\Lambda_1 \approx 2.27$ for $F=10$.

\subsection{Results on the Lorenz-96 Model}
\label{sec:observable:lorenz96:results}

\begin{figure}
\begin{subfigure}[t]{0.475\textwidth}
\centering
\includegraphics[height=4.5cm]{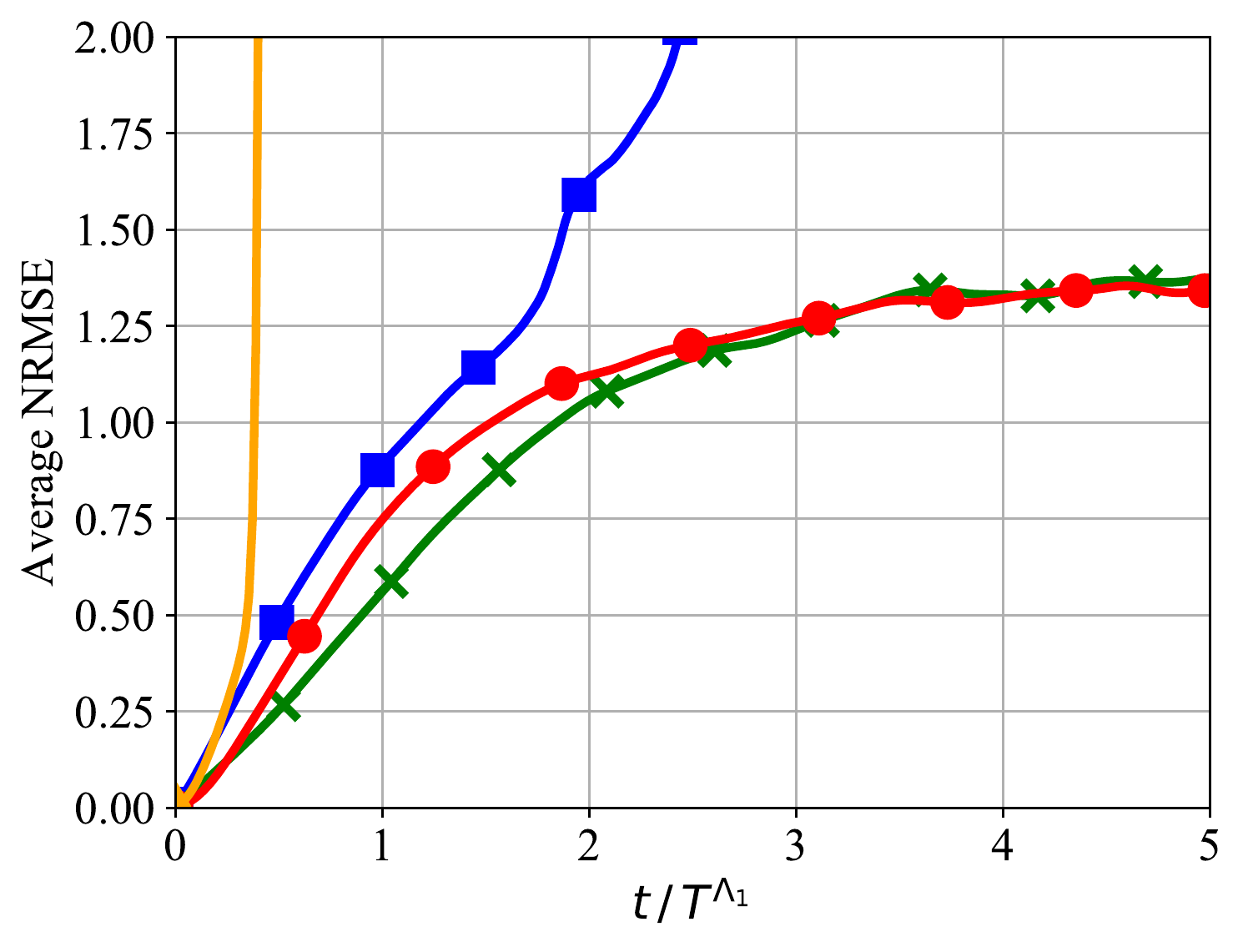}
\caption{Reduced order observable ($\bm{d_o=35}$), $F=8$}
\label{fig:L96F8GP40R40:RMNSE_BBO_NAP_2_M_RDIM_35_5000_2000}
\end{subfigure}
\hfill
\begin{subfigure}[t]{0.475\textwidth}
\centering
\includegraphics[height=4.5cm]{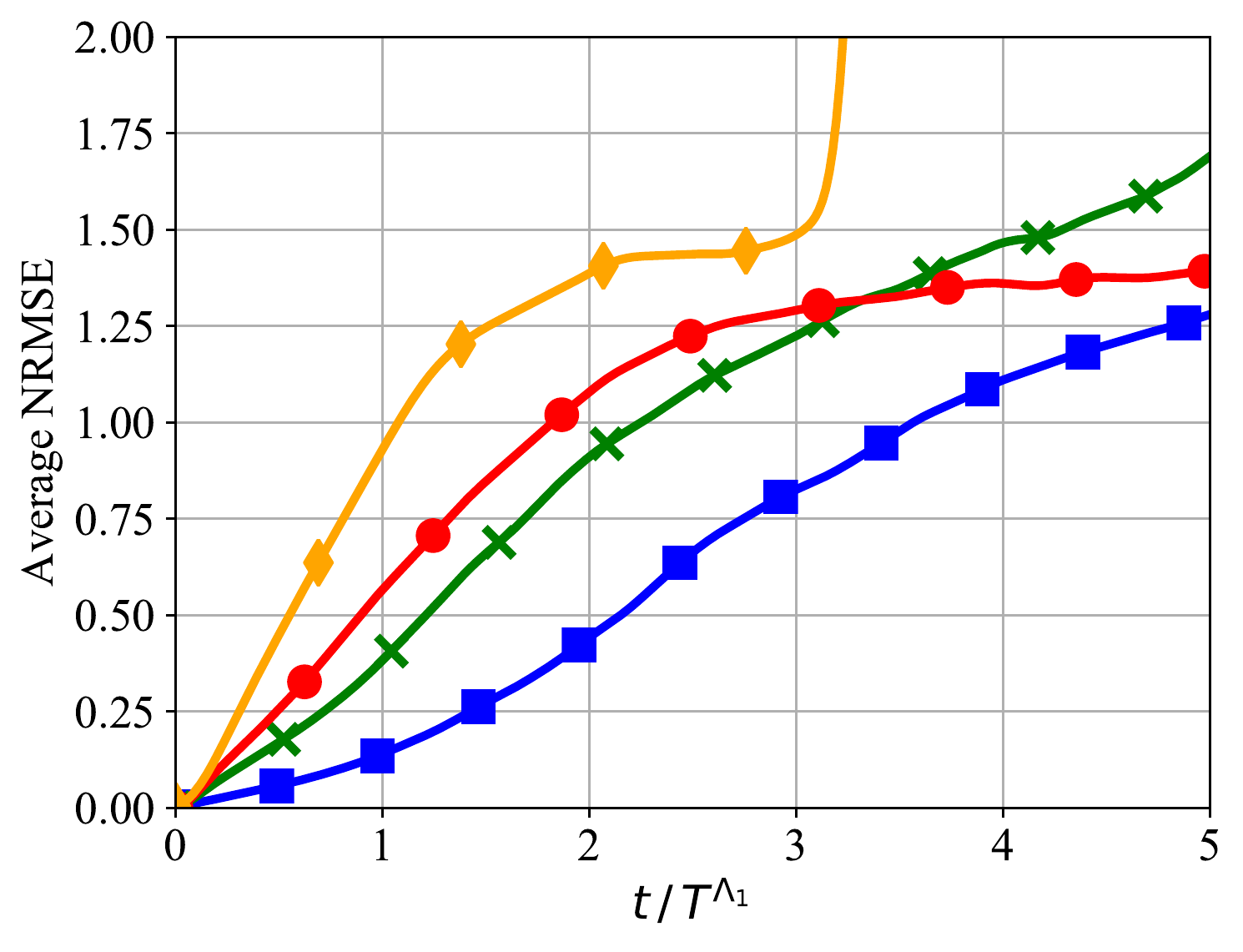}
\caption{Full state ($\bm{d_o=40}$), $F=8$}
\label{fig:L96F8GP40R40:RMNSE_BBO_NAP_2_M_RDIM_40_5000_2000}
\end{subfigure}
\begin{subfigure}[t]{0.475\textwidth}
\centering
\includegraphics[height=4.5cm]{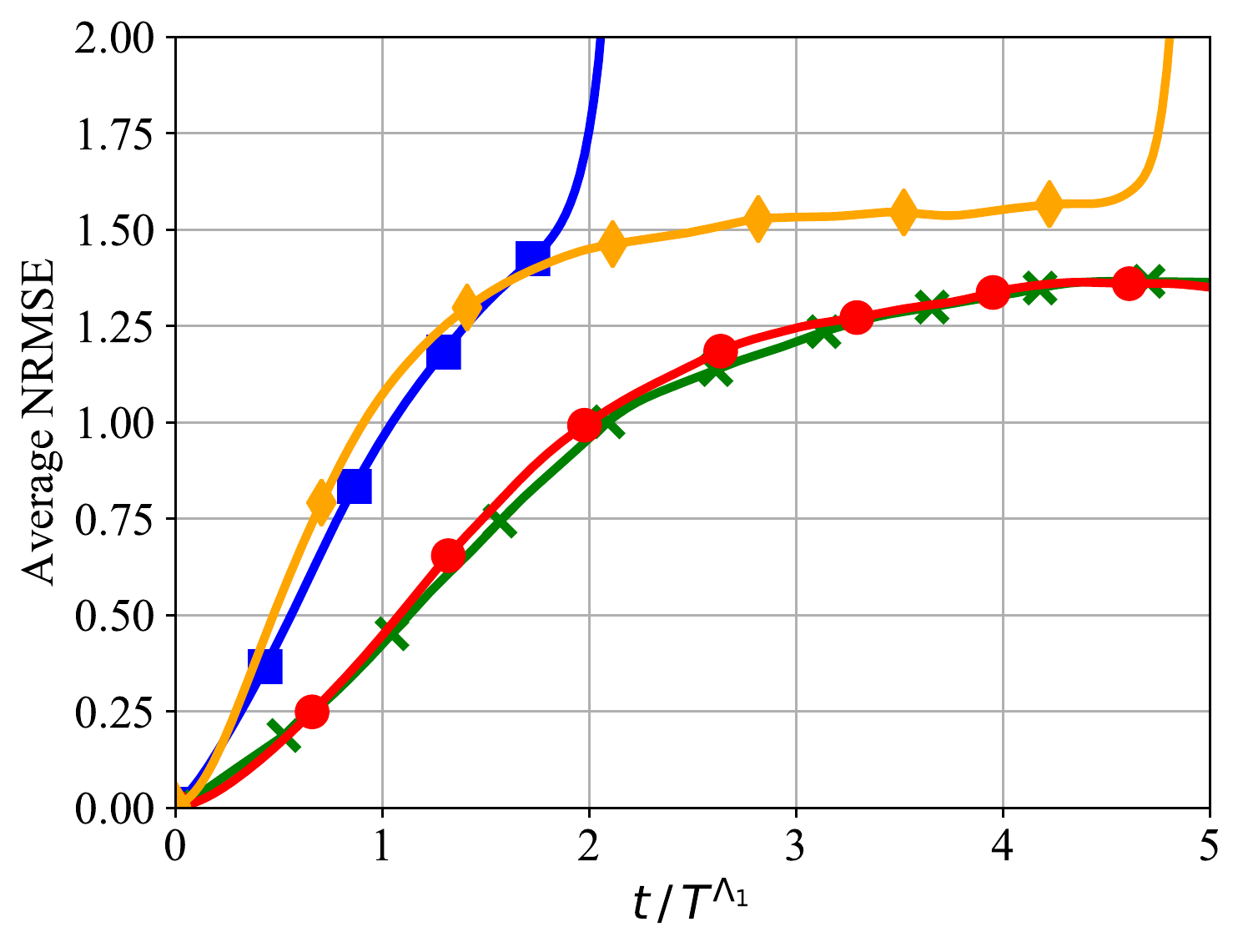}
\caption{Reduced order observable ($\bm{d_o=35}$), $F=10$}
\label{fig:L96F10GP40R40:RMNSE_BBO_NAP_2_M_RDIM_35_5000_2000}
\end{subfigure}
\hfill
\begin{subfigure}[t]{0.475\textwidth}
\centering
\includegraphics[height=4.5cm]{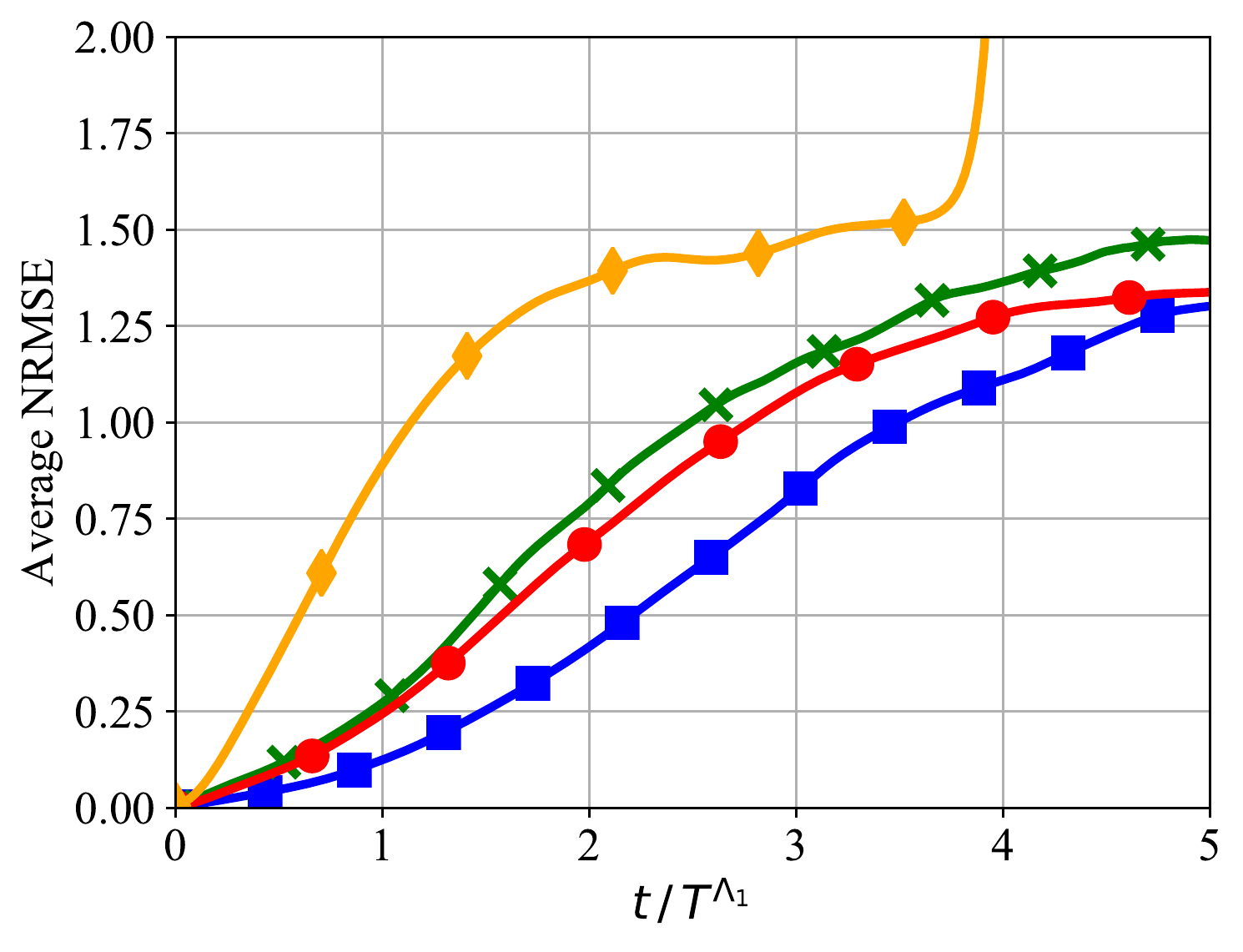}
\caption{Full state ($\bm{d_o=40}$), $F=10$}
\label{fig:L96F10GP40R40:RMNSE_BBO_NAP_2_M_RDIM_40_5000_2000}
\end{subfigure}
\caption{The evolution of the NRMSE error (average over $100$ initial conditions) of the model with the highest VPT for each architecture in the Lorenz-96 with $F\in \{8,10\}$ and $d_o \in \{35,40\}$.
Reservoir computers show remarkable predictive capabilities when the full state is observed, surpassing all other models (plots \textbf{(b)} and \textbf{(d)}).
Predictions of Unitary networks diverge from the attractor in all scenarios, while iterative forecasts of RC suffer from instabilities when only partial information of a reduced order observable is available.
In contrast, GRU and LSTM show stable behavior and superior performance in the reduced order scenario (plots \textbf{(a)} and \textbf{(c)}).
\\
RC \protect \bluelineRectangle;
GRU \protect \greenlineX;
LSTM \protect \redlineCircle;
Unit \protect \orangelineDiamond;
}\label{fig:L96F8GP40R40:RMNSE_BBO_NAP_2_MODELS}
\end{figure}

The evolution of the NRMSE of the model with the largest VPT of each architecture for $F\in \{8, 10\}$ is plotted in \Cref{fig:L96F8GP40R40:RMNSE_BBO_NAP_2_MODELS} for two values of the dimension of the observable $d_o \in \{35,40 \}$, where $d_o=40$ corresponds to full state information.
Note that the observable is given by first transforming the state to its SVD modes and then keeping the $d_o$ most energetic ones.
As indicated by the slopes of the curves, models predicting the observable containing full state information ($d_o=40$) exhibit a slightly slower NRMSE increase compared to models predicting in the reduced order state, as expected.

When the full state of the system is observed, the predictive performance of RC is superior to that of all other models.
Unitary networks diverge from the attractor in both reduced order and full space in both forcing regimes $F \in \{8,10\}$.
This divergence (inability to reproduce the long-term climate of the dynamics) stems from the iterative propagation of the forecasting error.
The issue has been also demonstrated in previous studies in both RC~\citep{Pathak2018b,lu2018} and RNNs~\citep{Vlachas2018}.
This is because the accuracy of the network for long-term climate modeling, depends not only on how well it approximates the dynamics on the attractor locally, but also on how it behaves near the attractor, where we do not have data.
As noted in Ref.~\citep{lu2018}, assuming that the network has a full Lyapunov spectrum near the attractor, if any of the Lyapunov exponents that correspond to infinitesimal perturbations transverse to the attractor phase space is positive, then the predictions of the network will eventually diverge from the attractor.
Empirically, the divergence effect can also be attributed to insufficient network size (model expressiveness) and training, or attractor regions in the state space that are underrepresented in the training data (poor sampling).
Even with a densely sampled attractor, during iterative forecasting in the test data, the model is propagating its own predictions, which might lead to a region near (but not on) the attractor where any positive Lyapunov exponent corresponding to infinitesimal perturbations transverse to the attractor will cause divergence.

In this work, we use $10^5$ samples to densely capture the attractor.
Still, RC suffers from the iterative propagation of errors leading to divergence especially in the reduced order forecasting scenario.
In order to alleviate the problem, a parallel scheme for RC is proposed in~\citep{Pathak2018b} that enables training of many reservoirs locally forecasting the state.
However, this method is limited to systems with local interactions in their state space.
In the case we discuss here the observable obtained by singular value decomposition does not fulfill this assumption.
In many systems the assumption of local interaction may not hold. 
GRU and LSTM show superior forecasting performance in the reduced order scenario setting in Lorenz-96 as depicted in \Cref{fig:L96F8GP40R40:RMNSE_BBO_NAP_2_M_RDIM_35_5000_2000}-\Cref{fig:L96F10GP40R40:RMNSE_BBO_NAP_2_M_RDIM_35_5000_2000}.
Especially in the case of $F=10$, the LSTM and GRU models are able to predict up to $2$ Lyapunov times ahead before reaching an NRMSE of $\epsilon=1$, compared to RC and Unitary RNNs that reach this error threshold in $1$ Lyapunov time.
However, it should be noted that the predictive utility of all models (considering an error threshold of $\epsilon=0.5$) is limited to one Lyapunov time when applied to reduced order data and up to two Lyapunov times in the full state.

\begin{figure}
\begin{subfigure}[t]{0.45\textwidth}
\centering
\includegraphics[width=0.9\textwidth]{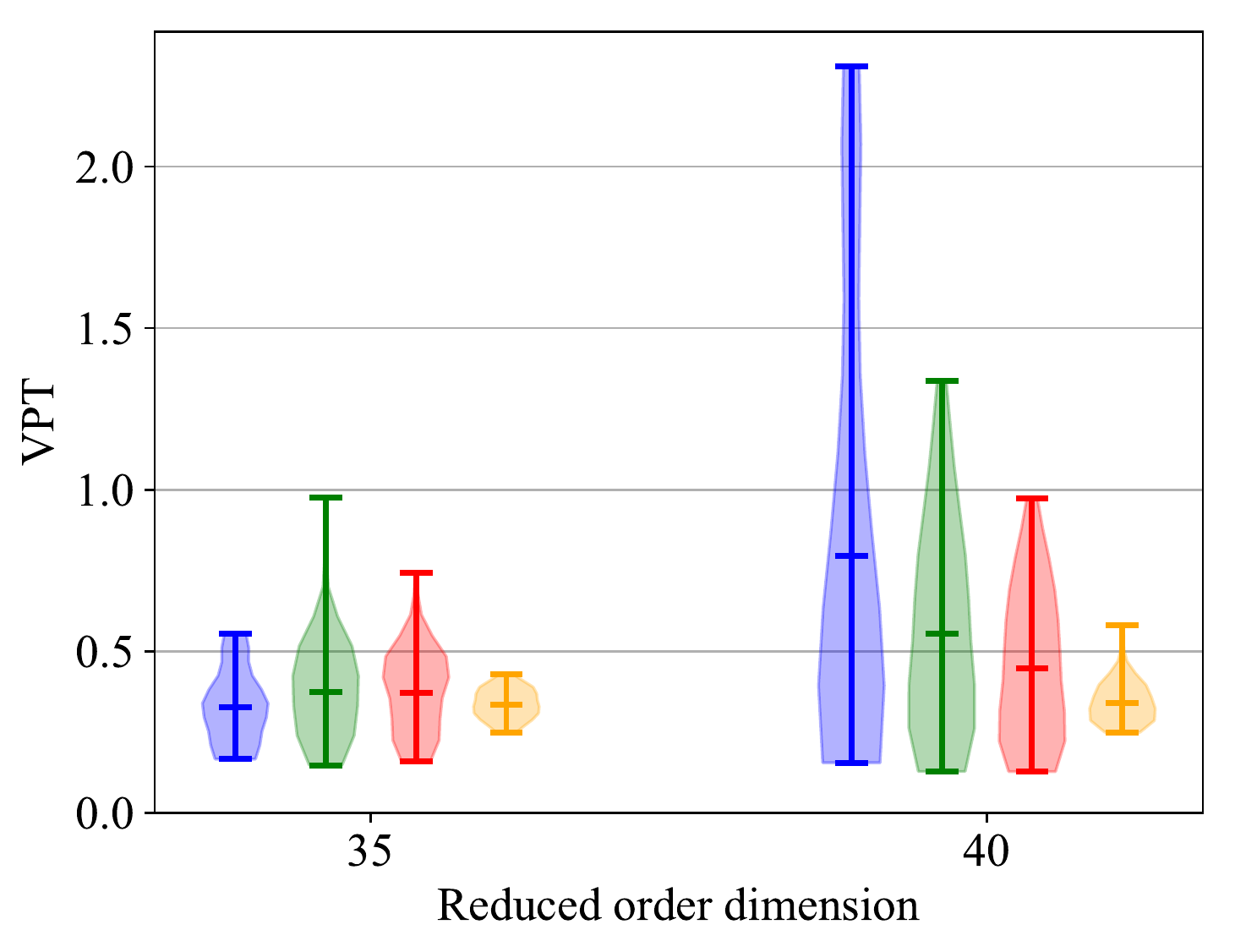}
\caption{Forcing regime $F=8$}
\end{subfigure}
\hfill
\begin{subfigure}[t]{0.45\textwidth}
\centering
\includegraphics[width=0.9\textwidth]{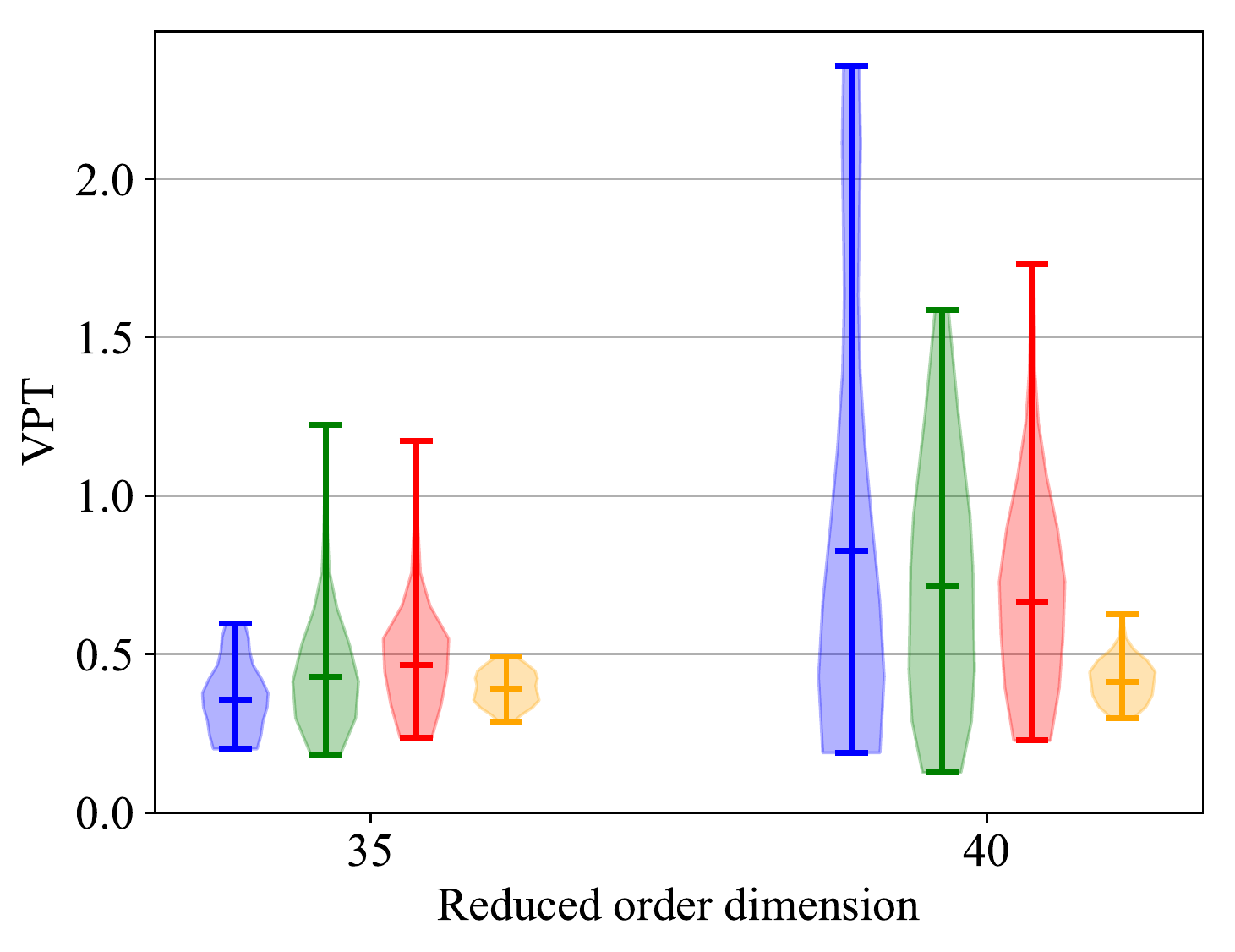}
\caption{Forcing regime $F=10$}
\end{subfigure}
\caption{
Violin plot showing the probability density of the VPT of all hyperparameter sets for each model for observable dimension $d_o=35$ and $d_o=40$ and forcing regimes \textbf{(a)} $F=8$ and \textbf{(b)} $F=10$ in the Lorenz-96.
\\
RC \protect \violinBlueRectangle;
GRU \protect \violinGreenRectangle;
LSTM \protect \violinRedRectangle;
Unit \protect \violinOrangeRectangle;
}
\label{fig:L96F8GP40R40:VPT_BBO_NAP_RDIM_MDLS_LEGEND}
\end{figure}

In order to analyze the sensitivity of the VPT to the hyperparameter selection, we present violin plots in \Cref{fig:L96F8GP40R40:VPT_BBO_NAP_RDIM_MDLS_LEGEND}, showing a smoothed kernel density estimate of the VPT values of all tested hyperparameter sets for $d_o=35$ and $d_o=45$ and $F=8$ and $F=10$.
The horizontal markers denote the maximum, average and minimum value.
Quantitative results for both $F \in \{8,10\}$ are provided on \Cref{tab:Lorenz96reducedresults}.

In the full state scenario ($d_o=40$) and forcing regime $F=8$, RC shows a remarkable performance with a maximum VPT $\approx 2.31$, while GRU exhibits a max VPT of $\approx1.34$.
The LSTM has a max VPT of $\approx 0.97$, while Unitary RNNs show the lowest forecasting ability with a max VPT of $\approx 0.58$.
From the violin plots in \Cref{fig:L96F8GP40R40:VPT_BBO_NAP_RDIM_MDLS_LEGEND} we notice that densities are wider at the lower part,  corresponding to many models (hyperparameter sets) having much lower VPT than the maximum, emphasizing the importance of tuning the hyperparameters.
Similar results are obtained for the forcing regime $F=10$.
One noticeable difference is that the LSTM exhibits a max VPT of $\approx 1.73$ which is higher than that of GRU which is $\approx1.59$.
Still, the VPT of RC in the full state scenario is $\approx2.35$ which is the highest among all models.

\begin{table*}
\caption{Maximum and average Valid Prediction Time (VPT) over all hyperparameter sets averaged over $100$ initial conditions sampled from the testing data for each model.}
\label{tab:Lorenz96reducedresults}
\includegraphics[width=0.6\textwidth]{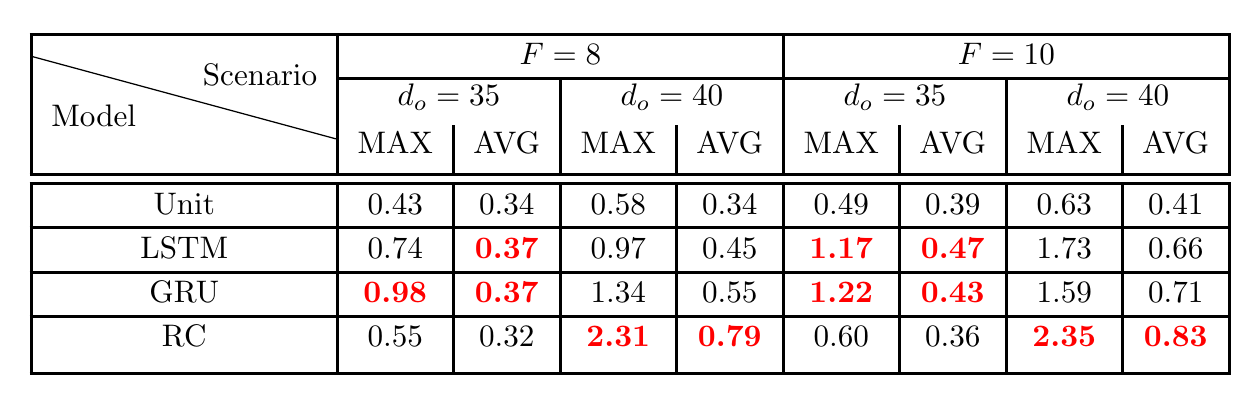}
\end{table*}

In contrast, in the case of $d_o=35$ where the models are forecasting on the reduced order space in the forcing regime $F=8$, GRU is superior to all other models with a maximum VPT $\approx 0.98$ compared to LSTM showing a max VPT $\approx 0.74$.
LSTM shows inferior performance to GRU which we speculate may be due to insufficient hyperparameter optimization.
Observing the results on $F=10$ justifies our claim, as indeed both the GRU and the LSTM show the highest VPT values of $\approx 1.22$ and $\approx 1.17$ respectively.
In both scenarios $F=8$ and $F=10$, when forecasting the reduced order space $d_o=35$, RC shows inferior performance compared to both GRU and LSTM networks with max VPT$\approx 0.55$ for $F=8$ and $\approx 0.60$ for $F=10$.
Last but not least, we observe that Unitary RNNs show the lowest forecasting ability among all models.
This may not be attributed to the expressiveness of Unitary networks, but rather to the difficulty on identifying the right hyperparameters~\citep{greff2016lstm}.
Note from the violin plots in \Cref{fig:L96F8GP40R40:VPT_BBO_NAP_RDIM_MDLS_LEGEND} that the violin plots in the reduced order state are much thinner at the top compared to the ones in the full state.
These results show that hyperparameter sets that achieve a high VPT in the reduced order space are more rare compared to the full state space.
This emphasizes that forecasting on the reduced order state is a more difficult task compared to the full state scenario and thus, identification of hyperparameters is more challenging.

In the following, we evaluate the ability of the trained networks to forecast the long-term statistics of the dynamical system.
In almost all scenarios and all cases considered in this work, forecasts of Unitary RNN networks fail to remain close to the attractor and diverge.
For this reason, we omit the results on these networks. 
 
\begin{figure*}
\hfill
\begin{subfigure}[t]{0.24\textwidth}
\centering
\includegraphics[width=\textwidth]{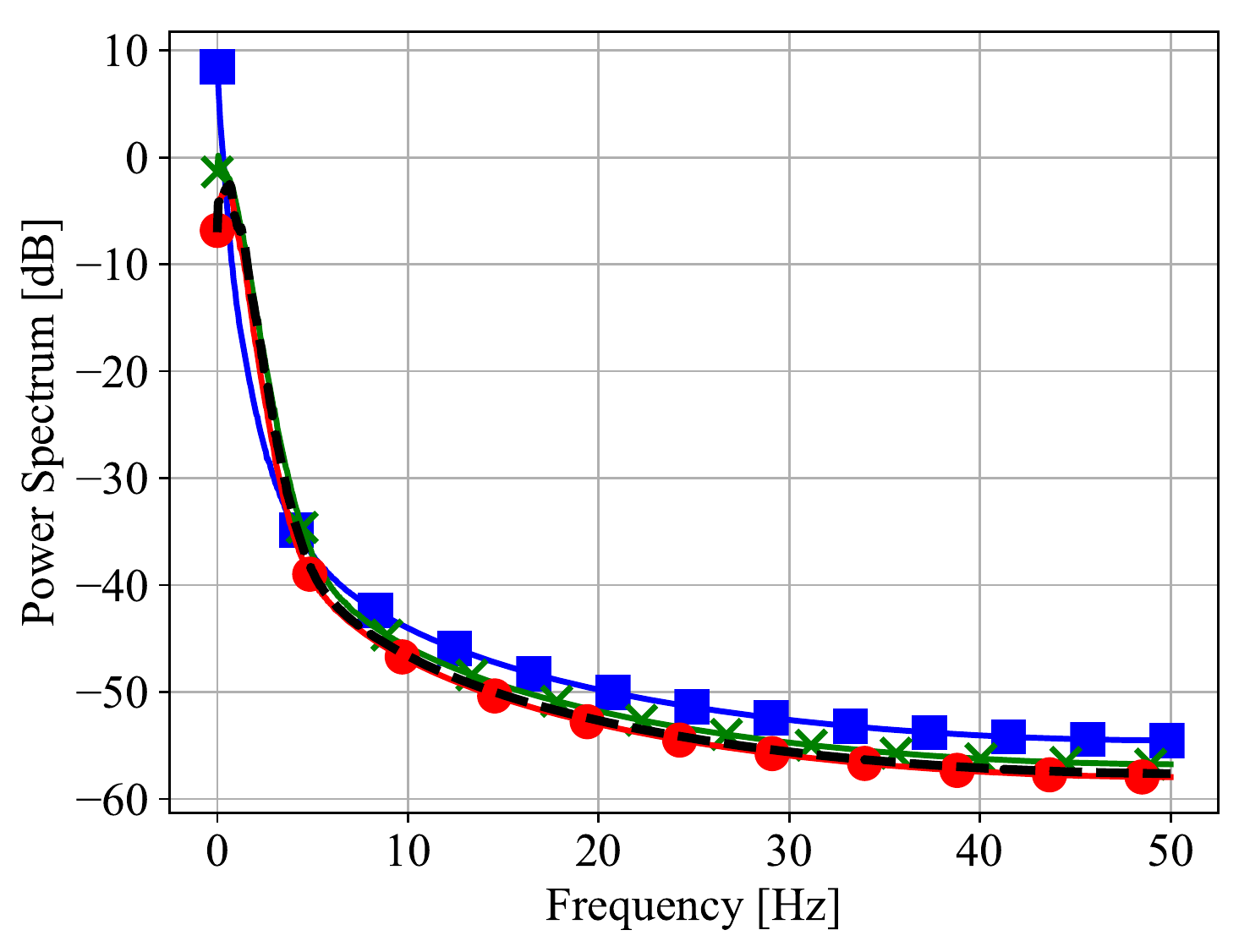}
\caption{
$d_o=35$, $F=8$.
}
\label{fig:L96F8GP40R40:POWSPEC_BBO_FREQERROR_TEST_M_RDIM_35_0}
\end{subfigure}
\hfill
\begin{subfigure}[t]{0.24\textwidth}
\centering
\includegraphics[width=\textwidth]{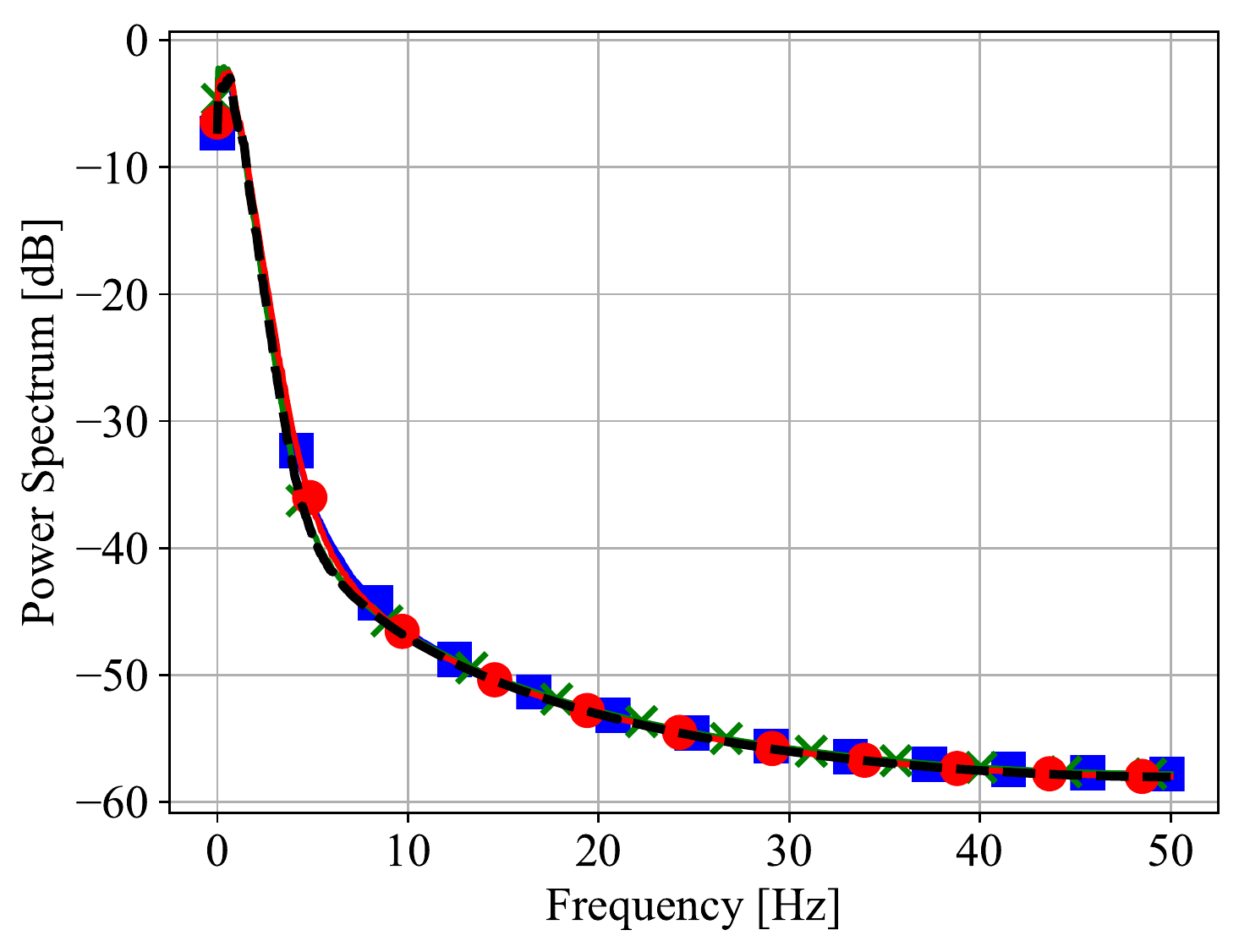}
\caption{
$d_o=40$, $F=8$.
}
\label{fig:L96F8GP40R40:POWSPEC_BBO_FREQERROR_TEST_M_RDIM_40_0}
\end{subfigure}
\hfill
\begin{subfigure}[t]{0.24\textwidth}
\centering
\includegraphics[width=\textwidth]{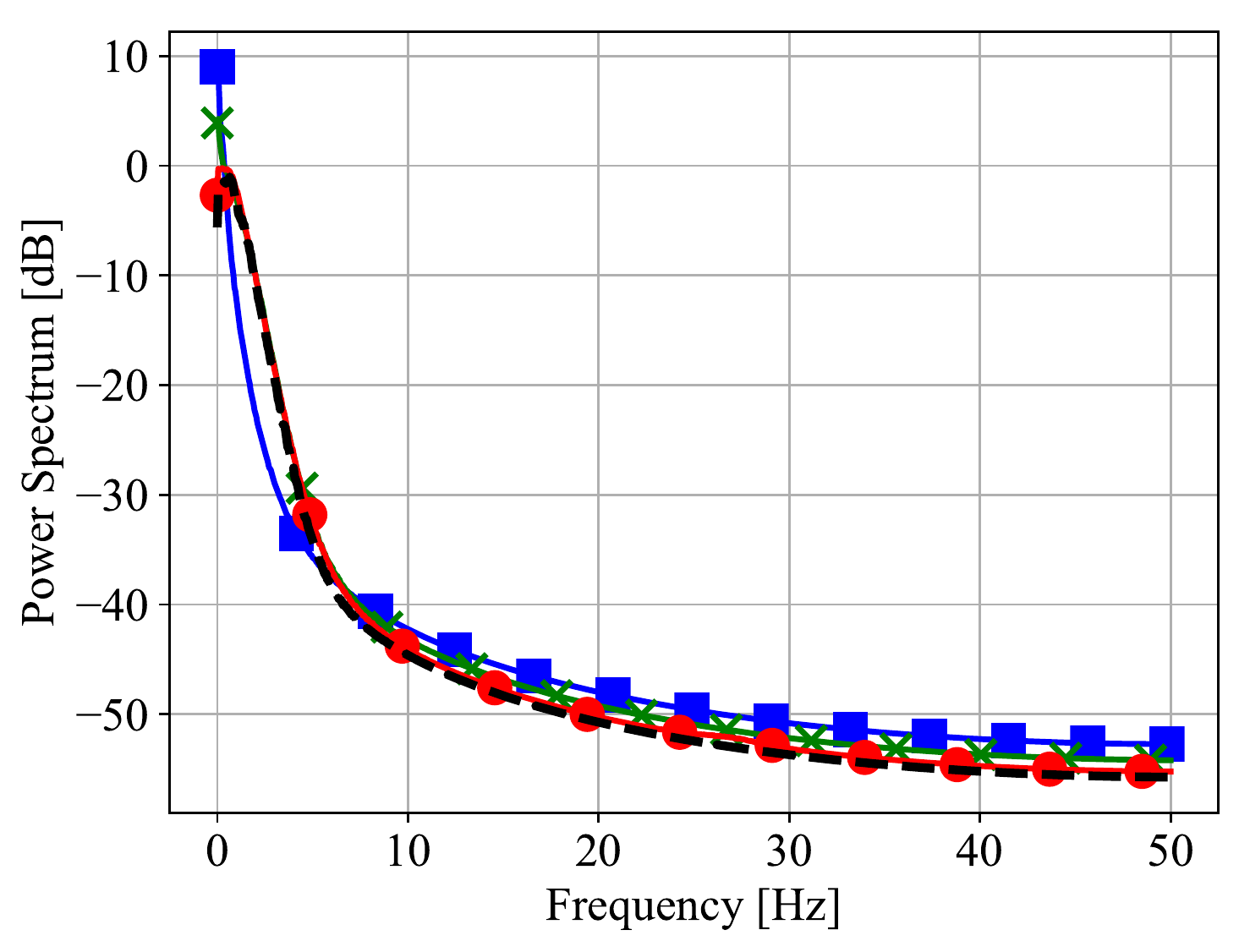}
\caption{
$d_o=35$, $F=10$.
}
\label{fig:L96F10GP40R40:POWSPEC_BBO_FREQERROR_TEST_M_RDIM_35_0}
\end{subfigure}
\hfill
\begin{subfigure}[t]{0.24\textwidth}
\centering
\includegraphics[width=\textwidth]{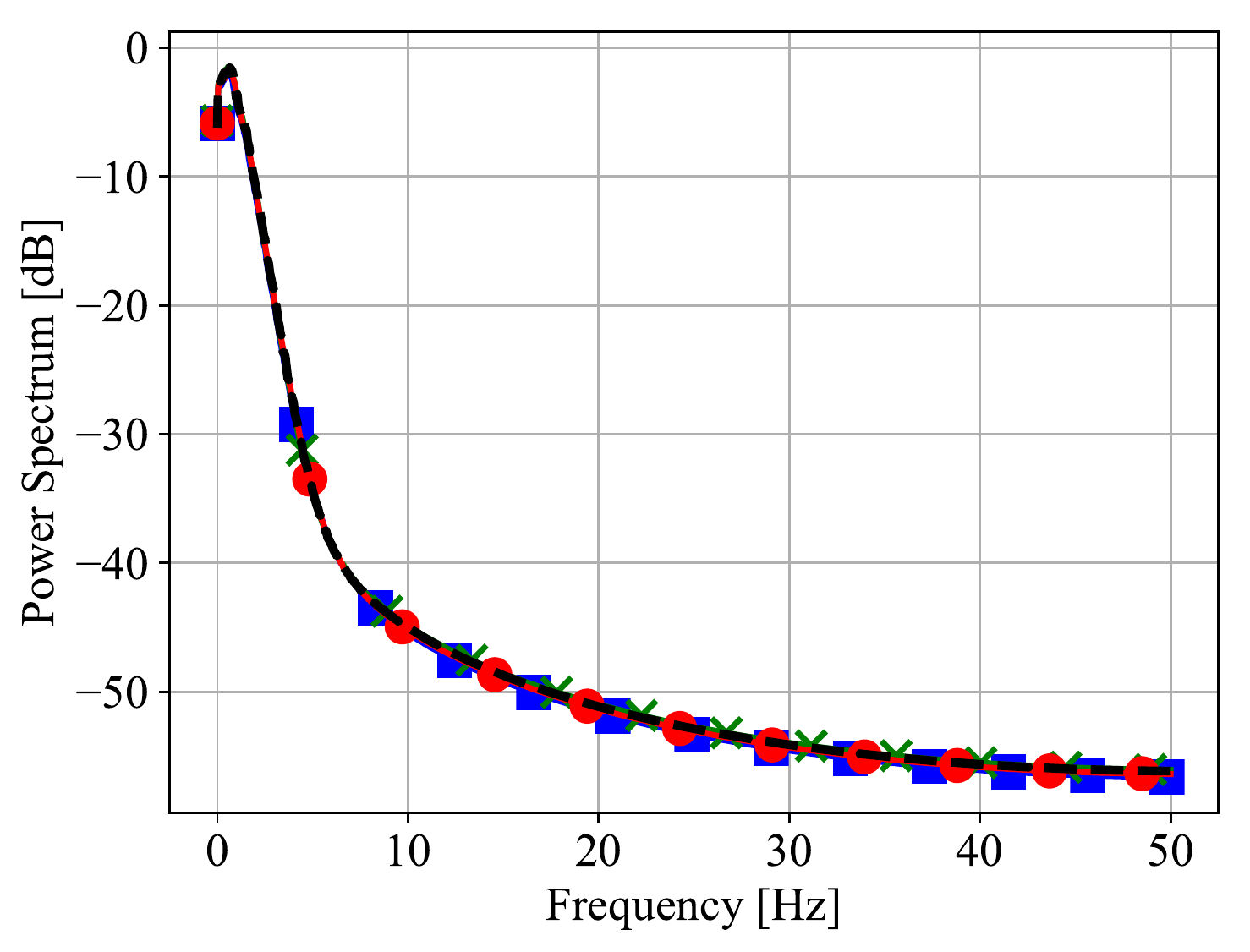}
\caption{
$d_o=40$, $F=10$.
}
\label{fig:L96F10GP40R40:POWSPEC_BBO_FREQERROR_TEST_M_RDIM_40_0}
\end{subfigure}
\caption{
Predicted power spectrum of the RC, GRU, and LSTM networks with the lowest spectrum error forecasting the dynamics of an observable consisting of the SVD modes of the Lorenz-96 system with forcing $F \in \{8,10\}$.
The observable consists of the $d_o=35$ most energetic modes or full state information $d_o=40$.
\textbf{(a)} Reduced order observable at forcing $F=8$.
\textbf{(b)} Full state observable at forcing $F=8$.
\textbf{(c)} Reduced order observable at forcing $F=10$.
\textbf{(d)} Full state observable at forcing $F=10$.
\\
RC \protect \bluelineRectangle;
GRU \protect \greenlineX;
LSTM \protect \redlineCircle;
Groundtruth \protect \blacklineDashed;
}
\label{fig:L96F8_10GP40R40powerspectrum}
\end{figure*}

We quantify the long-term behavior in terms of the power spectrum of the predicted dynamics and its difference with the true spectrum of the testing data.
In \Cref{fig:L96F8_10GP40R40powerspectrum}, we plot the power spectrum of the predicted dynamics from the model (hyperparameter set) with the lowest power spectrum error for each architecture for $d_o \in \{35, 40 \}$ and $F \in \{8, 10 \}$ against the ground-truth spectrum computed from the testing data  (dashed black line).
In the full state scenario in both forcing regimes (\Cref{fig:L96F8GP40R40:POWSPEC_BBO_FREQERROR_TEST_M_RDIM_40_0}, \Cref{fig:L96F10GP40R40:POWSPEC_BBO_FREQERROR_TEST_M_RDIM_40_0}), all models match the true statistics in the test dataset, as the predicted power spectra match the ground-truth.
These results imply that RC is a powerful predictive tool in the full order state scenario, as RC models both capture the long-term statistics and have the highest VPT among all other models analyzed in this work.
However, in the case of a reduced order observable, the RC cannot match the statistics.
In contrast, GRU and LSTM networks achieve superior forecasting performance while matching the long-term statistics, even at this challenging setting of a chaotic system with reduced order information.

\begin{figure*}
\begin{subfigure}[t]{0.32\textwidth}
\includegraphics[width=.95\textwidth]{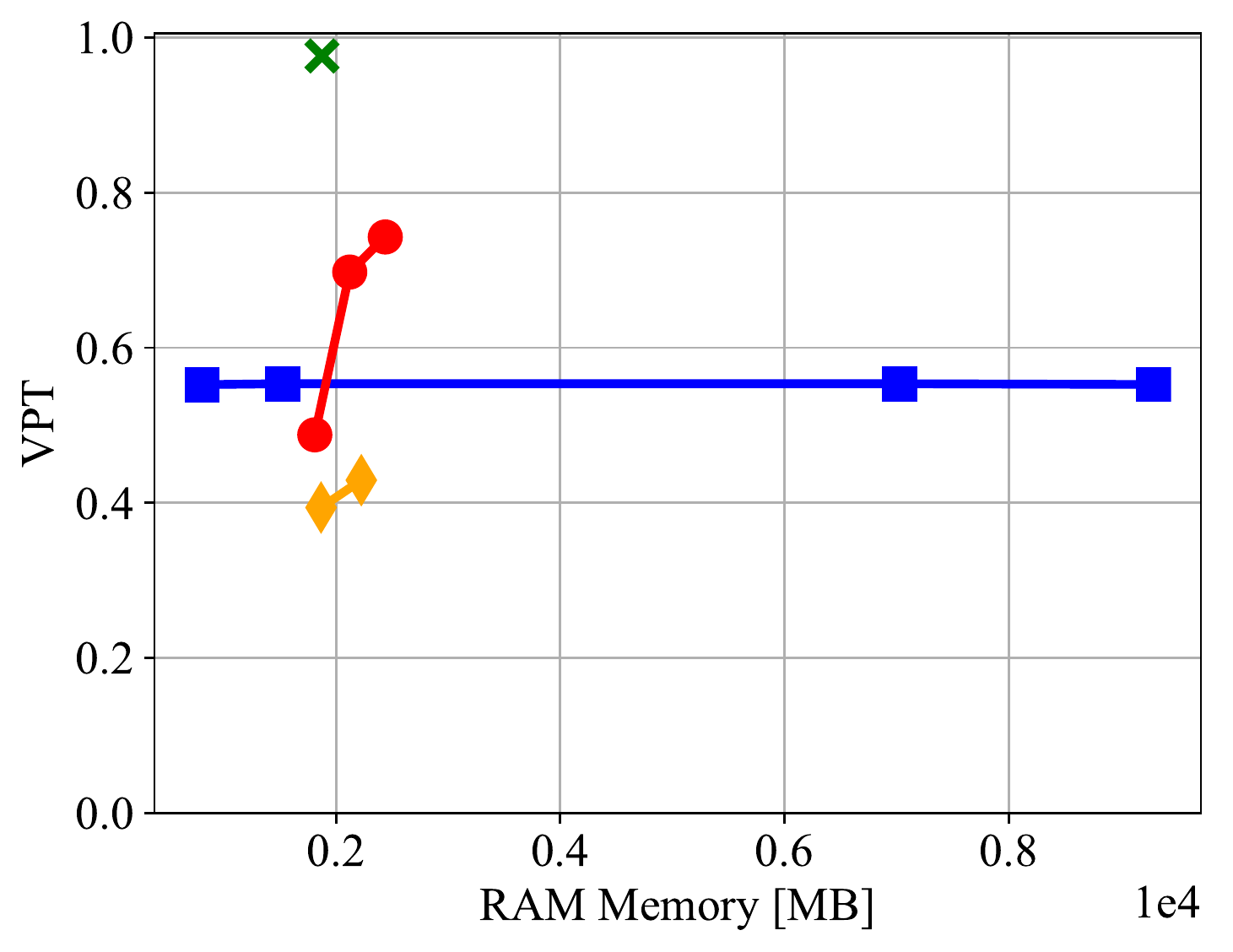}
\caption{
VPT w.r.t. RAM memory for $d_o=35$.
}
\label{fig:L96F8GP40R40:NAP_2_RAM_RDIM35}
\end{subfigure}
\hfill
\begin{subfigure}[t]{0.32\textwidth}
\includegraphics[width=.95\textwidth]{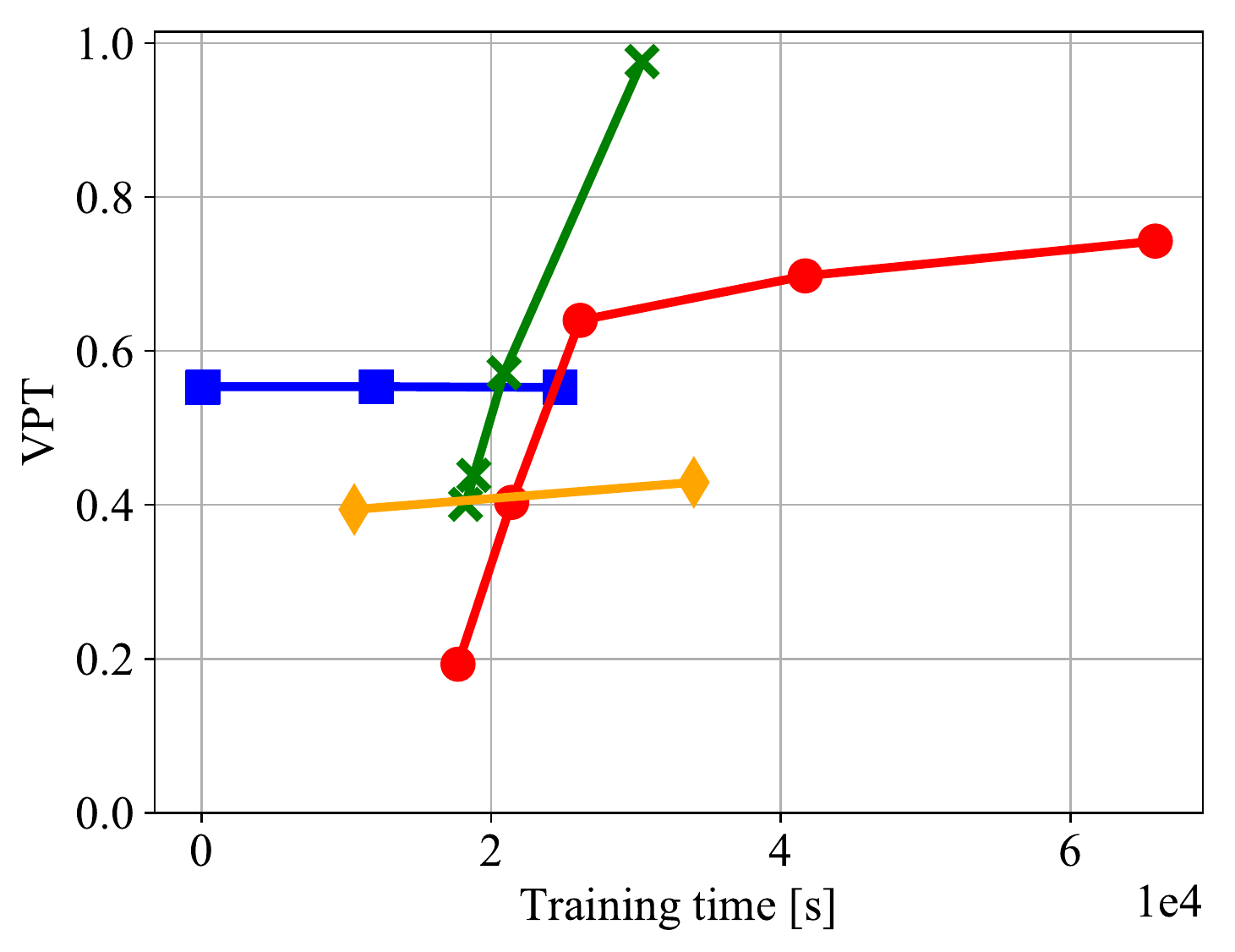}
\caption{
VPT w.r.t. the total training time for $d_o=35$.
}
\label{fig:L96F8GP40R40:NAP_2_TRAINTIME_RDIM35}
\end{subfigure}
\hfill
\begin{subfigure}[t]{0.32\textwidth}
\includegraphics[width=.95\textwidth]{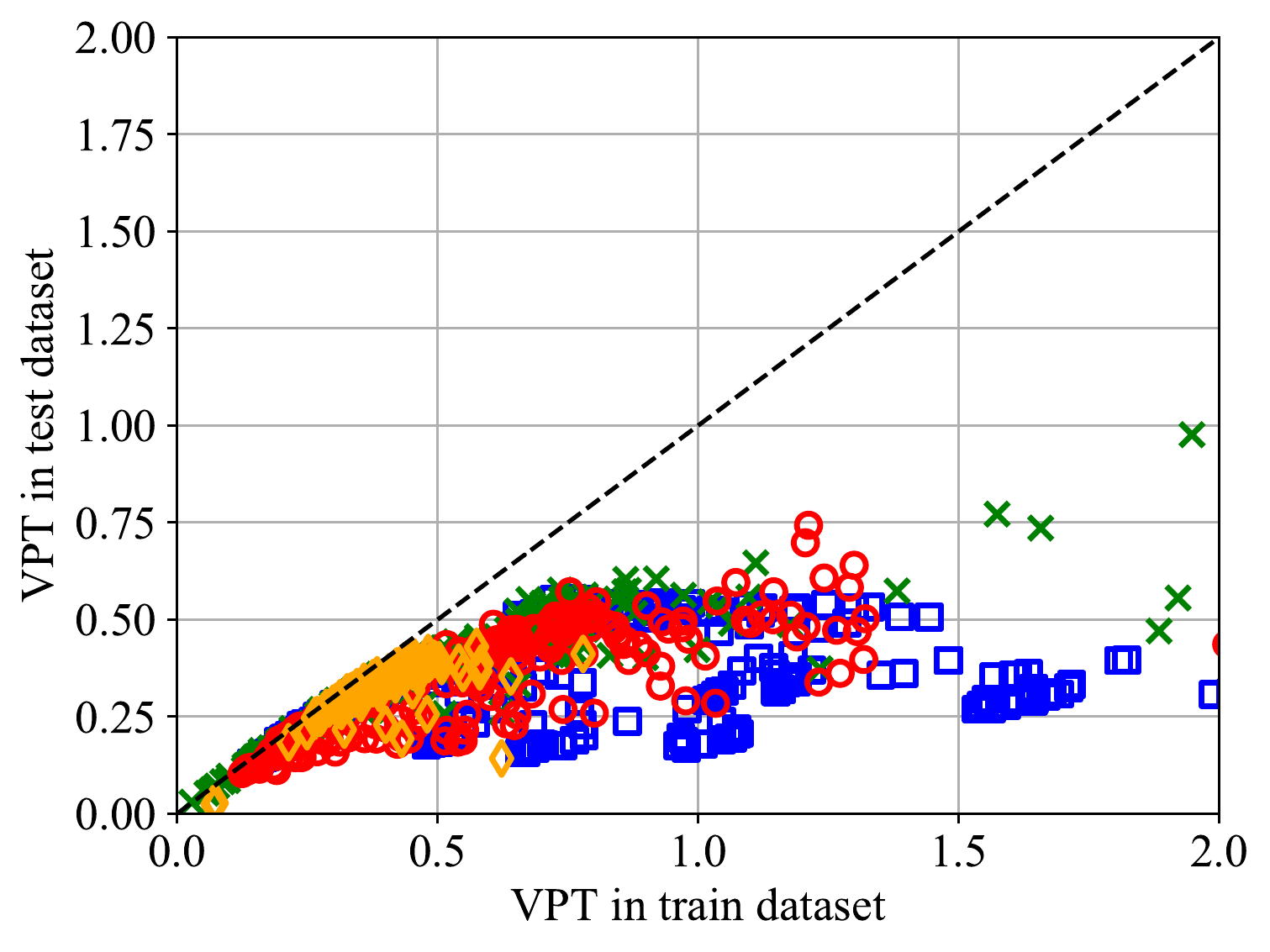}
\caption{
VPT in test data w.r.t. VPT in the train data for $d_o=35$.
}
\label{fig:L96F8GP40R40:OFSP_BBO_NAP_RDIM_35}
\end{subfigure}
\hfill
\begin{subfigure}[t]{0.32\textwidth}
\includegraphics[width=.95\textwidth]{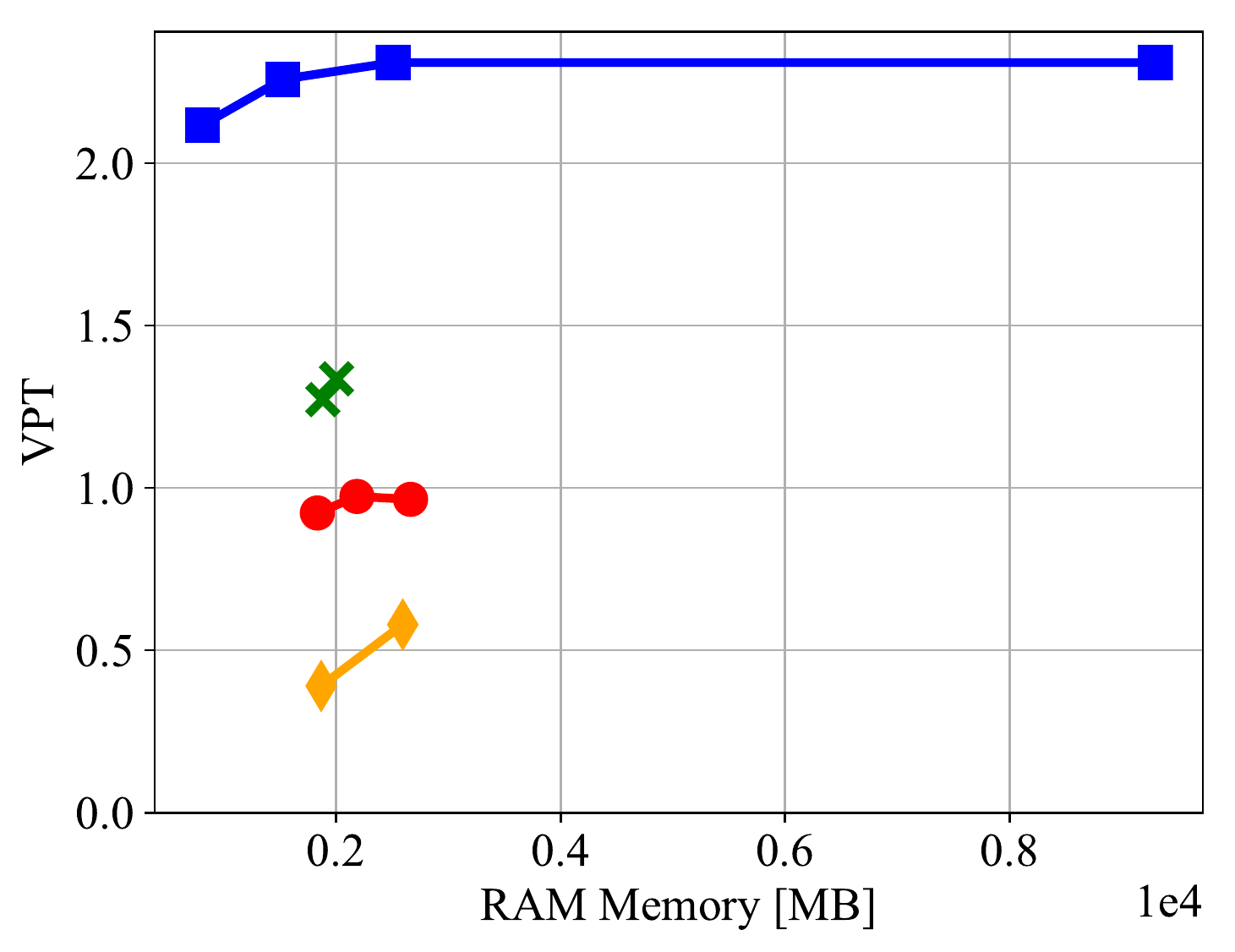}
\caption{
VPT w.r.t. RAM memory for $d_o=40$.
}
\label{fig:L96F8GP40R40:NAP_2_RAM_RDIM40}
\end{subfigure}
\hfill
\begin{subfigure}[t]{0.32\textwidth}
\includegraphics[width=.95\textwidth]{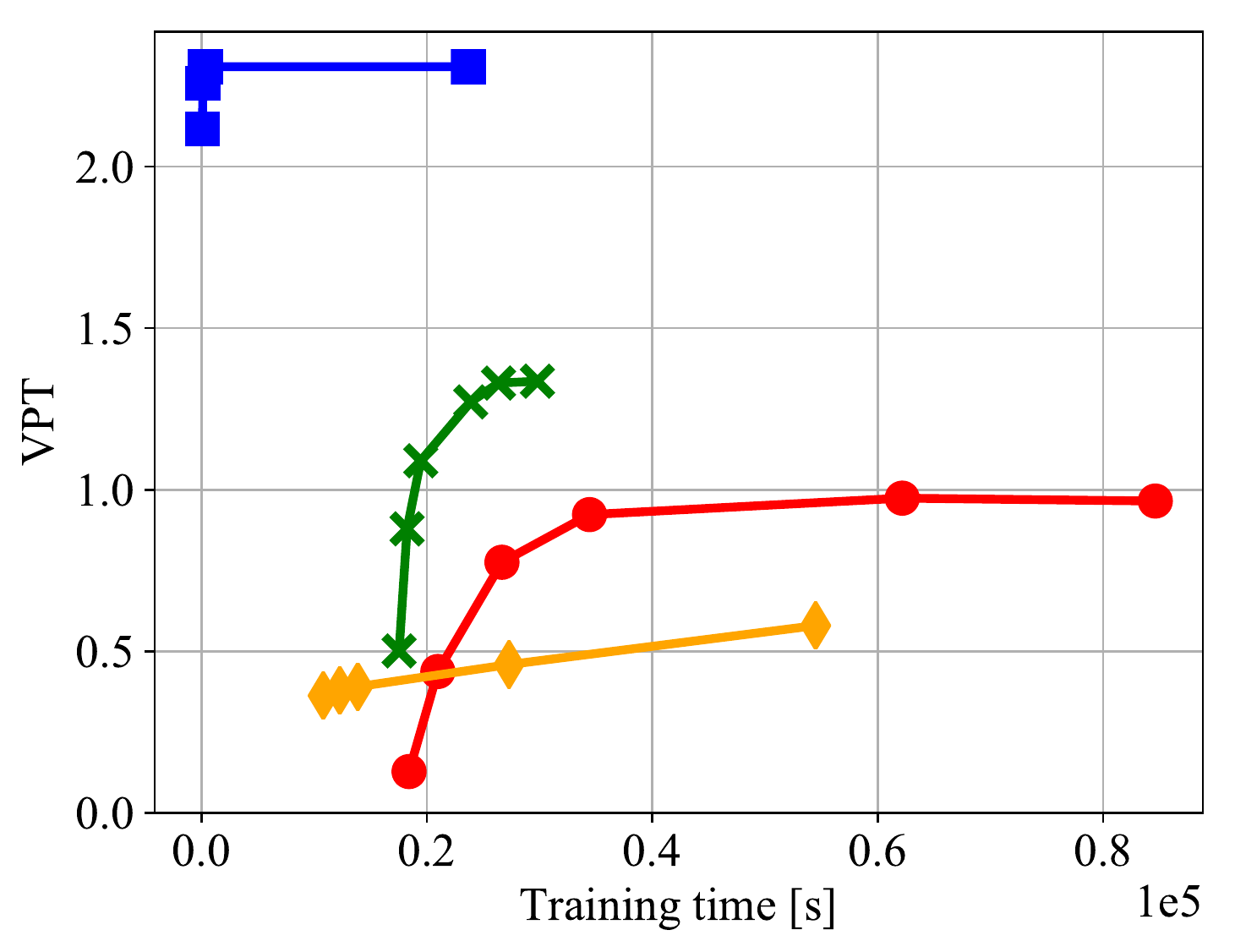}
\caption{
VPT w.r.t. the total training time for $d_o=40$.
}
\label{fig:L96F8GP40R40:NAP_2_TRAINTIME_RDIM40}
\end{subfigure}
\hfill
\begin{subfigure}[t]{0.32\textwidth}
\includegraphics[width=.95\textwidth]{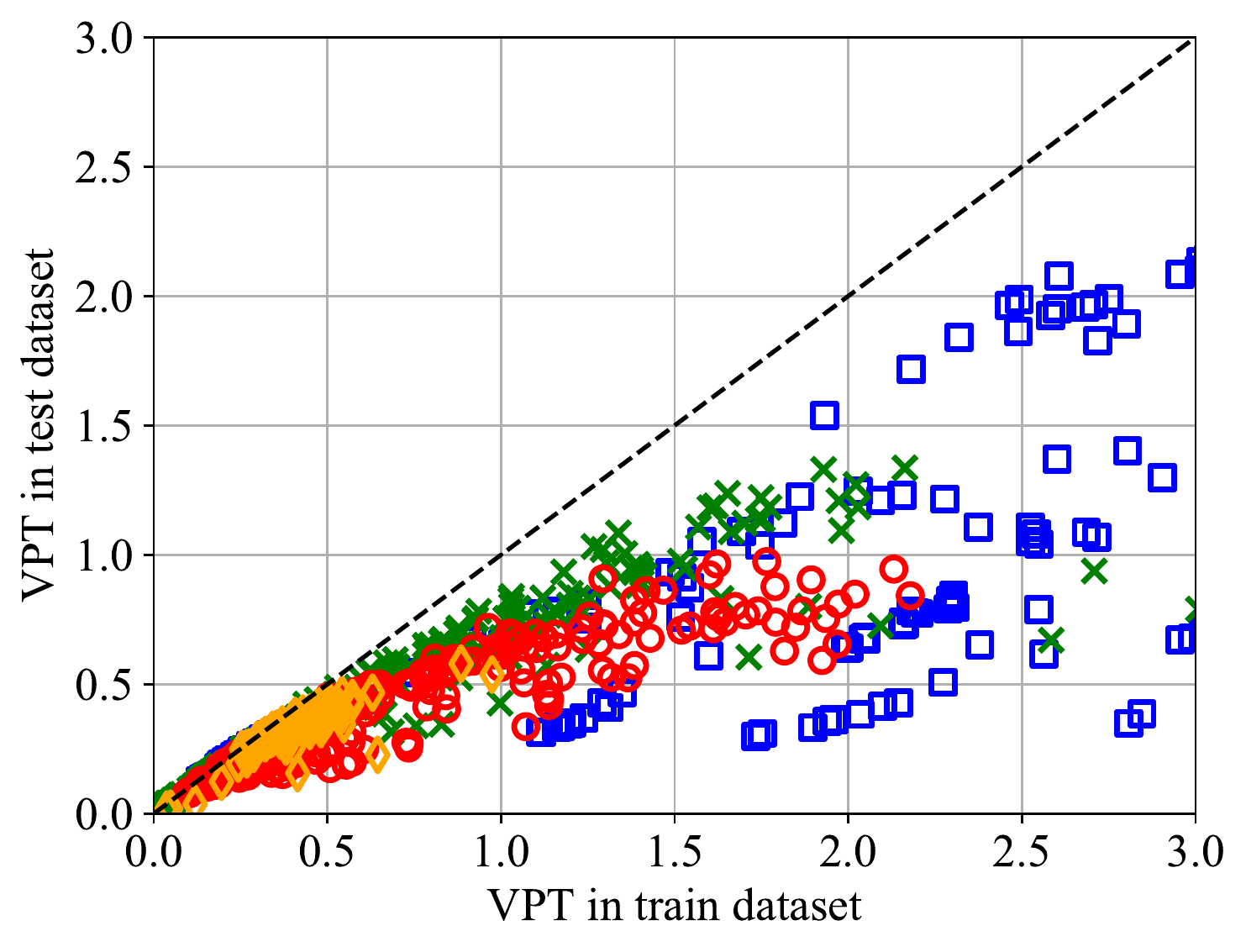}
\caption{
VPT in test data w.r.t. VPT in the train data for $d_o=40$.
}
\label{fig:L96F8GP40R40:OFSP_BBO_NAP_RDIM_40}
\end{subfigure}
\caption{
Forecasting results on the dynamics of an observable consisting of the SVD modes of the Lorenz-96 system with $F=8$ and state dimension $40$.
The observable consists of the $d_o \in \{35, 40 \}$ most energetic modes.
(a), (d) Valid prediction time (VPT) plotted w.r.t. the required RAM memory for dimension $d_o\in \{ 35, 40\}$.
(b), (e) VPT plotted w.r.t. total training time for dimension $d_o\in \{ 35, 40\}$.
(c), (f) VPT measured from $100$ initial conditions sampled from the \textbf{test} data plotted against the VPT from $100$ initial conditions sampled from the \textbf{training} data for each model for $d_o\in \{ 35, 40\}$.
In the reduced order space ($d_o=35$) RCs tend to overfit easier compared to GRUs/LSTMs that utilize validation-based early stopping.
In the full order space  ($d_o=40$) , RCs demonstrate excellent generalization ability and high forecasting accuracy.
\\
RC \protect \bluelineRectangle (or \protect \blueRectangle) ;
GRU \protect \greenlineX (or \protect \greenCross) ;
LSTM \protect \redlineCircle (or  \protect\redCircle) ;
Unit \protect \orangelineDiamond (or \protect \orangeDiamond) ;
Ideal \protect \blacklineDashed;
}
\label{fig:L96F8GP40R40VPTparametersplot}
\end{figure*}

An important aspect of machine learning models is their scalability to high-dimensional systems and their requirements in terms of training time and memory utilization.
Large memory requirements and/or high training times might hinder the application of the models in challenging scenarios, like high-performance applications in climate forecasting~\citep{kurth2018exascale}.
In \Cref{fig:L96F8GP40R40:NAP_2_RAM_RDIM35} and \Cref{fig:L96F8GP40R40:NAP_2_RAM_RDIM40}, we present a Pareto front of the VPT with respect to the CPU RAM memory utilized to train the models with the highest VPT for each architecture for an input dimensions of $d_o=35$ (reduced order) and $d_o=40$ (full dimension) respectively.
\Cref{fig:L96F8GP40R40:NAP_2_TRAINTIME_RDIM35} and \Cref{fig:L96F8GP40R40:NAP_2_TRAINTIME_RDIM40}, 
show the corresponding Pareto fronts of the VPT with respect to the training time.
In case of the full state space ($d_o=40$), the RC is able to achieve superior VPT with smaller memory usage and vastly smaller training time than the other methods.
However, in the case of reduced order information ($d_o=35$), the BPTT algorithms (GRU and LSTM) are superior to the RC even when the latter is provided with one order of magnitude more memory.

Due to the fact that the RNN models are \textbf{learning} the recurrent connections, they are able to reach higher VPT when forecasting in the reduced order space without the need for large models.
In contrast, in RC the maximum reservoir size (imposed by computer memory limitations) may not be sufficient to capture the dynamics  of high-dimensional systems with reduced order information and non-local interactions.
We argue that this is the reason why the RC models do not reach the performance of GRU/LSTM trained with Back-propagation (see \Cref{fig:L96F8GP40R40:NAP_2_RAM_RDIM35}).

At the same time, letting memory limitations aside, training of RC models requires the solution of a linear system of equations $\bm{H} \bm{W}_{out}^T=\bm{Y}$, with $\bm{H} \in \mathbb{R}^{d_N \times d_h}$, $\bm{W}_{out}^T \in \mathbb{R}^{d_h \times d_o}$ and $\bm{Y} \in \mathbb{R}^{d_N \times d_o}$ (see Appendix A).
The Moore-Penrose method of solving this system, scales cubically with the reservoir size as it requires the inversion of a matrix with dimensions $d_h \times d_h$.
We also tried an approximate iterative method termed LSQR based on diagonalization, without any significant influence on the training time. 
In contrast, the training time of an RNN is very difficult to estimate a priori, as convergence of the training method depends on initialization and various other hyperparameters and are not necessarily dependent on the size.
That is why we observe a greater variation of the training time of RNN models.
Similar results are obtained for $F=10$, the interested reader is referred to the appendix.

In the following, we evaluate to which extend the trained models overfit to the training data.
For this reason, we measure the VPT in the training dataset and plot it against the VPT in the test dataset for every model we trained.
This plot provides insight on the \textbf{generalization} error of the models.
The results are shown in \Cref{fig:L96F8GP40R40:OFSP_BBO_NAP_RDIM_35}, and \Cref{fig:L96F8GP40R40:OFSP_BBO_NAP_RDIM_40} for $d_o=35$ and $d_o=40$.
Ideally a model architecture that guards effectively against overfitting, exhibits a low generalization error, and should be represented by a point in this plot that is close to the identity line (zero generalization error).
As the expressive power of a model increases, the model may fit better to the training data, but bigger models are more prone to memorizing the training dataset and overfitting (high generalization error).
Such models would be represented by points on the right side of the plot.
In the reduced order scenario, GRU and LSTM models lie closer to the identity line than RC models, exhibiting lower generalization errors.
This is due to the validation-based early stopping routine utilized in the RNNs that guards effectively against overfitting.

We may alleviate the overfitting in RC by tuning the Tikhonov regularization parameter ($\eta$). However, this requires to rerun the training for every other combination of hyperparameters.
For the four tested values $\eta \in \{10^{-3},$ $10^{-4},$ $10^{-5},$ $10^{-6}\}$ of the Tikhonov regularization parameter the RC models tend to exhibit higher generalization error compared to the RNNs trained with BBTT.
We also tested more values $\eta \in \{10^{-1},$ $10^{-2},$ $10^{-3},$ $10^{-4},$ $10^{-5},$ $10^{-6},$ $10^{-7},$ $10^{-8} \}$, while keeping fixed the other hyperparameters, without any observable differences in the results.

However, in the full-order scenario, the RC models achieve superior forecasting accuracy and generalization ability as clearly depicted in \Cref{fig:L96F8GP40R40:OFSP_BBO_NAP_RDIM_40}.
Especially the additional regularization of the training procedure introduced by adding Gaussian noise in the data was decisive to achieve this result.

\begin{figure*}
\centering
\begin{subfigure}[t]{0.8\textwidth}
\centering
\includegraphics[width=.98\textwidth]{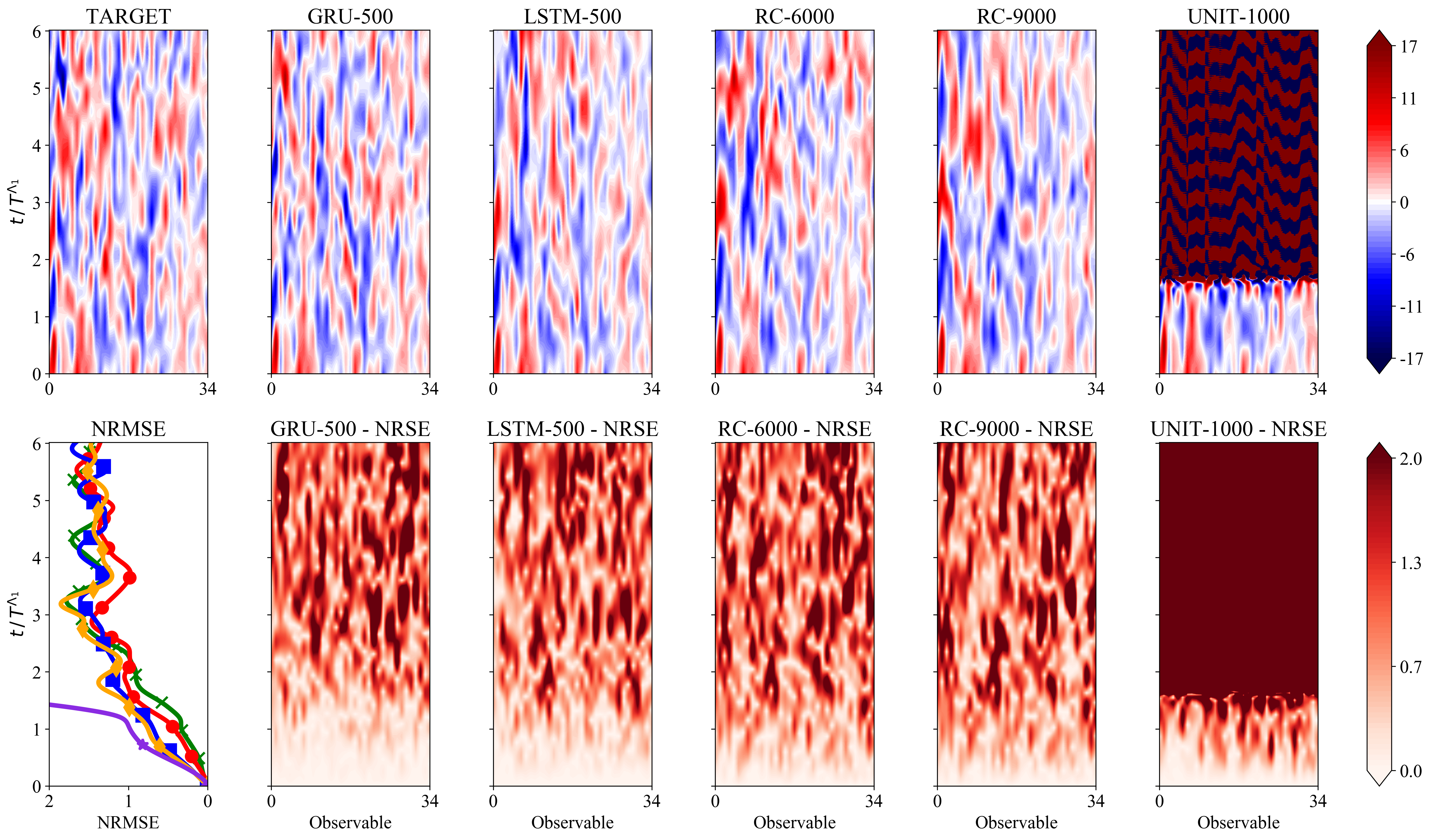}
\caption{Input dimension (dimension of the observable in the reduced order space)  $\bm{d_o=35}$}
\label{fig:L96F8GP40R40:CONTOUR_BBO_NAP_2_M_RDIM_35_IC45}
\end{subfigure}
\hfill
\begin{subfigure}[t]{0.8\textwidth}
\centering
\includegraphics[width=.98\textwidth]{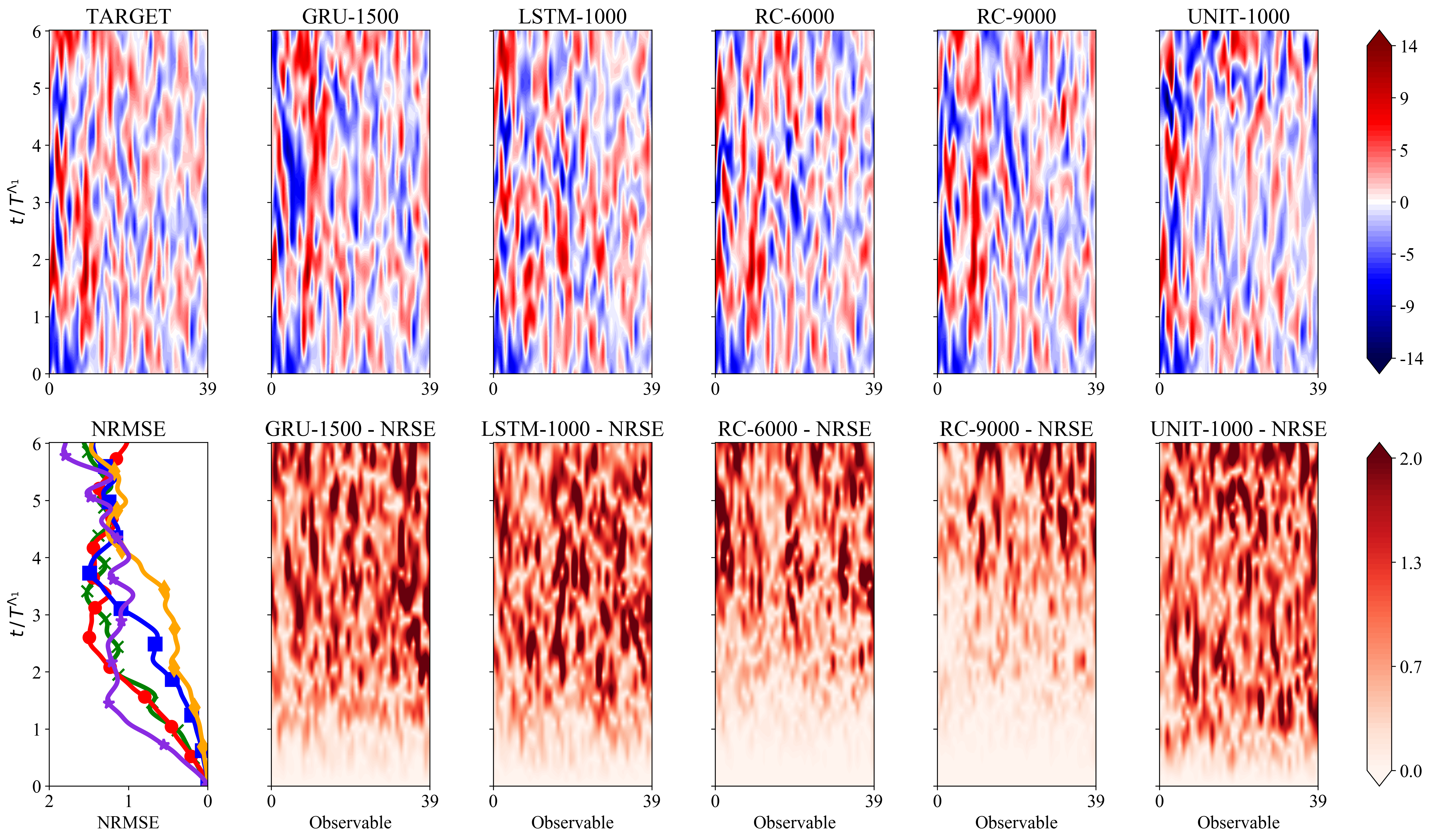}
\caption{Input dimension (dimension of the observable in the reduced order space)  $\bm{d_o=40}$ (full state)}
\label{fig:L96F8GP40R40:CONTOUR_BBO_NAP_2_M_RDIM_40_IC10}
\end{subfigure}
\caption{
Contour plots of a spatio-temporal forecast on the SVD modes of the Lorenz-96 system with $F=8$ in the testing dataset with GRU, LSTM, RC and a Unitary network along with the true (target) evolution and the associated NRSE contours for the reduced order observable \textbf{(a)} $d_o=35$ and the full state \textbf{(b)} $d_o=40$.
The evolution of the component average NRSE (NMRSE) is plotted to facilitate comparison.
Unitary networks suffer from propagation of forecasting error and eventually their forecasts diverge from the attractor.
Forecasts in the case of an observable dimension $d_o=40$ diverge slower as the dynamics are deterministic.
In contrast, forecasting the observable with $d_o=35$ is challenging due to both \textbf{(1)} sensitivity to initial condition and \textbf{(2)} incomplete state information that requires the capturing of temporal dependencies.
In the full-state setting, RC models achieve superior forecasting accuracy compared to all other models.
In the challenging reduced order scenario, LSTM and GRU networks demonstrate a stable behavior in iterative prediction and reproduce the long-term statistics of the attractor.
In contrast, in the reduced order scenario RC suffer from frequent divergence (refer to the appendix). 
\\
GRU \protect \greenlineX;
LSTM \protect \redlineCircle;
RC-6000 \protect \bluelineRectangle;
RC-9000 \protect \orangelineDiamond; 
Unit \protect \bluevioletlineStar;
}\label{fig:L96F8GP40R40:CONTOUR_BBO_NAP_2_MODELS}
\end{figure*}

An example of an iterative forecast in the test dataset, is illustrated in \Cref{fig:L96F8GP40R40:CONTOUR_BBO_NAP_2_MODELS} for $F=8$ and $d_o \in \{35,40\}$.

\clearpage

%% file: sections/sections-5-parallel.tex
\section{Parallel Forecasting Leveraging Local Interactions}
\label{sec:parallel}

\begin{figure*}
\centering
\includegraphics[width=0.6\linewidth]{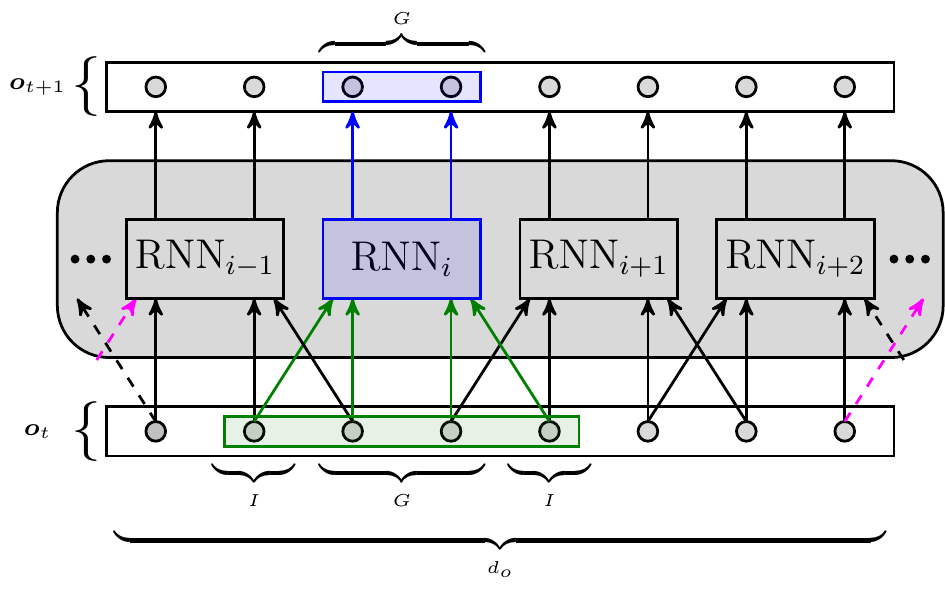}
\caption{Illustration of the parallel architecture for a group size of $G=2$ and an interaction length of $I=1$.
The network consists of multiple RNNs with different parameters.
Each RNN is trained to forecast the evolution of $G$ elements of the observable.
Additional information of $I$ elements from each neighboring network (left and right) are provided as additional input to capture local correlations.
}
\label{fig:parallel}
\end{figure*}

In spatially extended dynamical systems the state space (e.g., vorticity, velocity field, etc.) is high-dimensional (or even infinite dimensional), since an adequately fine grid is needed to resolve the relevant spatio-temporal scales of the dynamics.
Even though RC and RNNs can be utilized for modeling and forecasting of these systems in the short-term, the RC and RNN methods described in \Cref{sec:methods} do not scale efficiently with the input dimension, i.e., as the dimensionality of the observable $\bm{o}_t \in \mathbb{R}^{d_o}$ increases.
Two limiting factor are  the required time and RAM memory to train the model.
As $d_o$ increases, the size $d_h$ of the reservoir network required to predict the system using only a single reservoir rises.
This implies higher training times and more computational resources (RAM memory), which render the problem intractable for large values of $d_o$.
The same applies for RNNs.
More limiting factors arise by taking the process of identification of optimal model hyperparameters into account, since loading, storing and processing a very large number of large models can be computationally infeasible.
However, these scaling problems for large systems can be alleviated in case the system is characterized by local state interactions or translationally invariant dynamics.
In the first case, as shown in \Cref{fig:parallel} the modeling and forecasting task can be parallelized by employing multiple individually trained networks forecasting locally in parallel exploiting the local interactions, while, if translation invariance also applies, the individual parallel networks can be identical and training of only one will be sufficient.
This parallelization concept is utilized in RC in~\citep{Pathak2018a,Parlitz2000}.
The idea dates back to local delay coordinates~\citep{Parlitz2000}.
The model shares ideas from convolutional RNN architectures~\citep{Tara2015,Shi2015} designed to capture local features that are translationally invariant in image and video processing tasks.
In this section, we extend this parallelization scheme to RNNs and compare the efficiency of parallel RNNs and RCs in forecasting the state dynamics of the Lorenz-96 model and Kuramoto-Sivashinsky equation discretized in a fine grid.

\subsection{Parallel Architecture}
\label{sec:parallel:architecture}

Assume that the observable is $\bm{o}_t\in \R^{d_o}$ and each element of the observable is denoted by $\bm{o}^i_t \in \R, \forall i \in \{ 1,\dots, d_o \}$.
In case of local interactions, the evolution of each element is affected by its spatially neighboring grid points.
The elements $\bm{o}^i$ are split into $N_g$ groups, each of which consisting of $G$ spatially neighboring elements such that $d_o=G N_g$.
The parallel model employs $N_g$ RNNs, each of which is utilized to predict a spatially local region of the system observable indicated by the $G$ group elements $\bm{o}^i$.
Each of the $N_g$ RNNs receives $G$ inputs $\bm{o}^i$ from the elements $i$ it forecasts in addition to $I$ inputs from neighboring elements on the left and on the right, where $I$ is the interaction length.
An example with $G=2$ and $I=1$ is illustrated in \Cref{fig:parallel}.

During the training process, the networks can be trained independently.
However, for long-term forecasting, a communication protocol has to be utilized as each network requires the predictions of neighboring networks to infer.
In the case of a homogeneous system, where the dynamics are translation invariant, the training process can be drastically reduced by utilizing one single RNN and training it on data from all groups.
The weights of this RNN are then copied to all other members of the network.
In the following we assume that we have no knowledge of the underlying data generating mechanism and its properties, so we assume the data is not homogeneous.

The elements of the parallel architecture are trained independently, while the MPI~\citep{Dalcin2011,dalcin2008,MPI} communication protocol is utilized to communicate the elements of the interaction for long-term forecasting.

\subsection{Results on the Lorenz-96}
\label{sec:parallel:lorenz96}

In this section, we employ the parallel architecture to forecast the state dynamics of the Lorenz-96 system explained in \Cref{sec:observable:lorenz96} with a state dimension of $d_o=40$.
Note that in contrast to \Cref{sec:observable:lorenz96:results}, we do not construct an observable and then forecast the reduced order dynamics.
Instead, we leverage the local interactions in the state space and employ an ensemble of networks forecasting the local dynamics.

Instead of a single RNN model forecasting the $d_o=40$ dimensional global state (composed of the values of the state in the $40$ grid nodes), we consider $N_g=20$ separate RNN models, each forecasting the evolution of a $G=2$ dimensional local state (composed of the values of the state in $2$ grid nodes).
In order to forecast the evolution of the local state, we take into account its interaction with $I=4$ grid nodes on its left and on its right.
The group size of the parallel models is thus $G=2$, while the interaction length is $I=4$.
As a consequence, each model receives at its input an $2 I + G=10$ dimensional state and forecasts the evolution of a local state composed from $2$ grid nodes. 
The size of the hidden state in RC is $d_h \in \{1000, 3000, 6000,12000 \}$.
Smaller networks of size $d_h \in \{100, 250, 500 \}$ are selected for GRU and LSTM.
The rest of the hyperparameters are given in the appendix.
Results for Unitary networks are omitted, as the identification of hyperparameters leading to stable iterative forecasting was computationally heavy and all trained models led to unstable systems that diverged after a few iterations.

\begin{figure*}
\begin{subfigure}[t]{0.3\textwidth}
\includegraphics[height=3.5cm]{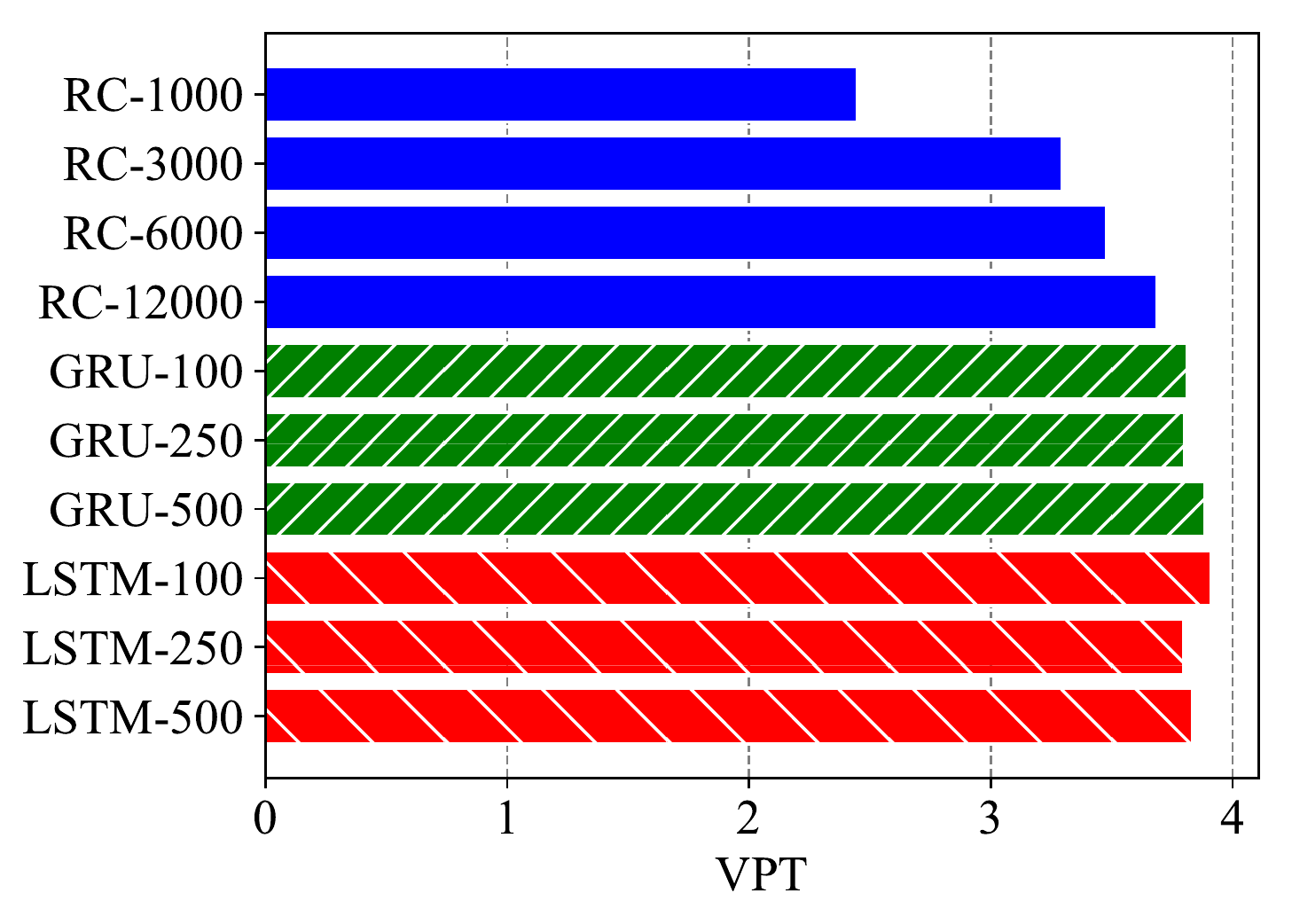}
\caption{Valid prediction time in the test dataset}
\label{fig:parallel:L96F8GP40:P_VPT_BAR_BBO_NAP}
\end{subfigure}
\hfill
\begin{subfigure}[t]{0.3\textwidth}
\includegraphics[height=3.5cm]{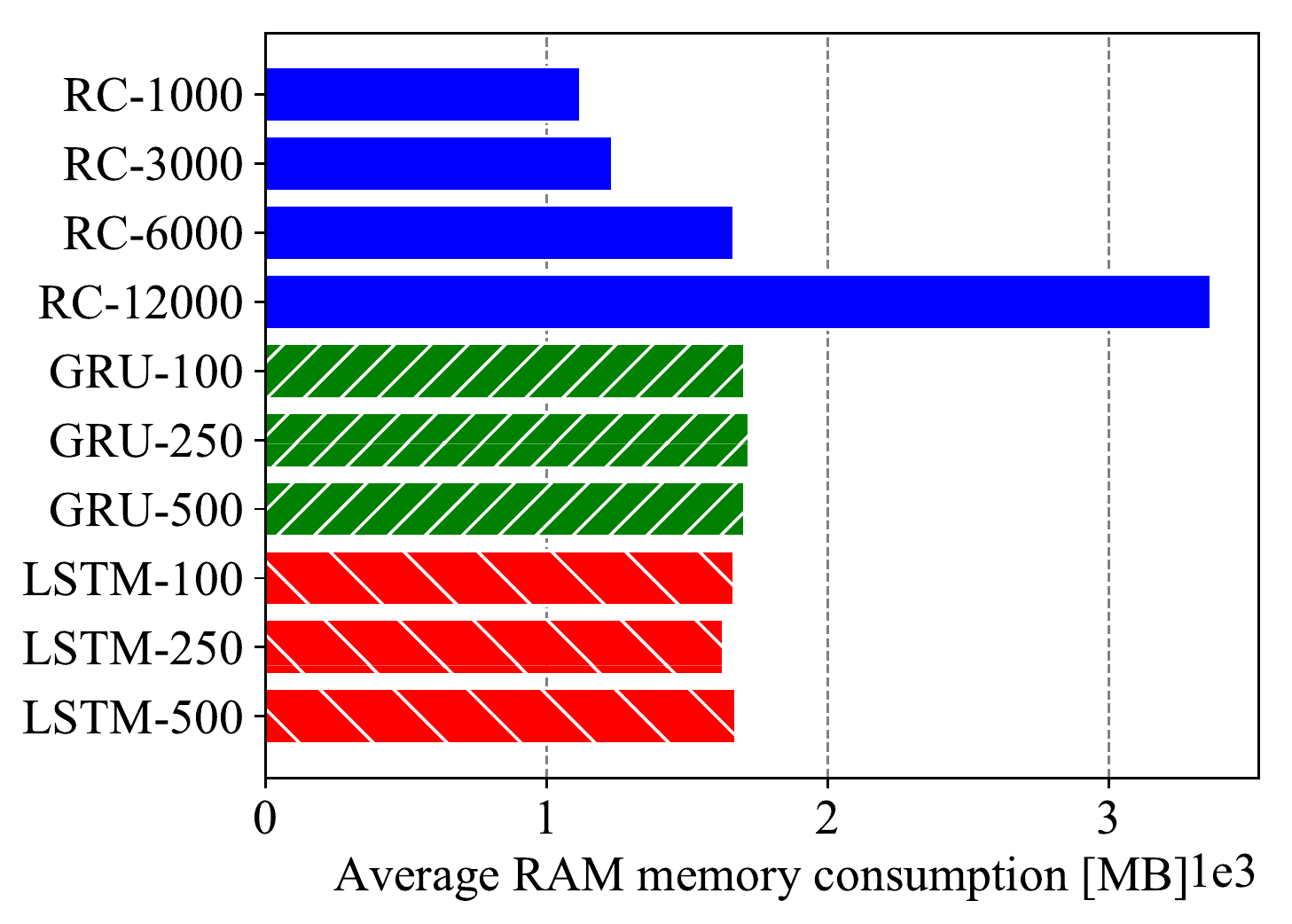}
\caption{Average RAM memory requirement}
\label{fig:parallel:L96F8GP40:P_RAM_AVERAGE_BAR_BBO_NAP}
\end{subfigure}
\hfill
\begin{subfigure}[t]{0.3\textwidth}
\includegraphics[height=3.5cm]{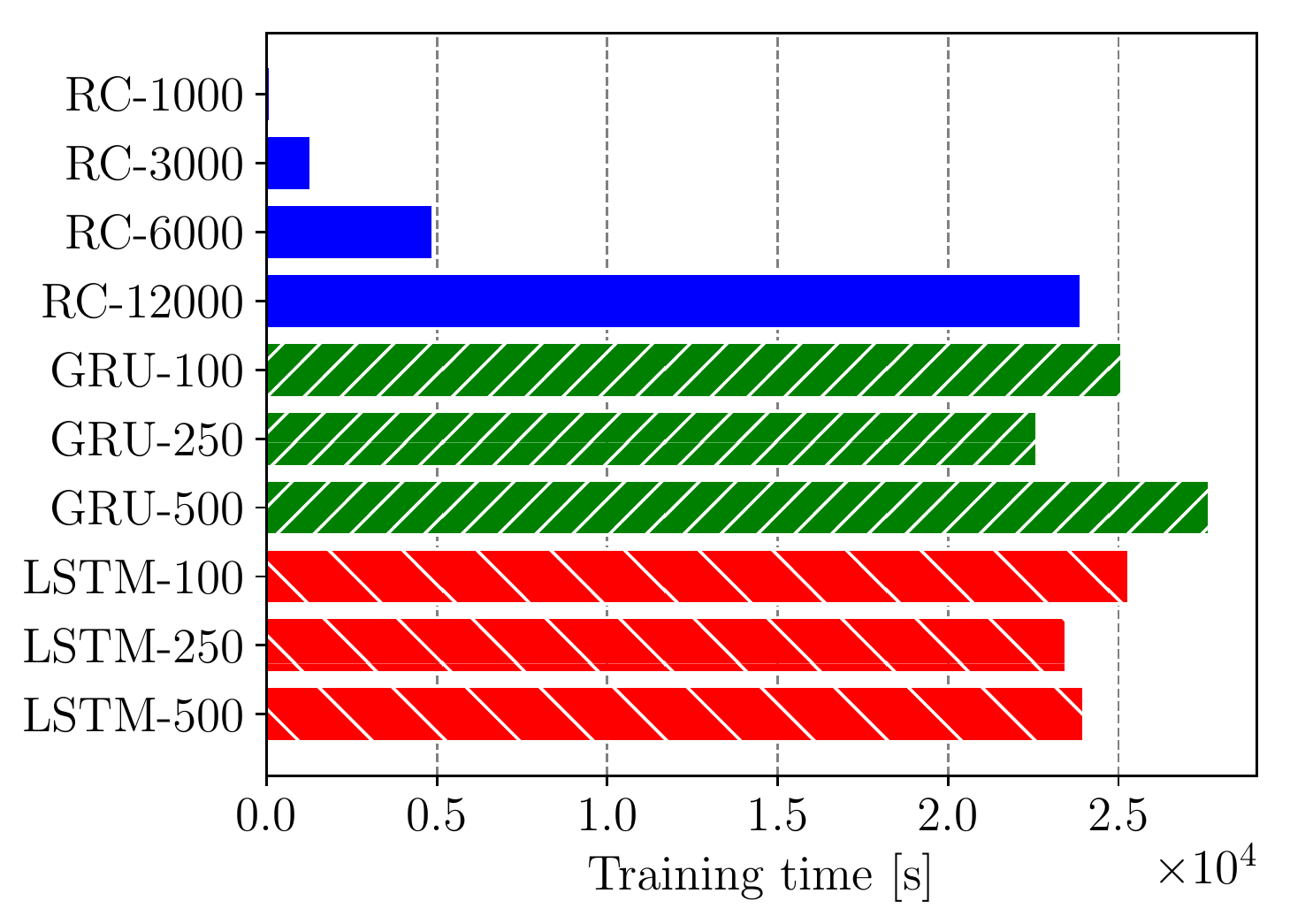}
\caption{Training time}
\label{fig:parallel:L96F8GP40:P_TRAINTIME_BAR_BBO_NAP}
\end{subfigure}
\caption{\textbf{(a)} Valid prediction time (VPT), \textbf{(b)} CPU memory utilization and \textbf{(c)} total training time of RNN parallel architectures with group size $G=2$ and an interaction length $I=4$  forecasting the dynamics of Lorenz-96 with state dimension $d_o=40$ (full state).
GRU and LSTM results do not depend significantly on network size.
RC with 3000 or 6000 nodes have slightly lower VPT, but require much less training time. Increasing RC size to more than 12000 nodes was not feasible due to memory requirements.
}
\label{fig:parallel:L96F8GP40:barplots}
\end{figure*}

In \Cref{fig:parallel:L96F8GP40:P_VPT_BAR_BBO_NAP}, we plot the VPT time of the RC and the BPTT networks.
We find that RNN trained by BPTT achieve comparable predictions with RC, albeit using much smaller number hidden nodes (between 100 and 500 for BPTT vs 6000 to 12000 for RC). We remark that RC with 3000 and 6000 nodes have slightly lower VPT than GRU and LSTM but require significantly lower training times as shown in  \Cref{fig:parallel:L96F8GP40:P_TRAINTIME_BAR_BBO_NAP}.
At the same time, using 12000 nodes for RC implies high RAM requirements, more than 3 GB per rank, as depicted in \Cref{fig:parallel:L96F8GP40:P_RAM_AVERAGE_BAR_BBO_NAP}.

\begin{figure*}
\begin{subfigure}[t]{0.45\textwidth}
\centering
\includegraphics[height=4.5cm]{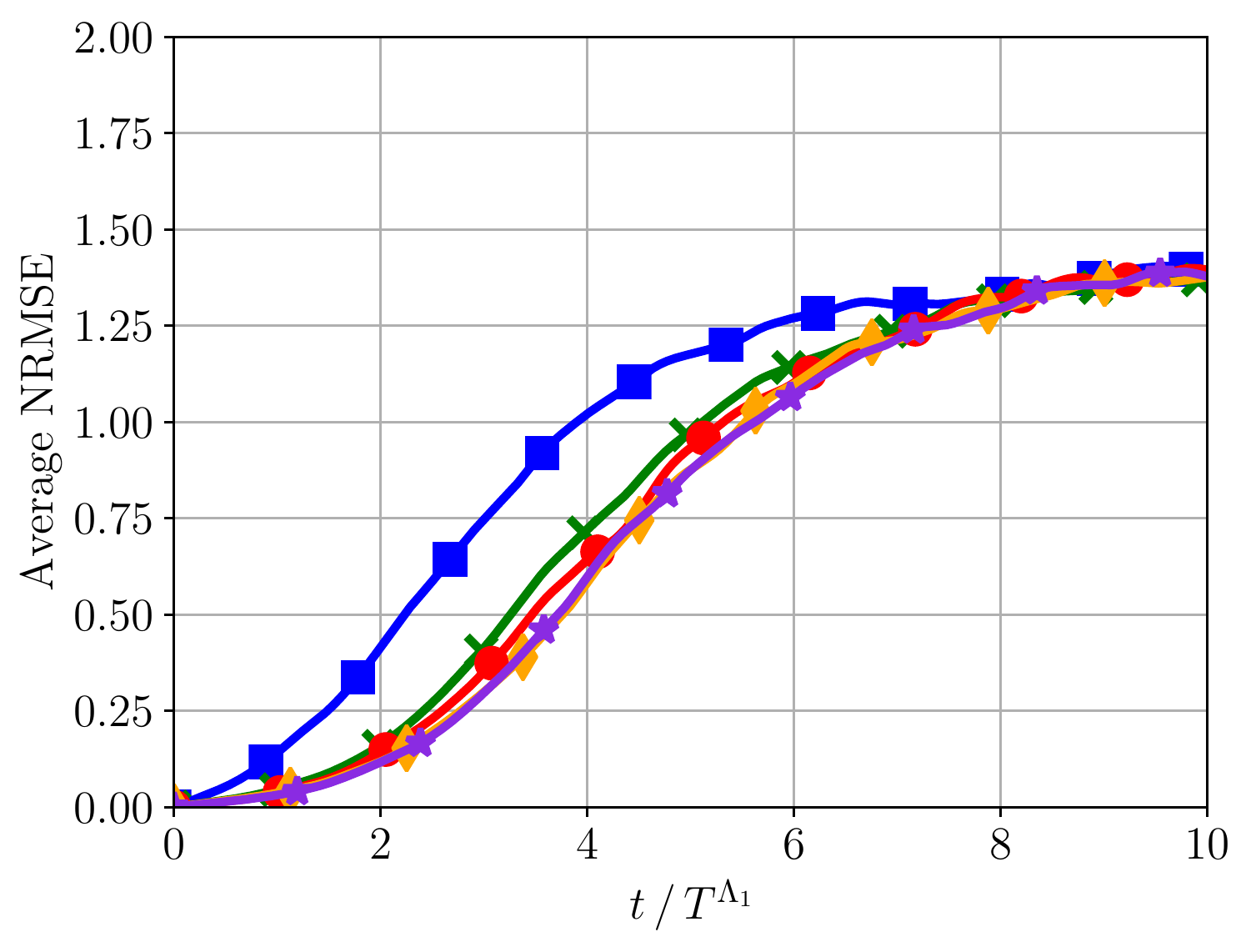}
\caption{NRMSE error evolution}
\label{fig:L96F8GP40:P_RMNSE_BBO_NAP_2_M_RDIM_40_0}
\end{subfigure}
\hfill
\begin{subfigure}[t]{0.45\textwidth}
\centering
\includegraphics[height=4.5cm]{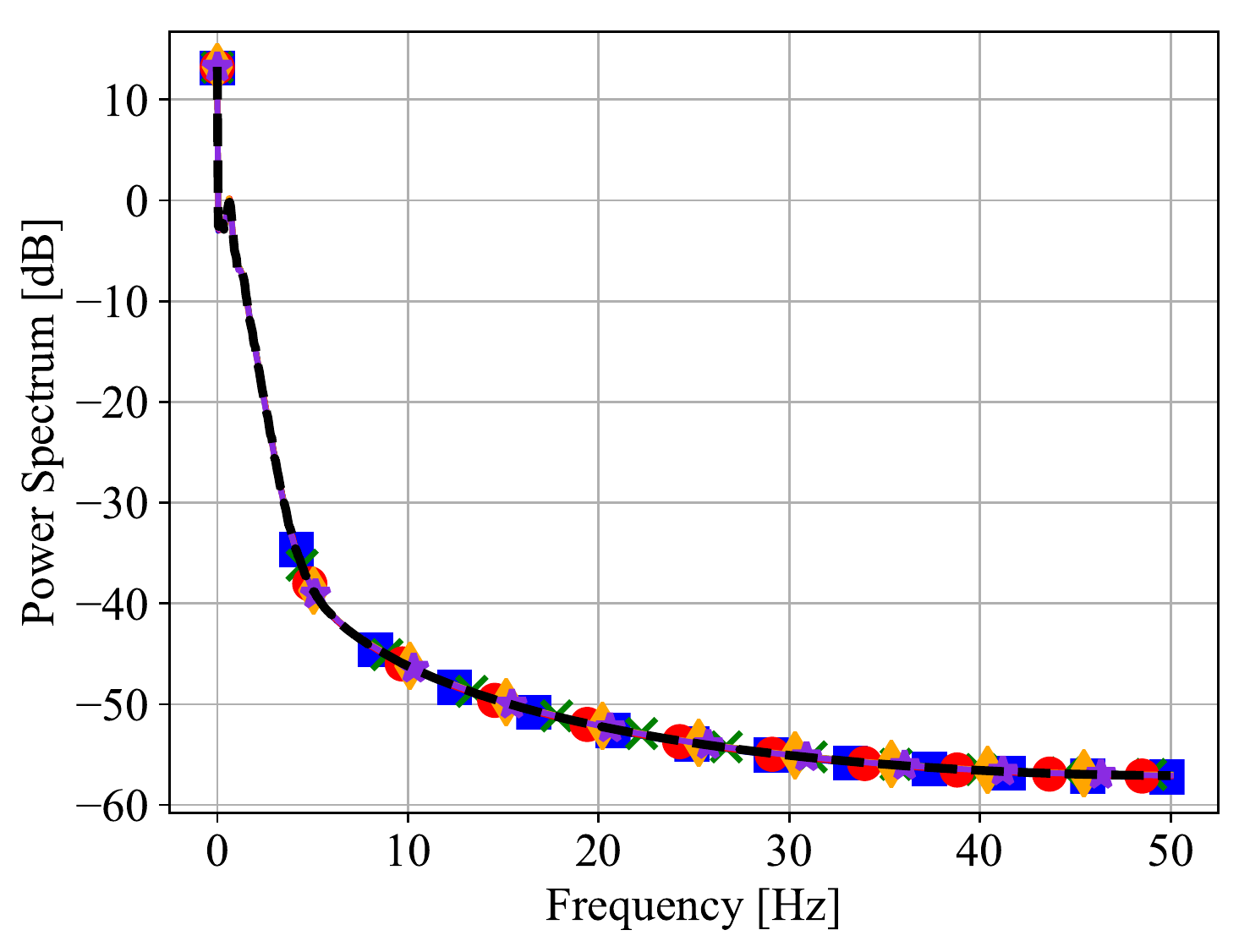}
\caption{Power spectrum}
\label{fig:L96F8GP40:P_POWSPEC_BBO_NAP_M_RDIM_40_0}
\end{subfigure}
\caption{ \textbf{(a)} The evolution of the NRMSE error (averaged over $100$ initial conditions) of different parallel models in the Lorenz-96 with state dimension $d_o=40$.
\textbf{(b)} The reconstructed power spectrum.
All models accurately capture the power spectrum.
RCs with $d_h \in \{ 6000, 12000 \}$ nodes are needed to match the predictive performance of an LSTM with $100$ nodes. 
\\
RC-1000 \protect \bluelineRectangle;
RC-6000 \protect \greenlineX;
RC-12000 \protect \redlineCircle;
GRU-500 \protect \orangelineDiamond;
LSTM-100 \protect \bluevioletlineStar;
Groundtruth \protect \blacklineDashed;
}
\label{fig:parallel:L96F8GP40:nrmsespectrum}
\end{figure*}

\begin{figure*}
\centering
\includegraphics[width=.98\textwidth]{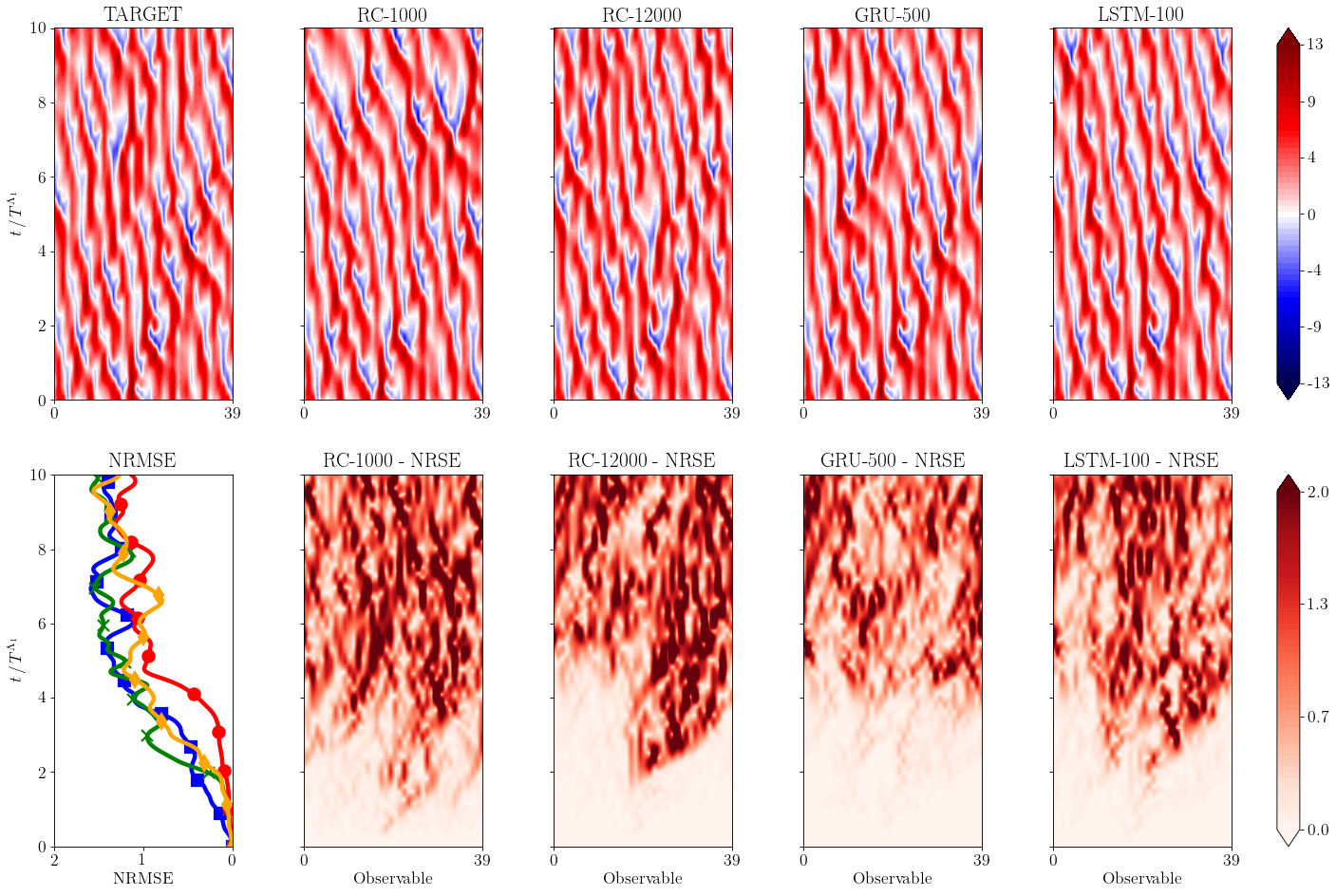}
\label{fig:L96F8GP40:P_CONTOUR_BBO_NAP_2_M_RDIM_40_IC10}
\caption{
Contour plots of a spatio-temporal forecast in the testing dataset with parallel GRU, LSTM, and RC networks along with the true (target) evolution and the associated NRSE contours in the Lorenz-96 system with the full state as an observable $d_o=40$.
The evolution of the component average NRSE (NMRSE) is plotted to facilitate comparison.
\\
RC-1000 \protect \bluelineRectangle;
RC-12000 \protect \greenlineX;
GRU-500 \protect \redlineCircle ;
LSTM-100 \protect \orangelineDiamond;
}
\label{fig:parallel:L96F8GP40:contour}
\end{figure*}

As elaborated in \Cref{sec:observable:lorenz96:results} and depicted in \Cref{fig:L96F8GP40R40:RMNSE_BBO_NAP_2_M_RDIM_35_5000_2000}, the VPT reached by large nonparallelized models that are forecasting the $40$ SVD modes of the system is approximately $1.4$.
We also verified that the nonparallelized models of \Cref{sec:observable:lorenz96} when forecasting the $40$ dimensional state containing local interactions instead of the $40$ modes of SVD, reach the same predictive performance.
Consequently, as expected the VPT remains the same whether we are forecasting the state or the SVD modes as the system is deterministic.
By exploiting the local interactions and employing the parallel networks, the VPT is increased from $\approx 1.4$ to $\approx 3.9$ as shown in \Cref{fig:parallel:L96F8GP40:P_VPT_BAR_BBO_NAP}.
The NRMSE error of the best performing hyperparameters is given in \Cref{fig:L96F8GP40:P_RMNSE_BBO_NAP_2_M_RDIM_40_0}.
All models are able to reproduce the climate as the reconstructed power spectrum plotted in \Cref{fig:L96F8GP40:P_POWSPEC_BBO_NAP_M_RDIM_40_0} matches the true one.
An example of an iterative prediction with LSTM, GRU and RC models starting from an initial condition in the test dataset is provided in \Cref{fig:parallel:L96F8GP40:contour}.

\newpage

\subsection{Kuramoto-Sivashinsky}
\label{sec:parallel:ks}

The Kuramoto-Sivashinsky (KS) equation is a nonlinear partial differential equation of fourth order that is used as a turbulence model for various phenomena.
It was derived by Kuramoto in~\citep{Kuramoto1978} to model the chaotic behavior of the phase gradient of a slowly varying amplitude in a reaction-diffusion type medium with negative viscosity coefficient.
Moreover, Sivashinsky~\citep{Sivashinsky1977} derived the same equations when studying the instantaneous instabilities in a laminar flame front.
For our study, we restrict ourselves to the one dimensional K-S equation
\begin{equation}
\begin{split}
&\frac{\partial u}{ \partial t} = - \nu \frac{\partial^4 u}{ \partial x ^4 } - \frac{\partial^2 u}{ \partial x ^2 } - u  \frac{\partial u}{ \partial x},\\
\end{split}
\label{eq:kuramoto}
\end{equation}
on the domain $\Omega=[0,L]$ with periodic boundary conditions $u(0,t)=u(L,t)$.
The dimensionless boundary size $L$ directly affects the dimensionality of the attractor.
For large values of $L$, the attractor dimension scales linearly with $L$~\citep{Manneville1984}.

In order to spatially discretize \Cref{eq:kuramoto} we select a grid size $\Delta x$ with $D=L/\Delta x + 1$ the number of nodes.
Further, we denote with $u_i=u(i \Delta x)$ the value of $u$ at node $i \in \{0,\dots,D-1\}$.
In the following, we select $\nu=1,L=200$, $\delta t=0.25$ and a grid of $d_o=512$ nodes.
We discretize \Cref{eq:kuramoto} and solve it using the fourth-order method for stiff PDEs introduced in~\citep{Kassam2005} up to $T=6 \cdot 10^4$.
This corresponds to $24 \cdot 10^4$ samples.
The first $4 \cdot 10^4$ samples are truncated to avoid initial transients.
The remaining data are divided to a training and a testing dataset of $10^5$ samples each.
The observable is considered to be the $d_o=512$ dimensional state.
The Lyapunov time $T^{\Lambda_1}$ of the system (see \Cref{sec:comparison}) is utilized as a reference timescale.
We approximate it with the method of Pathak~\citep{Pathak2018a} for $L=200$ and it is found to be $T^{\Lambda_1} \approx 0.094$.

\subsection{Results on the Kuramoto-Sivashinsky Equation}
\label{sec:ks:results}

In this section, we present the results of the parallel models in the Kuramoto-Sivashinsky equation.
The full system state is used as an observable, i.e., $d_o=512$.
The group-size of the parallel models is set to $G=8$, while the interaction length is $I=8$.
The total number of groups is $N_g=64$.
Each member forecasts the evolution of $8$ state components, receiving at the input $24$ components in total.
The size of the reservoir in RC is $d_h \in \{500, 1000, 3000 \}$.
For GRU and LSTM networks we vary $d_h \in \{100, 250, 500 \}$.
The rest of the hyperparameters are given in the appendix.
Results on Unitary networks are omitted, as the configurations tried in this work led to unstable models diverging after a few time-steps in the iterative forecasting procedure.

\begin{figure*}
\begin{subfigure}[t]{0.3\textwidth}
\includegraphics[height=3.5cm]{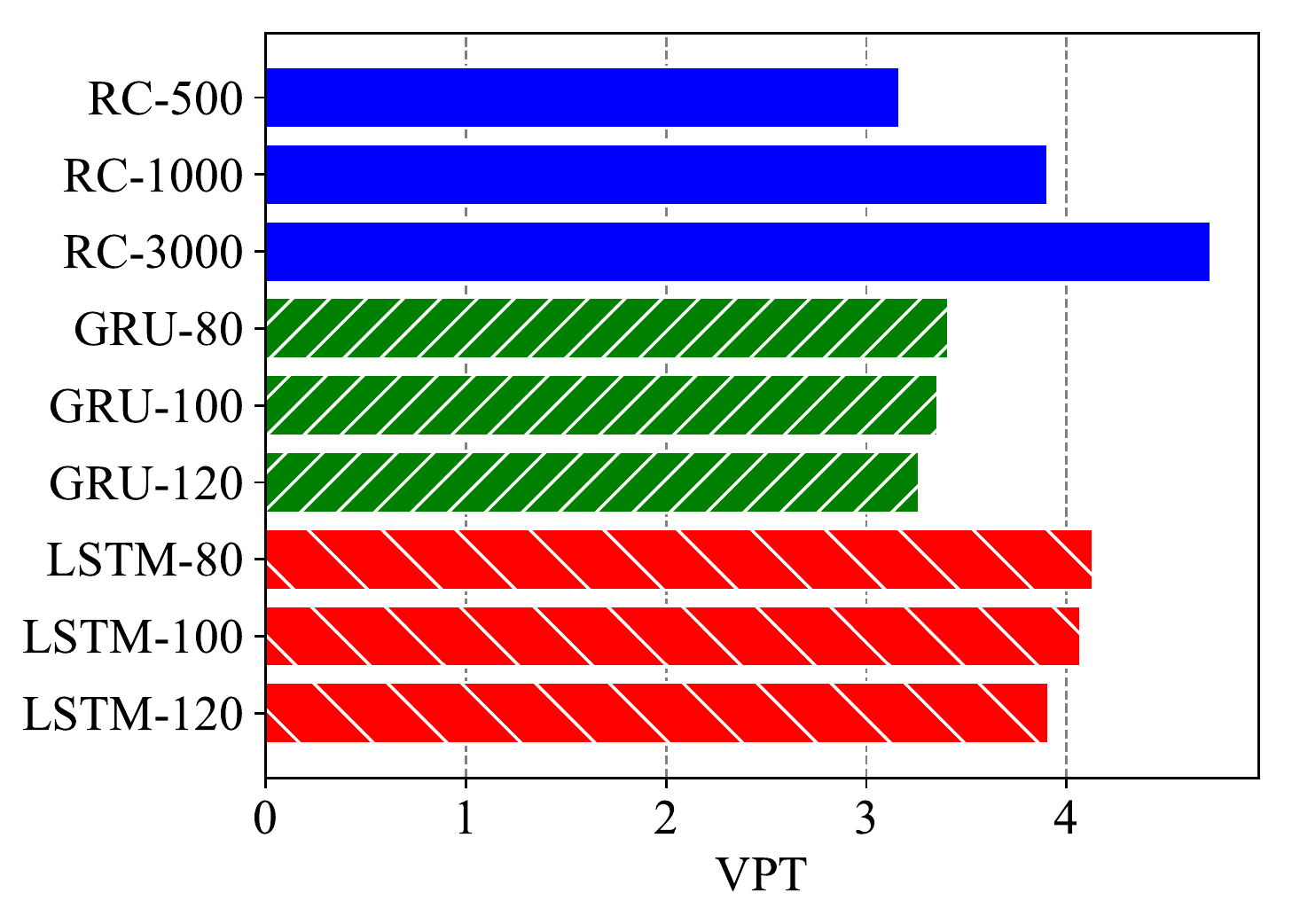}
\caption{Valid prediction time in the test dataset}
\label{fig:parallel:KSGP512:P_VPT_BAR_BBO_NAP}
\end{subfigure}
\hfill
\begin{subfigure}[t]{0.3\textwidth}
\includegraphics[height=3.5cm]{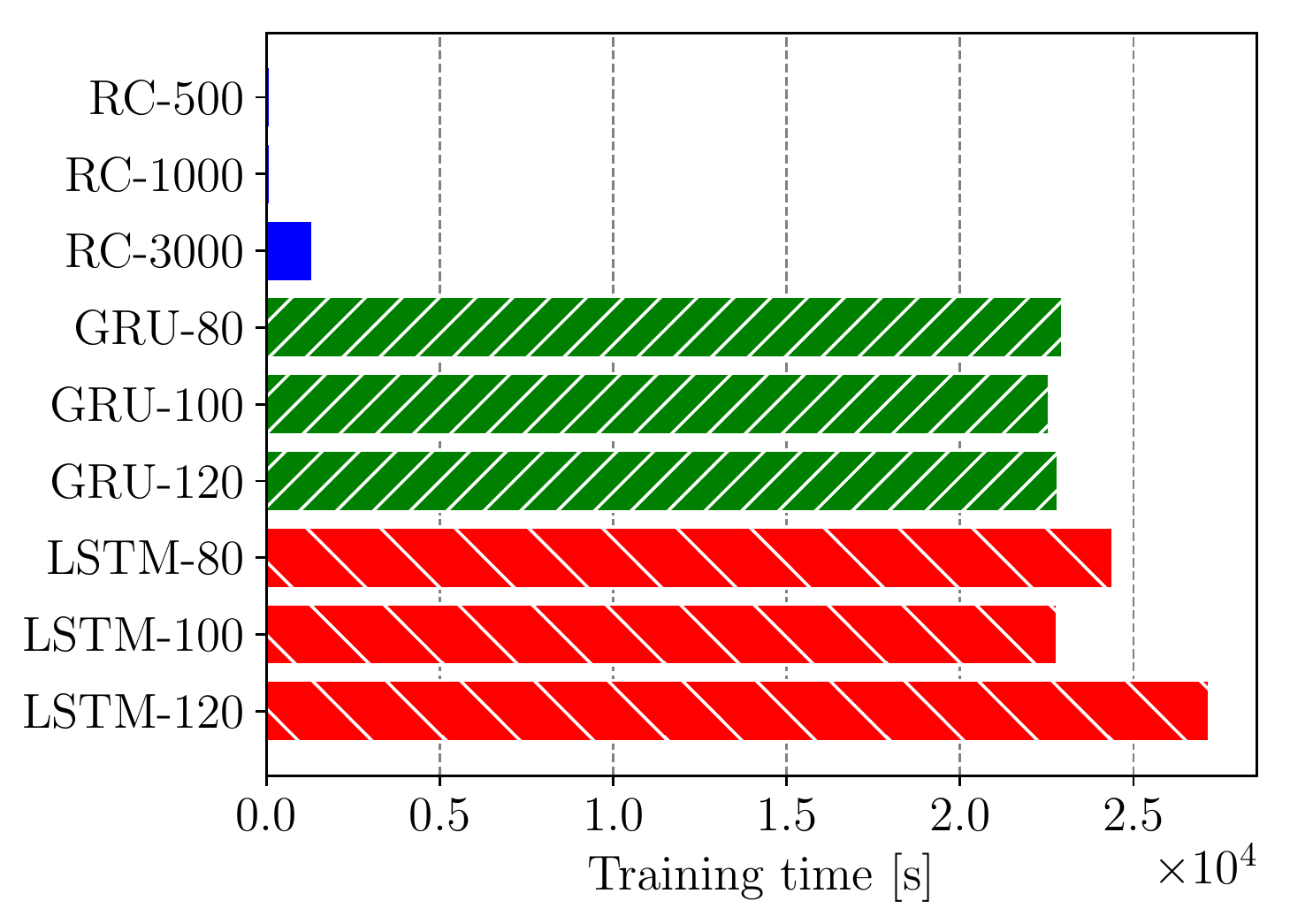}
\caption{Training time}
\label{fig:parallel:KSGP512:P_TRAINTIME_BAR_BBO_NAP}
\end{subfigure}
\hfill
\begin{subfigure}[t]{0.3\textwidth}
\includegraphics[height=3.5cm]{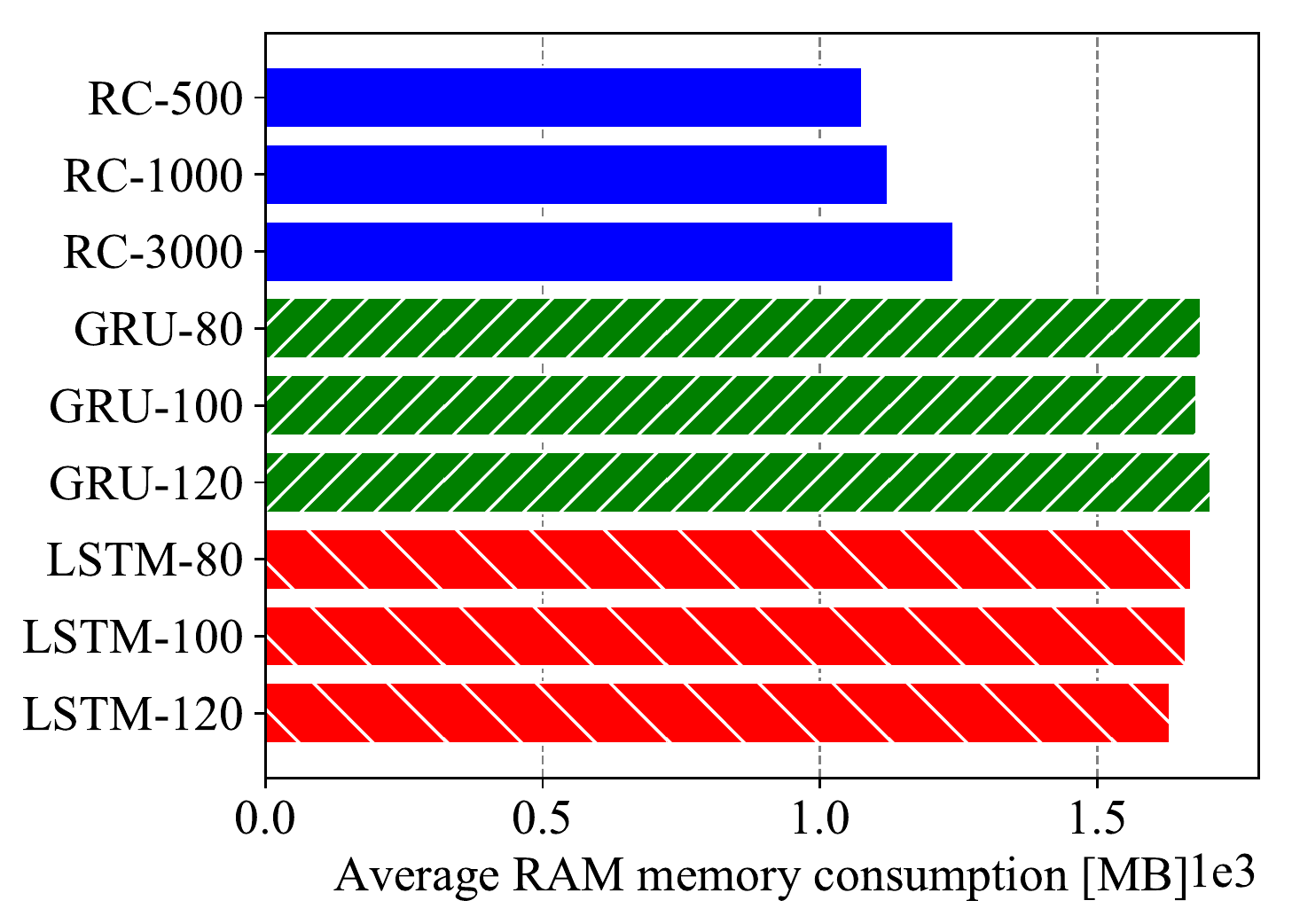}
\caption{Average RAM memory requirement}
\label{fig:parallel:KSGP512:P_RAM_AVERAGE_BAR_BBO_NAP}
\end{subfigure}
\caption{\textbf{(a)} Valid prediction time (VPT), \textbf{(b)} total training time, and \textbf{(c)} CPU memory utilization of parallel RNN architectures with group size $G=8$ and an interaction length $I=8$  forecasting the dynamics of Kuramoto-Sivashinsky equation with state dimension $d_o=512$.
}
\label{fig:parallel:KSGP512:barplots}
\end{figure*}

The results are summed up in the bar-plots in \Cref{fig:parallel:KSGP512:barplots}.
In \Cref{fig:parallel:KSGP512:P_VPT_BAR_BBO_NAP}, we plot the VPT time of the models.
LSTM models reach VPTs of $\approx 4$, while GRU show an inferior predictive performance with VPTs of $\approx 3.5$.
An RC with $d_h=500$ reaches a VPT of  $\approx 3.2$, and an RC with $1000$ modes reaches the VPT of LSTM models with a VPT of $\approx 3.9$.
Increasing the reservoir capacity of the RC to $d_h=3000$ leads to a model exhibiting a VPT of $\approx 4.8$.
In this case, the large RC model shows slightly superior performance to GRU/LSTM.
The low performance of GRU models can be attributed to the fact that in the parallel setting the probability that any RNN may converge to bad local minima rises, with a detrimental effect on the total predictive performance of the parallel ensemble.
In case of spatially translational invariant systems, we could alleviate this problem by using one single network.
Still, training the single network to data from all spatial locations would be expensive.

As depicted in \Cref{fig:parallel:KSGP512:barplots}, the reservoir size of $3000$ is enough for RC to reach and surpass the predictive performance of RNNs utilizing a similar amount of RAM memory and a much lower amount of training time as illustrated in \Cref{fig:parallel:KSGP512:P_TRAINTIME_BAR_BBO_NAP}.

\begin{figure*}
\begin{subfigure}[t]{0.45\textwidth}
\centering
\includegraphics[height=4.5cm]{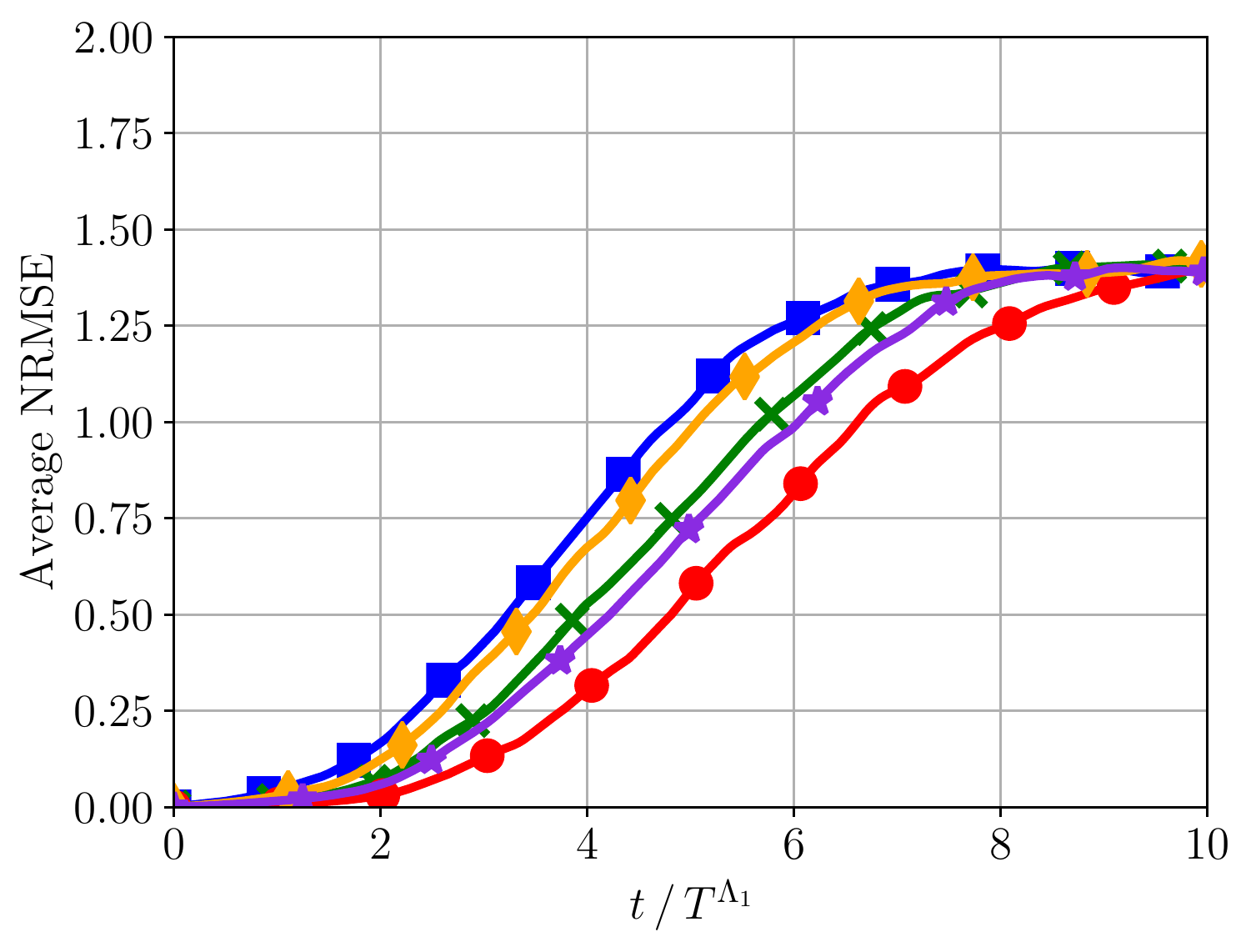}
\caption{Average number of accurate test-set predictions}
\label{fig:KSGP512:P_RMNSE_BBO_NAP_2_M_RDIM_512_0}
\end{subfigure}
\hfill
\begin{subfigure}[t]{0.45\textwidth}
\centering
\includegraphics[height=4.5cm]{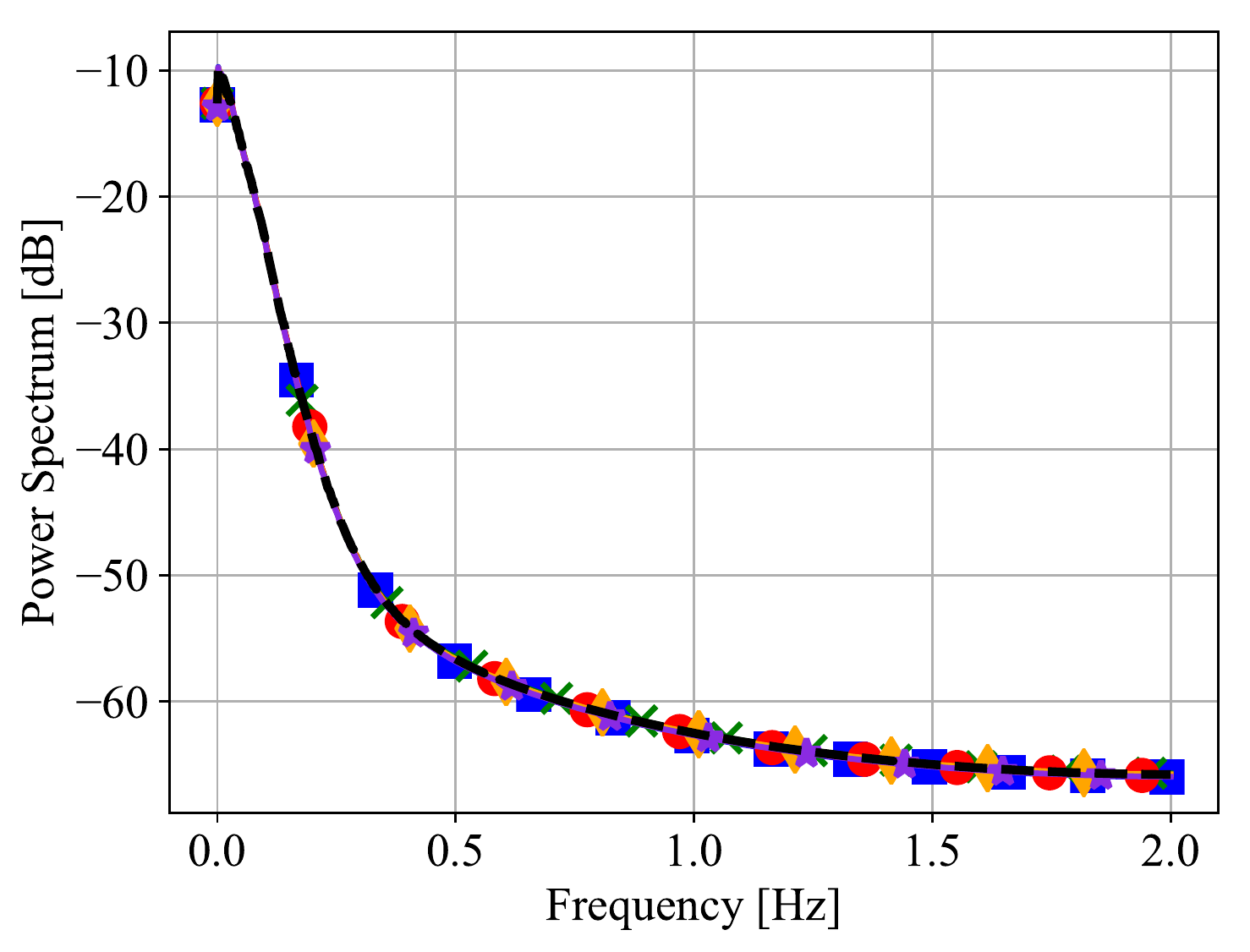}
\caption{CPU memory utilization}
\label{fig:KSGP512:P_POWSPEC_BBO_NAP_M_RDIM_512_0}
\end{subfigure}
\caption{ \textbf{(a)} The evolution of the NRMSE error (averaged over $100$ initial conditions) of different parallel models in the Kuramoto-Sivashinsky equation with state dimension $d_o=512$.
\textbf{(b)} The power spectrum.
All models capture the statistics of the system.
\\
RC-500 \protect \bluelineRectangle;
RC-1000 \protect \greenlineX;
RC-3000 \protect \redlineCircle;
GRU-80 \protect \orangelineDiamond;
LSTM-80 \protect \bluevioletlineStar;
Groundtruth \protect \blacklineDashed;
}
\label{fig:parallel:KSGP512:nrmsespectrum}
\end{figure*}

\begin{figure*}
\centering
\includegraphics[width=.98\textwidth]{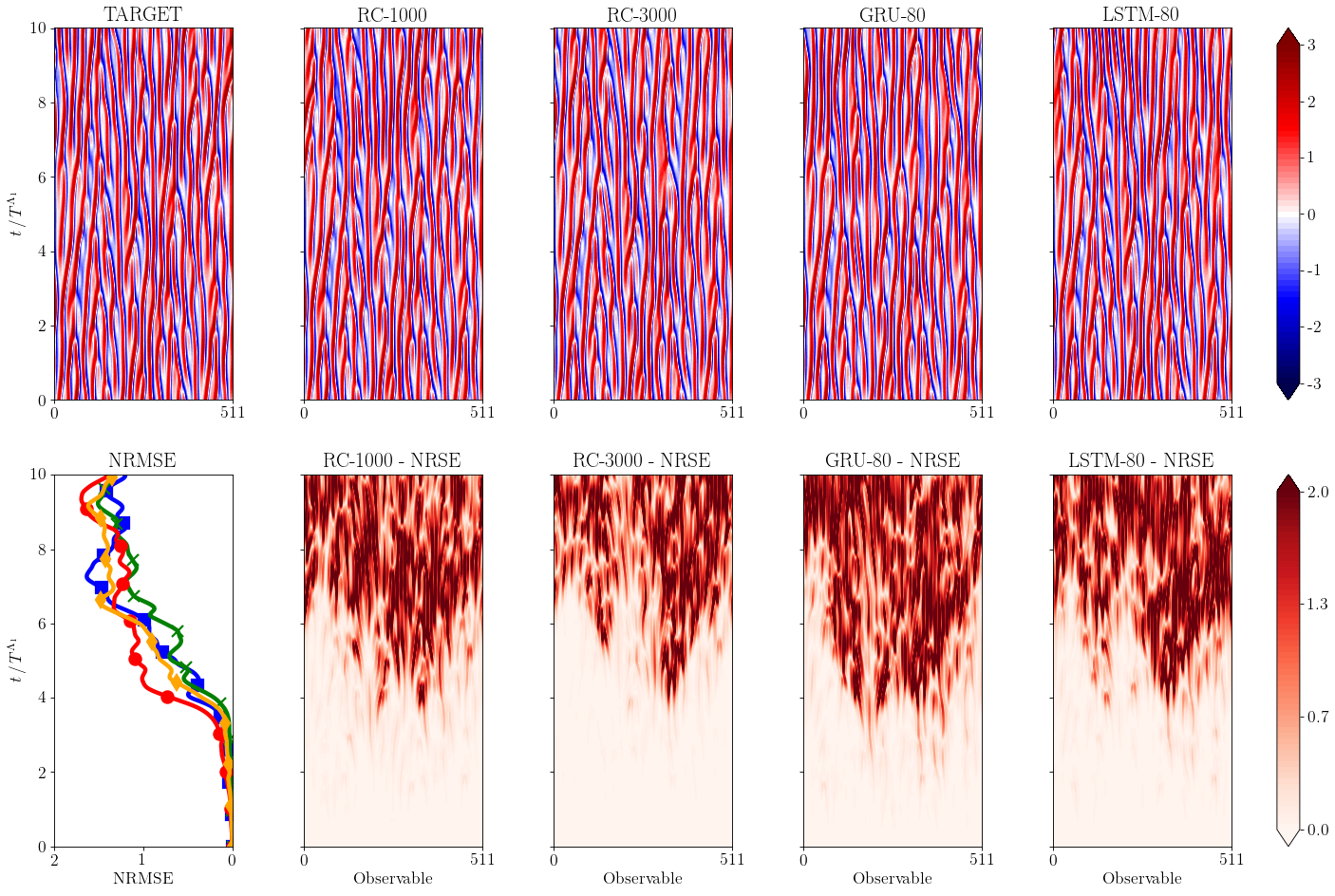}
\label{fig:KSGP512:P_CONTOUR_BBO_NAP_2_M_RDIM_512_IC20}
\caption{
Contour plots of a spatio-temporal forecast on the Kuramoto-Sivashinsky equation in the testing dataset with parallel GRU, LSTM, and RC networks along with the true (target) evolution and the associated NRSE contours in the Kuramoto-Sivashinsky system with the full state as an observable $d_o=512$.
The evolution of the component average NRSE (NMRSE) is plotted to facilitate comparison.
\\
RC-1000 \protect \bluelineRectangle; 
RC-3000 \protect \greenlineX; 
GRU-80 \protect \redlineCircle; 
LSTM-80 \protect \orangelineDiamond;
Groundtruth \protect \blacklineDashed;
}
\label{fig:parallel:KSGP512:contour}
\end{figure*}

The evolution of the NRMSE is given in \Cref{fig:KSGP512:P_RMNSE_BBO_NAP_2_M_RDIM_512_0}.
The predictive performance of a small LSTM network with $80$ hidden units, matches that of a large RC with $1000$ hidden units.
In \Cref{fig:KSGP512:P_POWSPEC_BBO_NAP_M_RDIM_512_0}, the power spectrum of the predicted state dynamics of each model is plotted along with the true spectrum of the equations.
The three models captured successfully the statistics of the system, as we observe a very good match.
An example of an iterative prediction with LSTM, GRU and RC models starting from an initial condition in the test dataset is provided in \Cref{fig:parallel:KSGP512:contour}.

\clearpage

%% file: sections/sections-6-lyapunov-spectrum.tex
\section{Calculation of Lyapunov Exponents in the Kuramoto-Sivashinsky Equation}
\label{sec:le}

The recurrent models utilized in this study can be used as surrogate models to calculate the Lyapunov exponents (LEs) of a dynamical system relying only on experimental time-series data.
The LEs characterize the rate of separation if positive (or convergence if negative) of trajectories that are initialized infinitesimally close in the phase space.
They can provide an estimate of the attractor dimension according to the Kaplan-Yorke formula~\citep{Kaplan1979}.
Early efforts to solve the challenging problem of data-driven LE identification led to local approaches~\citep{Wolf1985,Sano1985} that are computationally inexpensive at the cost of requiring a large amount of data.
Other approaches fit a global model to the data~\citep{Maus2013} and calculate the LE spectrum using the Jacobian algorithm.
These approaches were applied to low-order systems.

A recent machine learning approach utilizes deep convolutional neural networks for LE and chaos identification, without estimation of the dynamics~\citep{Makarenko2018}.
An RC-RNN approach capable of uncovering the whole LE spetrum in high-dimensional dynamical systems is proposed in~\citep{Pathak2018a}.
The method is based on the training of a surrogate RC model to forecast the evolution of the state dynamics, and the calculation of the Lyapunov spectrum of the hidden state of this surrogate model.
The RC method demonstrates excellent agreement for all positive Lyapunov exponents and many of the negative exponents for the KS equation with $L=60$~\citep{Pathak2018a}, alleviating the problem of spurious Lyapunov exponents of delay coordinate embeddings~\citep{Dechert1996}. 
We build on top of this work and demonstrate that a GRU trained with BPTT can reconstruct the Lyapunov spectrum accurately with lower error for all positive Lyapunov exponents at the cost of higher training times.

The Lyapunov spectrum of the KS equation is computed by solving the KS equations in the Fourier space with a fourth order time-stepping method called ETDRK4~\citep{Kassam2005} and utilizing a QR decomposition approach as in~\citep{Pathak2018a}.
The Lyapunov spectrum of the RNN and RC surrogate models is computed based on the Jacobian of the hidden state dynamics along a reference trajectory, while Gram-Schmidt orthonormalization is utilized to alleviate numerical divergence.
We employ a GRU-RNN over LSTM-RNN, due to the fact that the latter has two coupled hidden states, rendering the computation of the Lyapunov spectrum mathematically more involved and computationally more expensive.
The interested reader can refer to the Appendix for the details of the method.
The identified maximum LE is $\Lambda_1 \approx0.08844$.
In this work, a large RC with $d_h=9000$ nodes is employed for LS calculation in the Kuramoto-Sivashinsky equation with parameter $L=60$ and $D=128$ grid points as in \citep{Pathak2018a}.
The largest LE identified in this case is $\Lambda_1 \approx 0.08378$ leading to a relative error of $5.3\%$.
In order to evaluate the efficiency of RNNs, we utilize a large GRU with $d_h=2000$ hidden units.
An iterative RNN roll-out of $N=10^4$ total time-steps was needed to achieve convergence of the spectrum.
The largest Lyapunov exponent identified by the GRU is $\Lambda_1 \approx 0.0849$ reducing the error to $\approx 4 \%$.
Both surrogate models identify the correct Kaplan-Yorke dimension $\text{KY}\approx15$, which is the largest LE such that $\sum_i \Lambda_i>0$.

\begin{figure*}
\centering
\begin{subfigure}[t]{0.45\textwidth}
\centering
\includegraphics[width=0.95\textwidth]{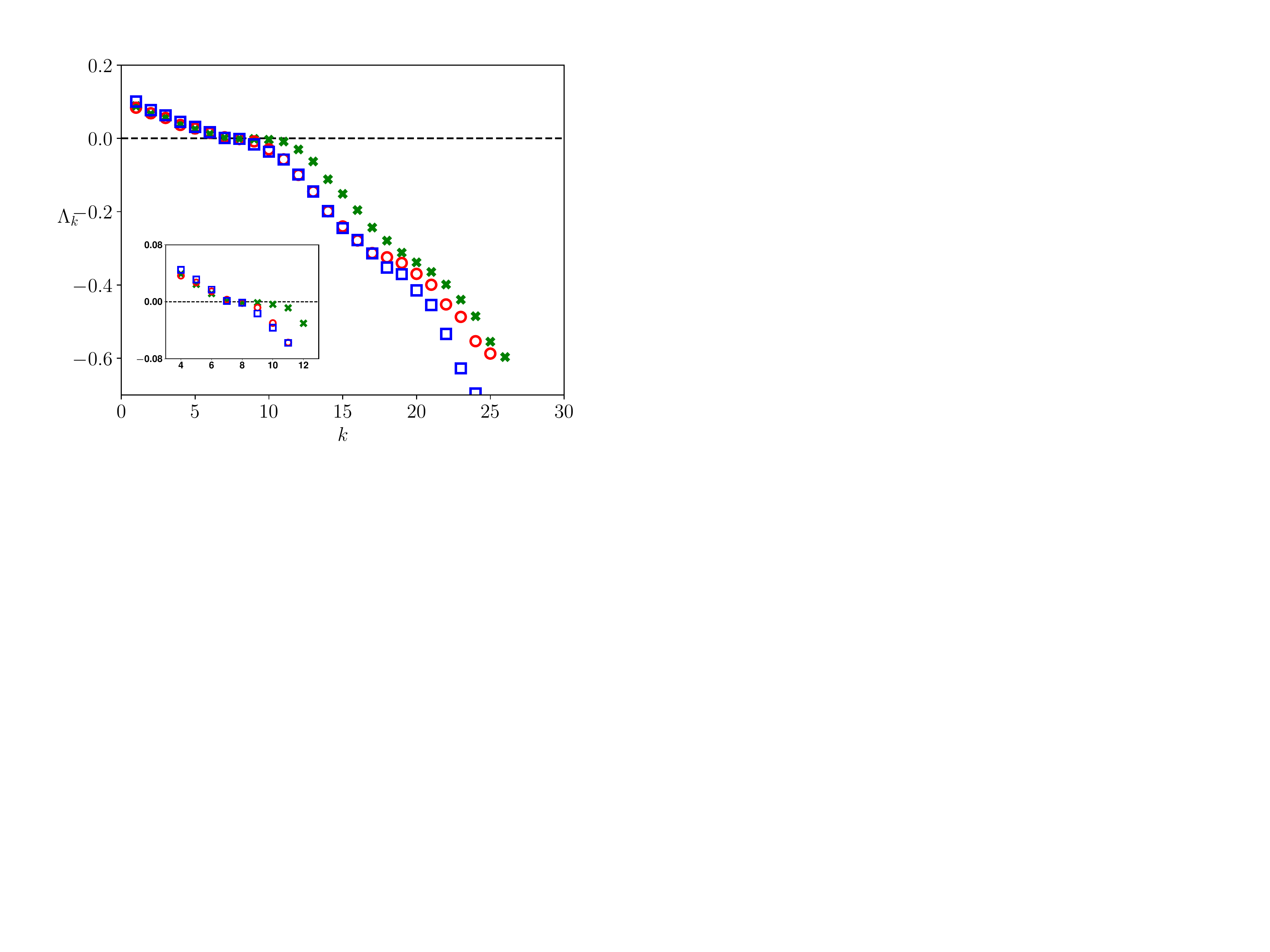}
\caption{Lyapunov spectrum.}
\label{fig:ks_le}
\vspace{0.5cm}
\end{subfigure}
\centering
\begin{subfigure}[t]{0.45\textwidth}
\centering
\includegraphics[width=0.95\textwidth]{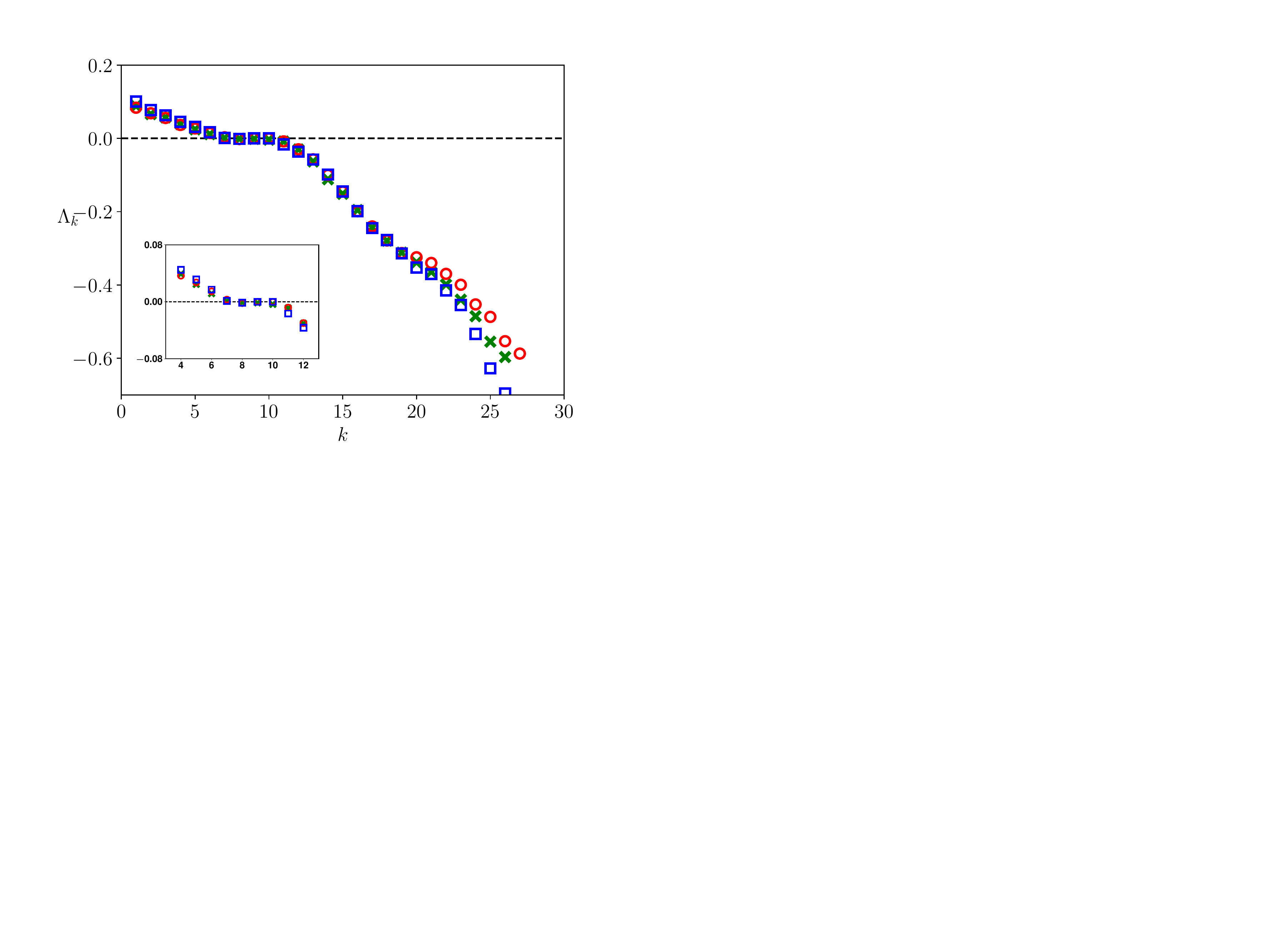}
\caption{Lyapunov spectrum augmented with $\Lambda_7,\Lambda_8$ set to zero.}
\label{fig:ks_le_augmented}
\vspace{0.5cm}
\end{subfigure}
\caption{
\textbf{(a)} Estimated Lyapunov exponents $\Lambda_k$ of the KS equation with $L=60$.
The true Lyapunov exponents are illustrated with green crosses, red circles are calculated with the RC surrogate, while the blue rectangles with GRU.
In \textbf{(b)} we augment the computed spectrums with the two zero Lyapunov exponents $\Lambda_7,\Lambda_8$.
Inset plots zoom in the zero crossing regions.
\\
True \protect \greenCross;
RC \protect \redCircle;
GRU \protect \blueRectangle;
}
\label{fig:ks_le_all}
\end{figure*}

The first $26$ Lyapunov exponents computed the GRU, RC as well as using the true equations of the Kuramoto-Sivashinsky are plotted in \Cref{fig:ks_le_all}.
We observe a good match between the positive Lyapunov exponents by both GRU and RC surrogates.
The positive Lyapunov exponents are characteristic of chaotic behavior.
However, the zero Lyapunov exponents $\Lambda_7$ and $\Lambda_8$ cannot be captured neither with RC nor with RNN surrogates.
This is also observed in RC in \citep{Pathak2018a}, and apparently the GRU surrogate employed in this work do not alleviate the problem.
In \Cref{fig:ks_le_augmented}, we augment the RC and the GRU spectrum with these two additional exponents to illustrate that there is an excellent agreement between the true LE and the augmented LS identified by the surrogate models.

\begin{figure*}
\centering
\begin{subfigure}[t]{0.45\textwidth}
\centering
\includegraphics[width=1\textwidth]{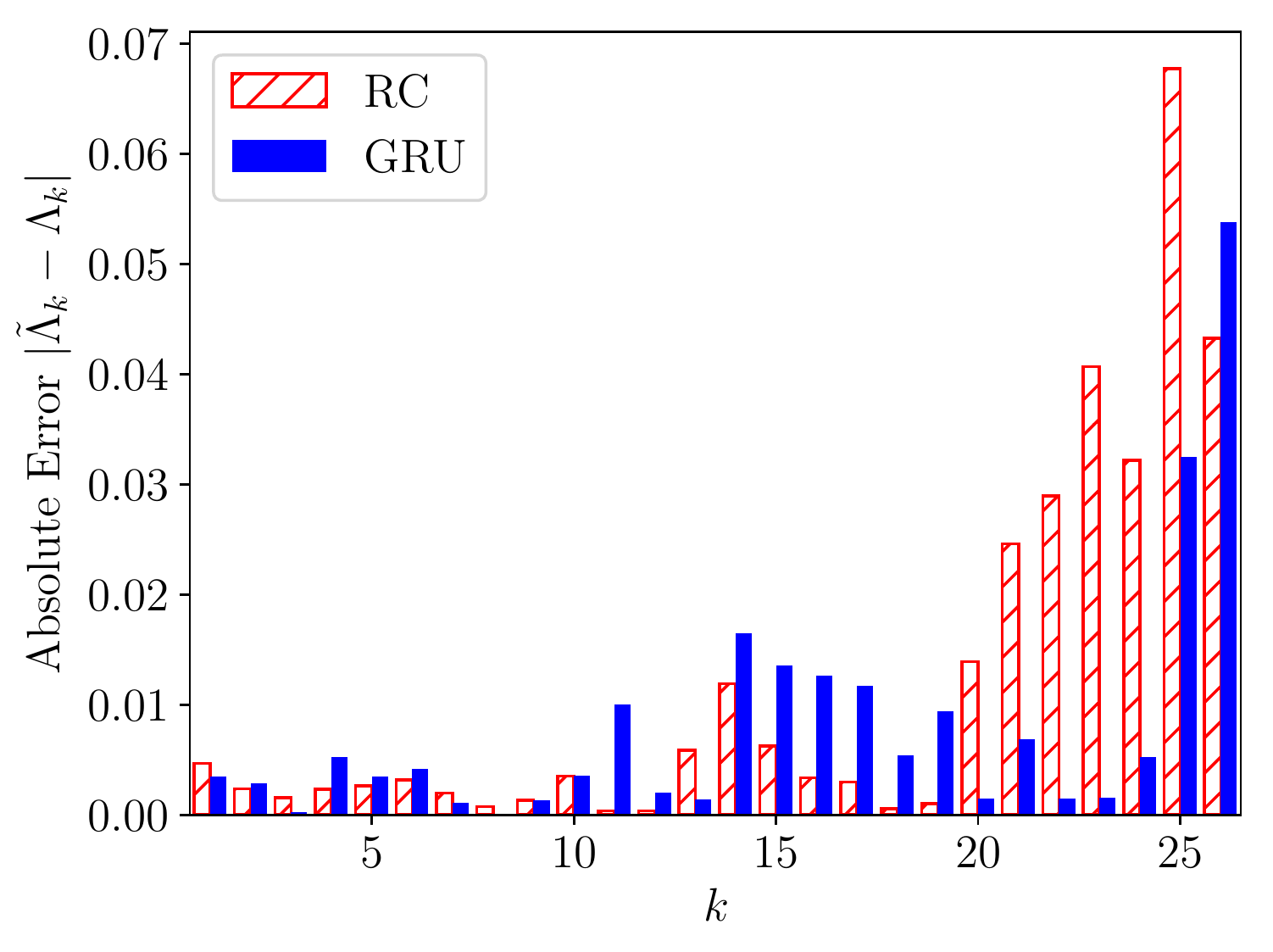}
\caption{Absolute error}
\label{fig:ks_le_abs_error}
\vspace{0.5cm}
\end{subfigure}
\centering
\begin{subfigure}[t]{0.45\textwidth}
\centering
\includegraphics[width=1\textwidth]{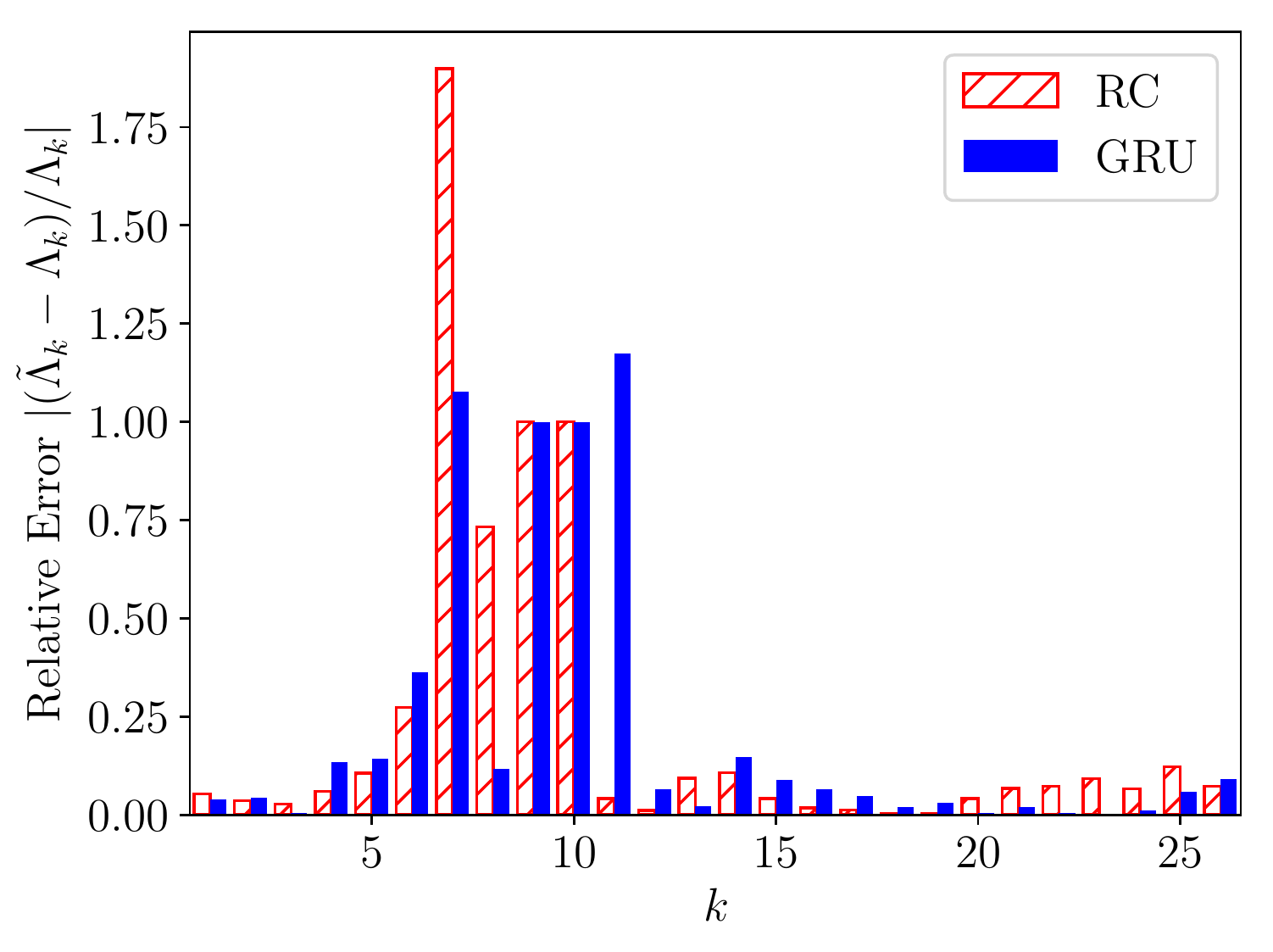}
\caption{Relative error}
\label{fig:ks_le_rel_error}
\vspace{0.5cm}
\end{subfigure}
\caption{
\textbf{(a)} Absolute and \textbf{(b)} Relative error of the LE spectrum of the KS equation with $L=60$.
The LE spectrum identified using the GRU shows a better agreement with the spectrum identified by the Kuramoto-Sivashinsky equations.
}
\label{fig:ks_error_all}
\end{figure*}

The relative and absolute errors in the spectrum calculation is illustrated in \Cref{fig:ks_error_all}.
After augmenting with these zero LE, we get a mean absolute error of $0.012$ for RC and $0.008$ for GRU.
The mean relative error is $0.23$ for RC, and $0.22$ for GRU.
As a conclusion, GRU in par with RC networks can be used to replicate the chaotic behavior of a reference system and calculate the Lyapunov spectrum accurately. 

\clearpage

%% file: sections/sections-7-conclusion.tex
\section{Conclusions}
\label{sec:conclusion}

In this work, we employed several variants of recurrent neural networks and reservoir computing to forecast the dynamics of chaotic systems. We present a comparative study based on their efficiency in capturing temporal dependencies, evaluate how they scale to systems with high-dimensional state space, and how to guard against overfitting. We highlight the advantages and limitations of these methods and elucidate their applicability to forecasting spatiotemporal dynamics.

We considered three different types of RNN cells that alleviate the well-known vanishing and exploding gradient problem in Back-propagation through time training (BPTT), namely LSTM, GRU and Unitary cells.
We benchmarked these networks against  reservoir computers with random hidden to hidden connection weights, whose training procedure amounts to least square regression on the output weights.

The efficiency of the models in capturing temporal dependencies in the reduced order state space is evaluated on the Lorenz-96 system in two different forcing regimes $F=\{ 8, 10 \}$, by constructing a reduced order observable using Singular Value Decomposition (SVD) and keeping the most energetic modes.
Even though this forecasting task is challenging due to (1) chaotic dynamics and (2) reduced order information, LSTM and GRU show superior forecasting ability to RC utilizing similar amounts of memory at the cost of higher training times.
GRU and LSTM models demonstrate stable behavior in the iterative forecasting procedure in the sense that the forecasting error usually does not diverge, in stark contrast to RC and Unitary forecasts.
Large RC models tend to overfit easier than LSTM/GRU models, as the latter are utilizing validation-based early stopping and regularization techniques (e.g., Zoneout, Dropout) that guard against overfitting which are not directly applicable to RC.
Validation in RC amounts to tuning an additional hyperparameter, the Tikhonov regularization.
However, RC shows excellent forecasting efficiency when the full state of the system is observed, outperforming all other models by a wide margin, while also reproducing the frequency spectrum of the underlying dynamics.

RNNs and RC both suffer from scalability problems in high-dimensional systems, as the required hidden state size $d_h$ to capture the high-dimensional dynamics can become prohibitively large especially with respect to the computational expense of training.
In order to scale the models to high-dimensional systems we employ a parallelization scheme that exploits the local interactions in the state of a dynamical system.
As a reference, we consider the Lorenz-96 system and the Kuramoto-Sivashinsky equation, and we train parallel RC, GRU, and LSTM models of various sizes.
Iterative forecasting with parallel Unitary models diverged after a few timesteps in both systems. 
Parallel GRU, LSTM and RC networks reproduced the long-term attractor climate, as well as the power spectrum of the state of the Lorenz-96 and the Kuramoto-Sivashinsky equation matched with the predicted ones.

In the Lorenz-96 and the Kuramoto-Sivashinsky equation, the parallel LSTM and GRU models exhibited similar predictive performance compared to the parallel RC.
The memory requirements of the models are comparable.
RC networks require large reservoirs with $1000-6000$ nodes per member to reach the predictive performance of parallel GRU/LSTM with a few hundred nodes, but their training time is significantly lower.


Last but not least, we evaluated and compared the efficiency of GRU and RC networks in capturing the Lyapunov spectrum of the KS equation.
The positive Lyapunov exponents are captured accurately by both RC and GRU.
Both networks cannot reproduce two zero LEs $\Lambda_7$ and $\Lambda_8$.
When these two are discarded from the spectrum, GRU and RC networks show comparable accuracy in terms of relative and absolute error of the Lyapunov spectrum.

Further investigation on the underlying reasons why the RNNs and RC cannot capture the zero Lyapunov exponents is a matter of ongoing work.
Another interesting direction could include studying the memory capacity of the networks.
This could offer more insight into which architecture and training method is appropriate for tasks with long-term dependencies.
Moreover, we plan to investigate a coupling of the two training approaches to further improve their predictive performance, for example a network can utilize both RC and LSTM computers to identify the input to output mapping.
While the weights of the RC are initialized randomly to satisfy the echo state property, the output weights alongside with the LSTM weights can be optimized by back-propagation.
This approach, although more costly, might achieve higher efficiency, as the LSTM is used as a residual model correcting the error that a plain RC would have.

Although we considered a batched version of RC training to reduce the memory requirements, further research is needed to alleviate the memory burden associated with the matrix inversion (see \Cref{app:rcmemoryefficient},\Cref{app:eq:rcmemoryefficient}) and the numerical problems associated with the eigenvalue decomposition of the sparse weight matrix.

Further directions could be the initialization of RNN weights with RC based heuristics based on the echo state property and fine-tuning with BPTT.
Another promising direction is to evaluate the models in terms of the amount of data needed to learn the system dynamics.
This is possible for the plain cell RNN, where the heuristics are directly applicable.
However, in more complex architectures like the LSTM or the GRU, more sophisticated initialization schemes that ensure some form of echo state property have to be investigated.
The computational cost of training networks of the size of RC with back-propagation is also challenging.
This hybrid training method is an interesting future direction.

One topic not covered in this work, is invertibility of the models, when forecasting the full state dynamics.
Non-invertible models like the RNNs trained in this work, may suffer from spurious dynamics not present the training data and the underlying governing equations~\citep{gicquel1998noninvertibility, frouzakis1997some}.
Invertible RNNs may constitute a promising alternative to further improve accurate short-term prediction and capturing of the long-term dynamics.

In conclusion, recurrent neural networks for data-driven surrogate modeling and forecasting of chaotic systems can efficiently be used to model high-dimensional dynamical systems, can be parallelized alleviating scaling problems and constitute a promising research subject that requires further analysis.

%% file: sections/sections-8-misc.tex
\section{Data and Code}
\label{sec:code}

The code and data will be available upon publication in the following link \href{https://github.com/pvlachas/RNN}{https://github.com/pvlachas/RNN} to assist reproducibility of the results.
The software was written in Python utilizing  Tensorflow~\citep{tensorflow} and Pytorch~\citep{Paszke2017}  for automatic differentiation and the design of the neural network architectures.

\section{Acknowledgments}
\label{sec:acknowledgments}

We thank Guido Novati for helpful discussions and valuable feedback on this manuscript.
Moreover, we would like to acknowledge the time and effort of Prof. Herbert Jaeger along with two anonymous reviewers whose insightful and thorough feedback led to substantial improvements of the manuscript.
TPS has been supported through the ARO-MURI grant W911NF-17-1-0306.
This work has been supported at UMD by DARPA under grant number DMS51813027.
We are also thankful to the Swiss National Supercomputing Centre (CSCS) providing the necessary computational resources under Projects s929.

%% file: appendix/appendix-0-rc-memory.tex
\section{Memory Efficient Implementation of RC Training}
\label{app:rcmemoryefficient}

In order to alleviate the RAM requirement for the computation of the RC weights we resort to a batched approach.
Assuming the hidden reservoir size is given by $\bm{h} \in \R^{d_h}$, by teacher forcing the RC network with true data from the system for $d_N$ time-steps and stacking the evolution of the hidden state in a single matrix we end up with matrix $\mathbf{H} \in \R^{d_N \times d_h}$.
Moreover, by stacking the target values, which are the input data shifted by one time-step, we end up in the target matrix $\mathbf{Y} \in \R^{d_N \times d_o}$, where $d_o$ is the dimension of the observable we are predicting.
In order to identify the output weights $W_{out} \in \R^{d_o \times d_h}$, we need to solve the linear system of $d_N \cdot d_o$ equations
\begin{equation}
\mathbf{H} W_{out}^T
=
\mathbf{Y}.
\label{eq:linearSystem}
\end{equation}

A classical way to solve this system of equations is based on the Moore-Penrose inverse (pseudo-inverse) computed using
\begin{equation}
W_{out} = \underbrace{\mathbf{Y}^T \mathbf{H}}_{\overline{\mathbf{Y}}} \Big( \underbrace{\mathbf{H}^T \mathbf{H}}_{\overline{\mathbf{H}}} + \eta \mathbf{I} \Big)^{-1}
\label{app:eq:rcmemoryefficient}
\end{equation}
where $\eta$ is the Tikhonov regularization parameter and $\mathbf{I}$ the unit matrix.
In our case $d_N$ is of the order of $10^5$ and $d_N>>d_h$.
To reduce the memory requirements of the training method, we compute the matrices $\overline{\mathbf{H}} = \mathbf{H}^T \mathbf{H} \in \R^{d_h \times d_h}$ and $\overline{\mathbf{Y}} = \mathbf{Y}^T \mathbf{H} \in \R^{d_o \times d_h}$ in a time-batched schedule.

Specifically, we initialize $\overline{\mathbf{Y}}=\bm{0}$ and $\overline{\mathbf{H}}=\bm{0}$.
Then every $d_{n}$ time-steps with $d_n << d_N$, we compute the batch matrix $\overline{\mathbf{H}_b} = \mathbf{H}_b^T \mathbf{H}_b \in \R^{d_h \times d_h}$, where $\mathbf{H}_b \in \R^{d_n \times d_h}$ is formed by the stacking the hidden state only for the last $d_n$ time-steps.
In the same way, we compute $\overline{\mathbf{Y}_b} = \mathbf{Y}_b^T \mathbf{H}_b \in \R^{d_o \times d_h}$, where $\mathbf{Y}_b \in \R^{d_n \times d_o}$ is formed by the stacking of the target data for the last $d_n$ time-steps.
After every batch computation we update our beliefs with $\overline{\mathbf{H}} \leftarrow \overline{\mathbf{H}} + \overline{\mathbf{H}_b}$ and $\overline{\mathbf{Y}} \leftarrow \overline{\mathbf{Y}} + \overline{\mathbf{Y}_b}$.

In addition, we also experimented with two alternative solvers for the linear system \Cref{eq:linearSystem} in the Lorenz-96.
We tried a dedicated regularized least-squares routine utilizing an iterative procedure (\verb!scipy.sparse.linalg.lsqr!) and a method based on stochastic gradient descent.
We considered the solver as an additional hyperparameter of the RC models.
After testing the solvers in Lorenz-96 systems, we found out that the method of pseudo-inverse provides the most accurate results.
For this reason, and to spare computational resources, we used this method for the Kuramoto-Sivashinsky system.

%% file: appendix/appendix-1-rc-noise-reg.tex
\section{Regularizing Training with Noise}

\begin{figure*}
\centering
\begin{subfigure}[t]{0.45\textwidth}
\centering
\includegraphics[height=3.5cm]{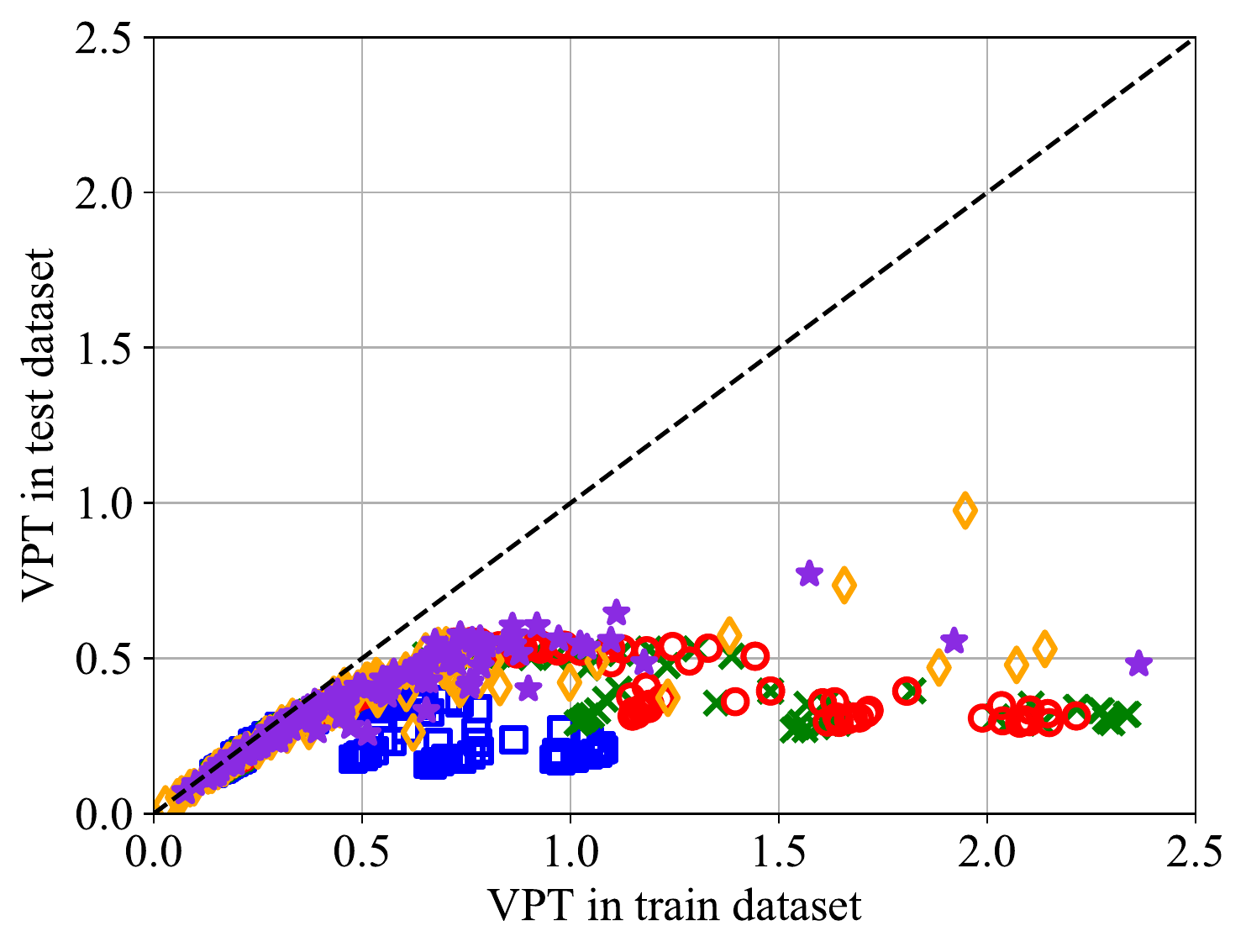}
\caption{$d_o=35$, $F=8$}
\label{fig:L96F8GP40R40:REG_NOISE_OFSP_BBO_NAP_RDIM_35}
\end{subfigure}
\begin{subfigure}[t]{0.45\textwidth}
\centering
\includegraphics[height=3.5cm]{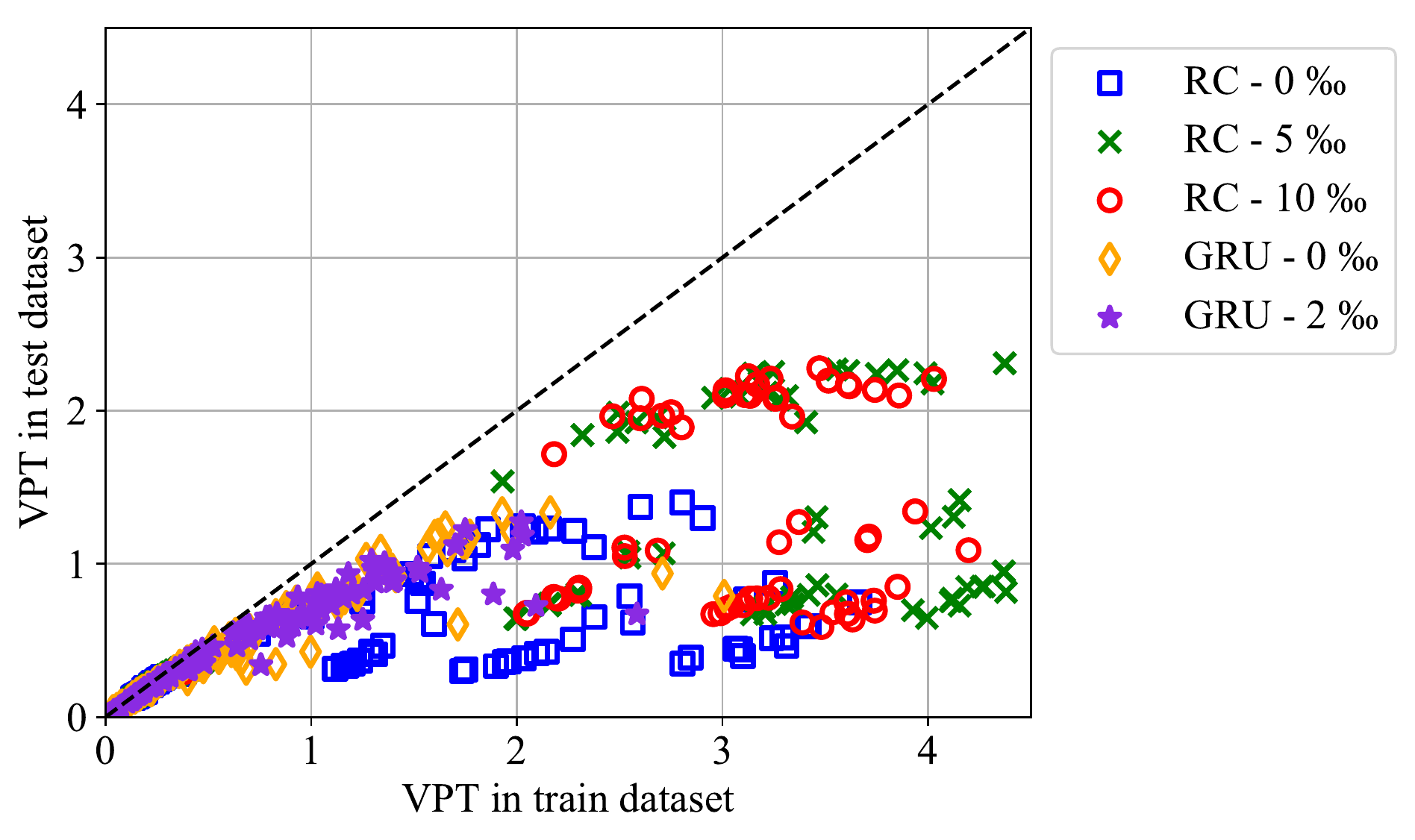}
\caption{$d_o=35$, $F=10$}
\label{fig:L96F8GP40R40:REG_NOISE_OFSP_BBO_NAP_RDIM_40_LEGEND}
\end{subfigure}
\begin{subfigure}[t]{0.45\textwidth}
\centering
\includegraphics[height=3.5cm]{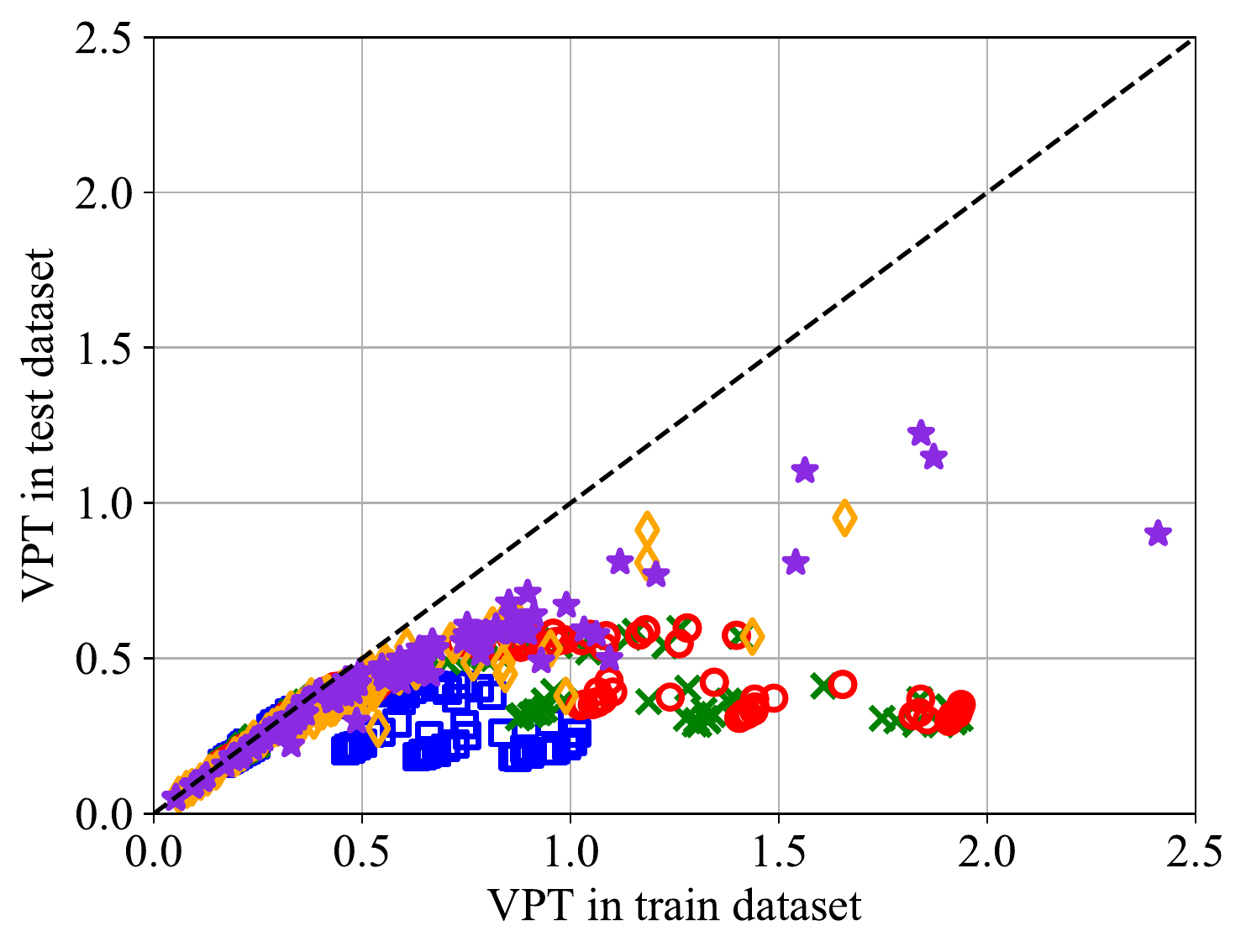}
\caption{$d_o=40$, $F=8$}
\label{fig:L96F10GP40R40:REG_NOISE_OFSP_BBO_NAP_RDIM_35}
\end{subfigure}
\begin{subfigure}[t]{0.45\textwidth}
\centering
\includegraphics[height=3.5cm]{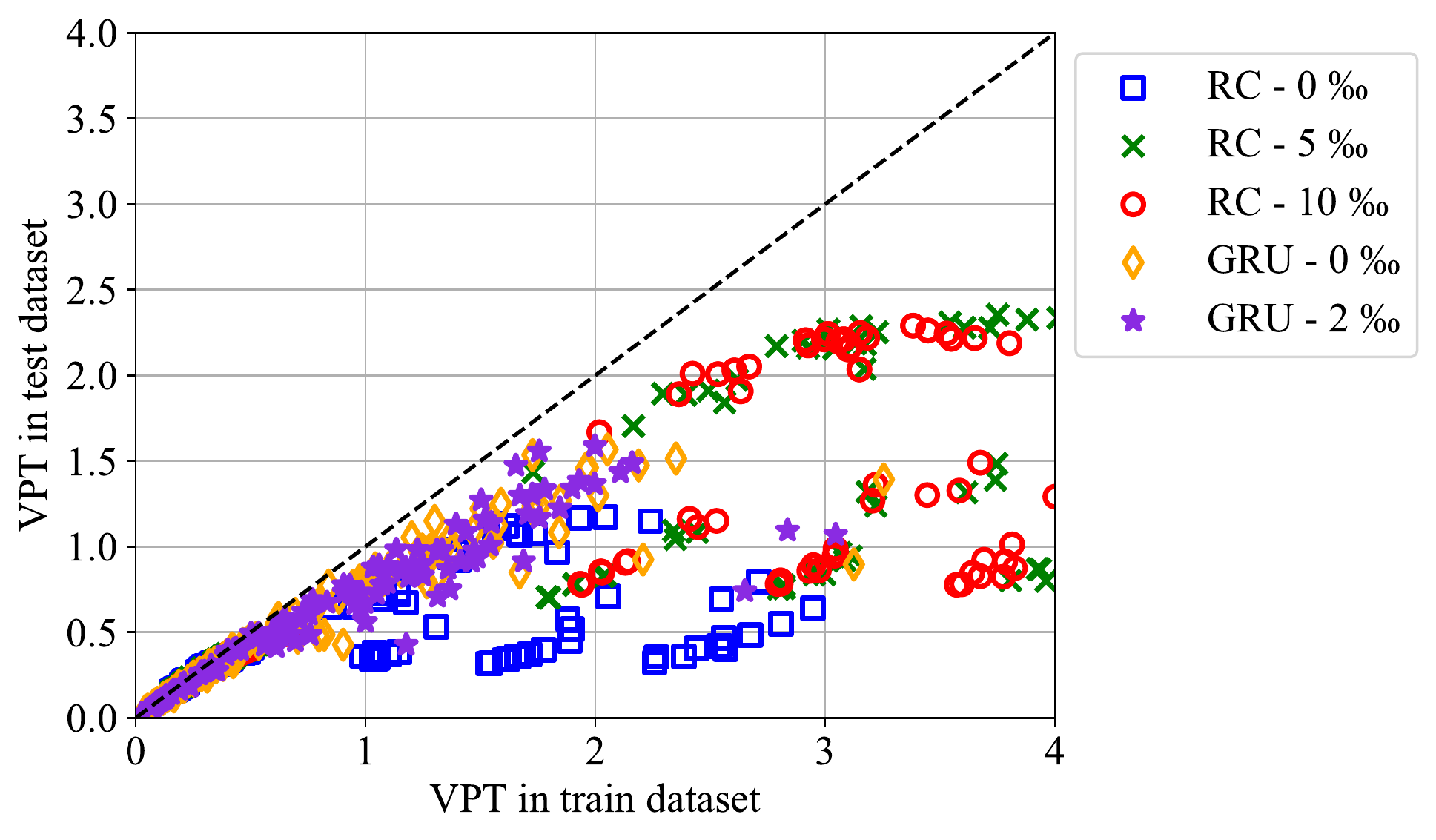}
\caption{$d_o=40$, $F=10$}
\label{fig:L96F10GP40R40:REG_NOISE_OFSP_BBO_NAP_RDIM_40_LEGEND}
\end{subfigure}
\caption{
VPT in the testing data plotted against VPT in the training data for RC and GRU models trained with added noise of different levels in the data.
Noise only slightly varies the forecasting efficiency.
In contrast, the effectiveness of RC in forecasting the full-order system is increased as depicted in plots \textbf{(b)} and \textbf{(d)}.
}
\label{fig:L96training_reg_noise}
\end{figure*}

In our study, we investigate the effect of noise to the training data.
In \Cref{fig:L96training_reg_noise}, we plot the Valid Prediction Time (VPT) in the testing data with respect to the VPT that each model achieves in the training data.
We find out that RC models trained with additional noise of $5-10$ \textperthousand  not only achieve better generalization, but their forecasting efficiency improves in both training and testing dataset.
Moreover, the effect of divergent predictions by iterative forecasts is alleviated significantly.
In contrast, adding noise does not seem to have an important impact on the performance of GRU models.

%% file: appendix/appendix-2-dim-red.tex
\clearpage
\section{Dimensionality Reduction with Singular Value Decomposition}

Singular Value Decomposition (SVD) can be utilized to perform dimensionality reduction in a dataset by identifying the modes that capture the highest variance in the data and then performing a projection on these modes.
Assuming that a data matrix is given by stacking the time-evolution of a state $\mathbf{u} \in \mathbb{D}$ as $U=[\mathbf{u}_1, \mathbf{u}_2, \dots, \mathbf{u}_N]$, where the index $N$ is the number of data samples.
By subtracting the temporal mean $\overline{\mathbf{u}}$ and stacking the data, we end up with the data matrix $\mathbf{U}\in \mathbb{R}^{T\times D}$.
Performing SVD on  $\mathbf{U}$ leads to 
\begin{equation}
\mathbf{U} = \mathbf{M} \mathbf{\Sigma} \mathbf{V}^T, \quad \mathbf{M} \in \mathbb{R}^{N\times N}, \quad \mathbf{\Sigma} \in \mathbb{R}^{N\times D}, \quad \mathbf{V} \in \mathbb{R}^{D \times D},
\label{eq:svd}
\end{equation}
with $\Sigma$ diagonal, with descending diagonal elements.
The columns of matrix $V$ are considered the modes of the SVD, while the square $D$ singular values of $\mathbf{\Sigma}$ correspond to the data variance explained by these modes.
This variance is also referred to as energy.
In order to calculate the percentage of the total energy the square of the singular value of each mode has to be divided by the sum of squares of the singular values of all modes.
In order to reduce the dimensionality of the dataset, we first have to decide on the reduced order dimension $r_{dim}<D$.
Then we identify the eigenvectors corresponding to the most high-energetic eigenmodes.
These are given by the first columns $\mathbf{V_r}$ of $\mathbf{V}$, i.e.,
$\mathbf{V}=[\mathbf{V_r}, \mathbf{V_{-r}}]$.
We discard the low-energetic modes $\mathbf{V_{-r}}$.
The dimension of the truncated eigenvector matrix is $\mathbf{V_r} \in \mathbb{R}^{D \times r_{dim}}$.
In order to reduce the dimensionality of the dataset, each vector $\mathbf{u} \in \mathbb{D}$ is projected to $\mathbf{u}_r \in r_{dim}$ by
\begin{equation}
\mathbf{c} = \mathbf{V_r}^T \mathbf{u}  , \quad  \mathbf{c}  \in \mathbb{R}^{r_{dim}}.
\end{equation}

\begin{figure*}
\centering
\begin{subfigure}[t]{0.3\textwidth}
\centering
\includegraphics[height=3.5cm]{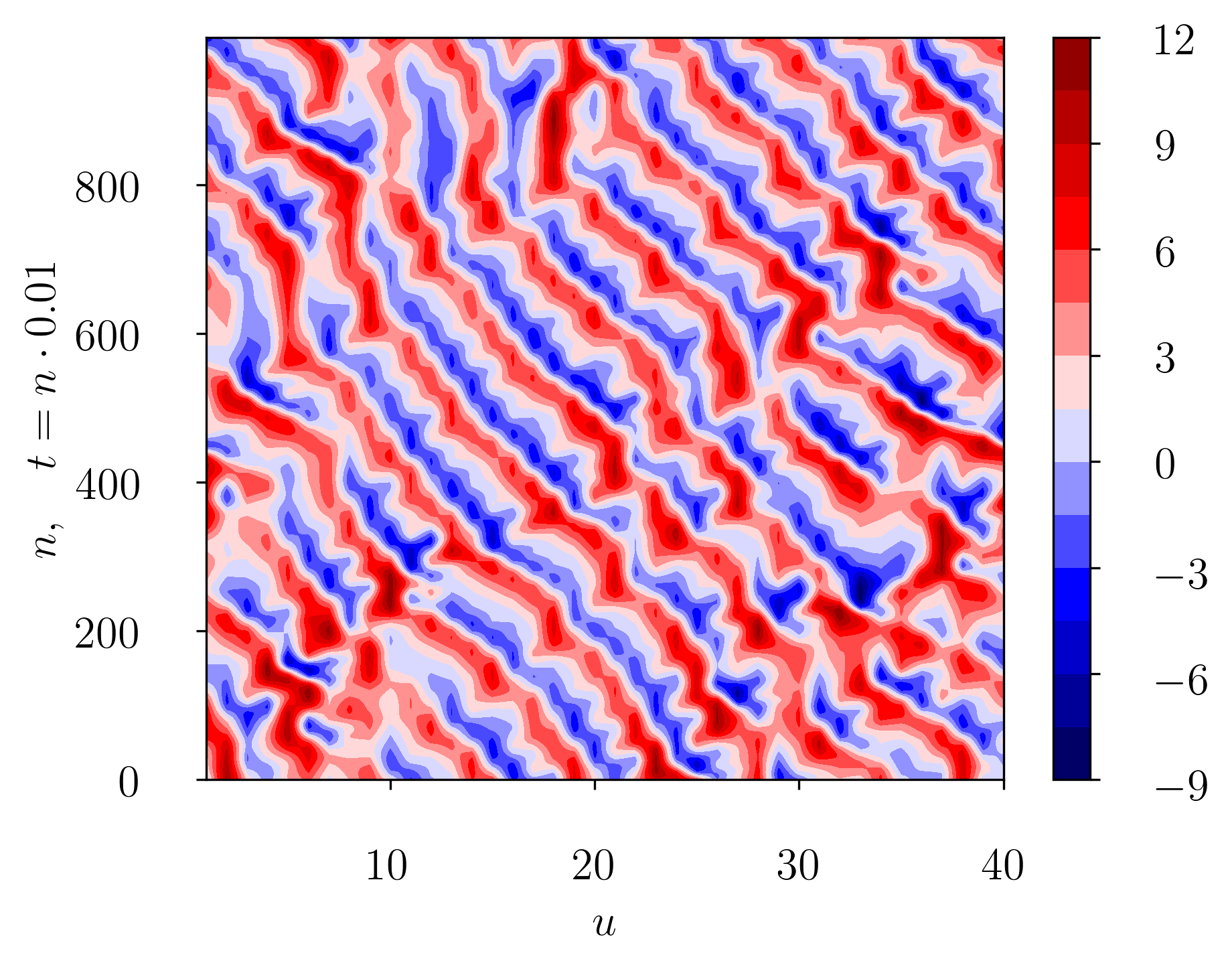}
\caption{Energy spectrum $F=8$}
\label{fig:L96Energy:Plot_U_F8}
\end{subfigure}
\begin{subfigure}[t]{0.3\textwidth}
\centering
\includegraphics[height=3.5cm]{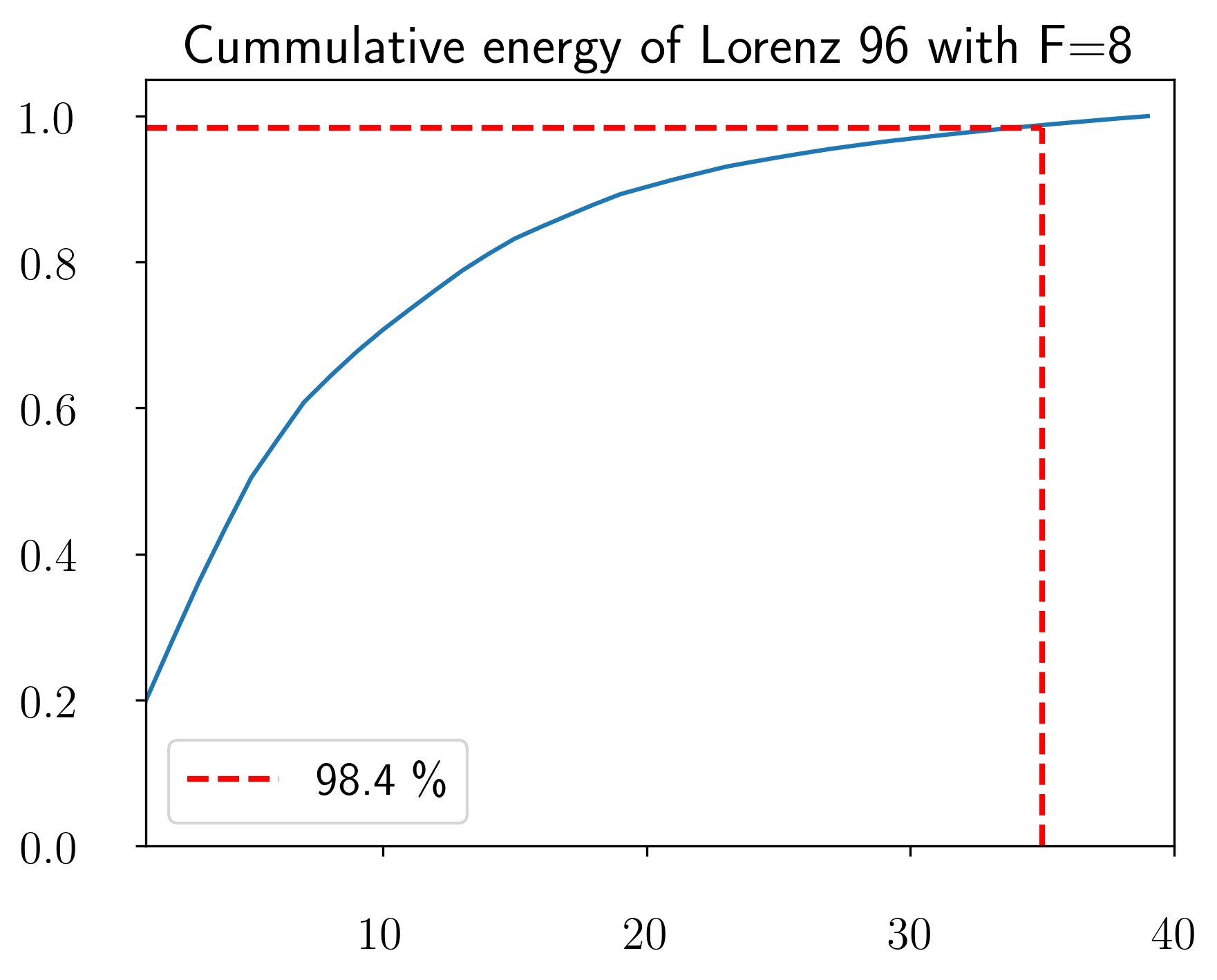}
\caption{State evolution $F=8$}
\label{fig:L96Energy:energy_F8}
\end{subfigure}
\begin{subfigure}[t]{0.3\textwidth}
\centering
\includegraphics[height=3.5cm]{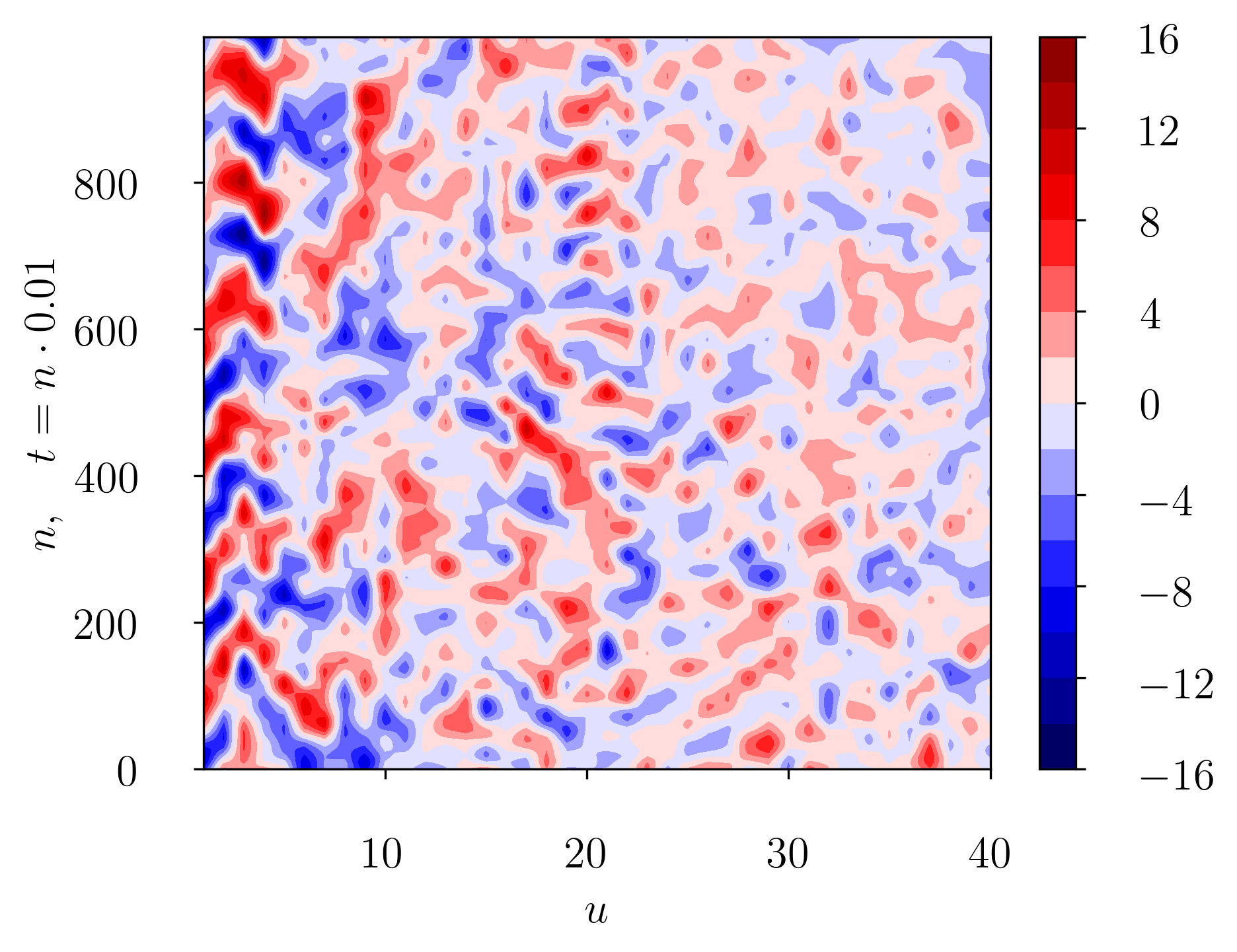}
\caption{SVD mode evolution $F=8$}
\label{fig:L96Energy:Plot_UR_F8}
\end{subfigure}
\begin{subfigure}[t]{0.3\textwidth}
\centering
\includegraphics[height=3.5cm]{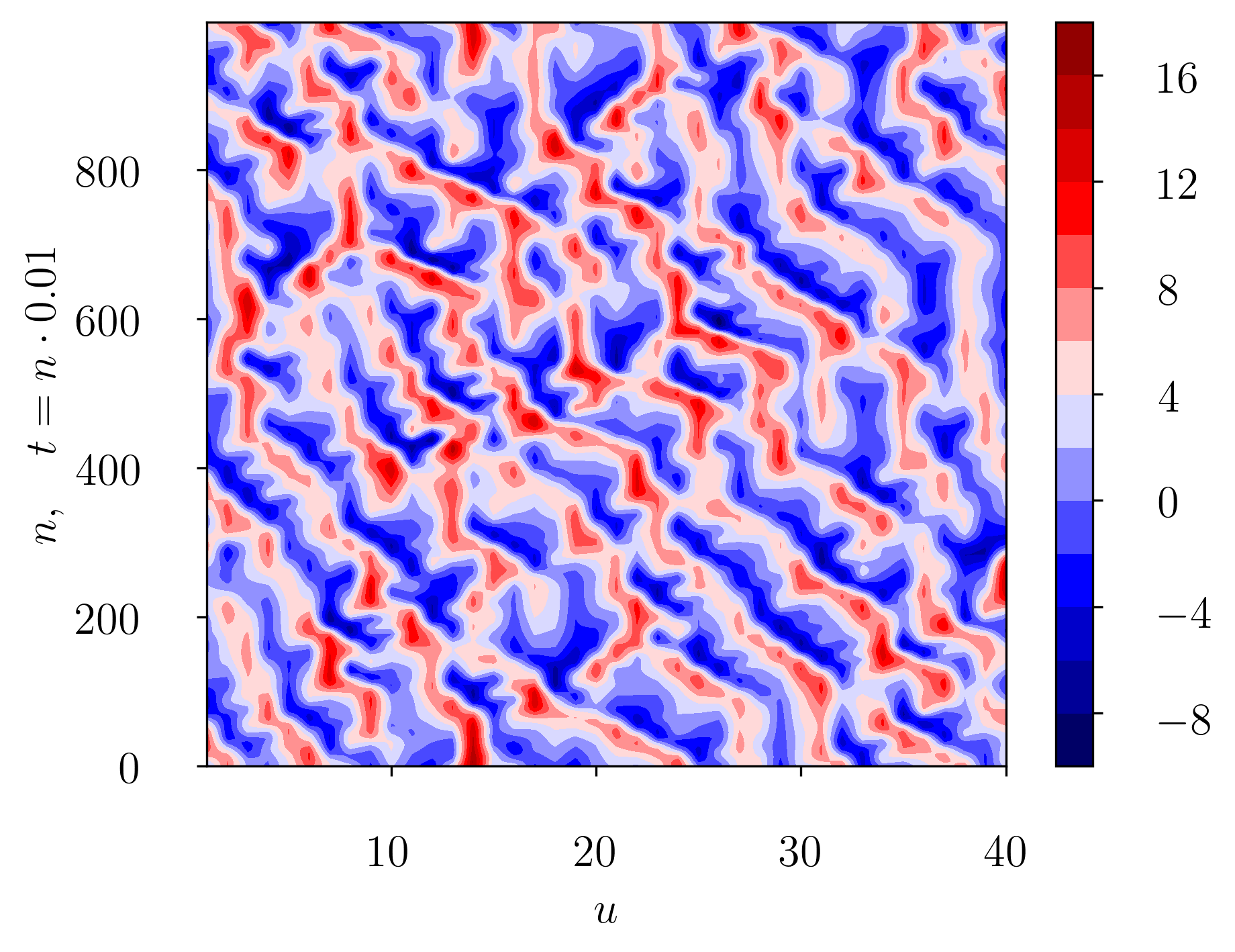}
\caption{Energy spectrum $F=10$}
\label{fig:L96Energy:Plot_U_F10}
\end{subfigure}
\begin{subfigure}[t]{0.3\textwidth}
\centering
\includegraphics[height=3.5cm]{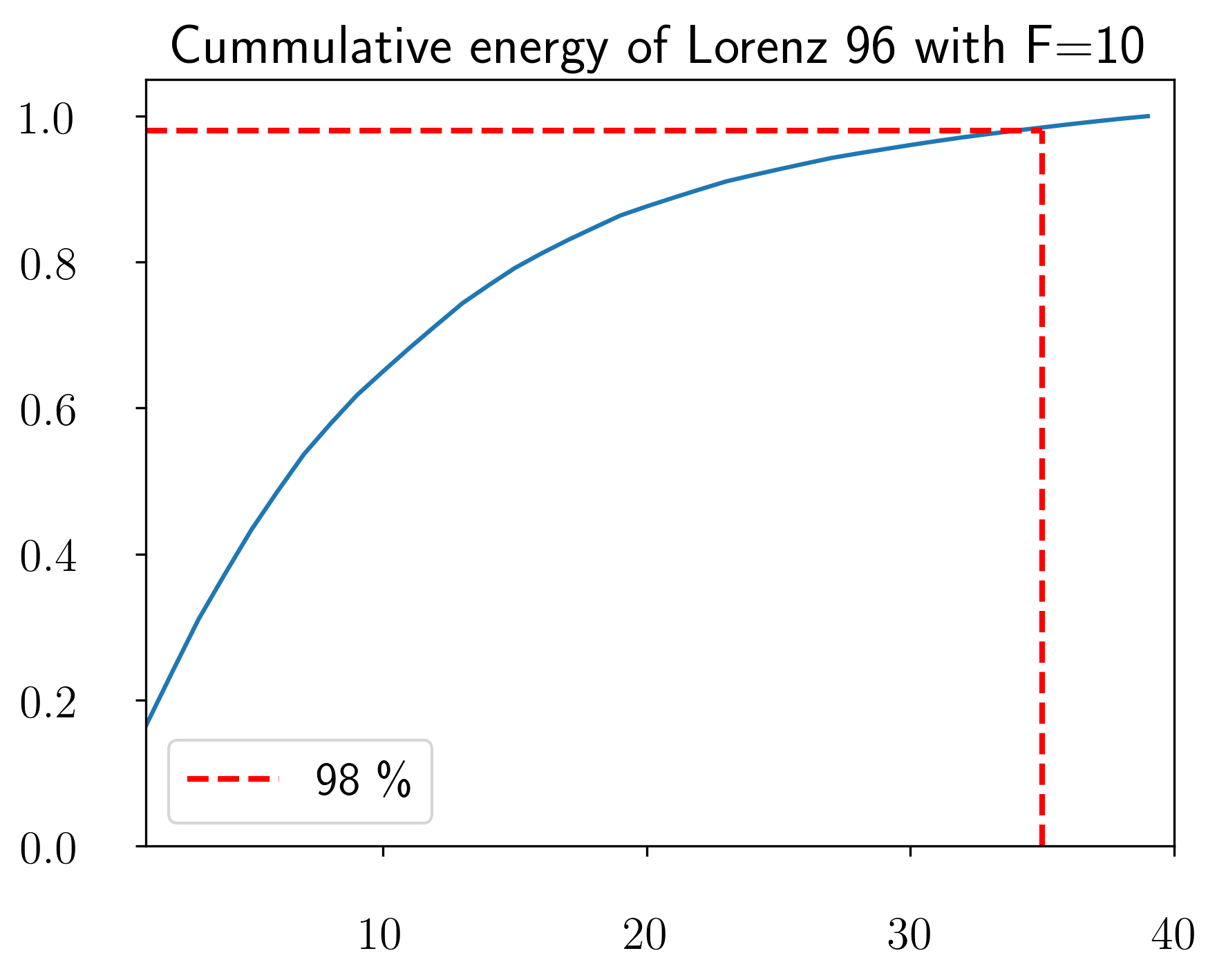}
\caption{State evolution $F=10$}
\label{fig:L96Energy:energy_F10}
\end{subfigure}
\begin{subfigure}[t]{0.3\textwidth}
\centering
\includegraphics[height=3.5cm]{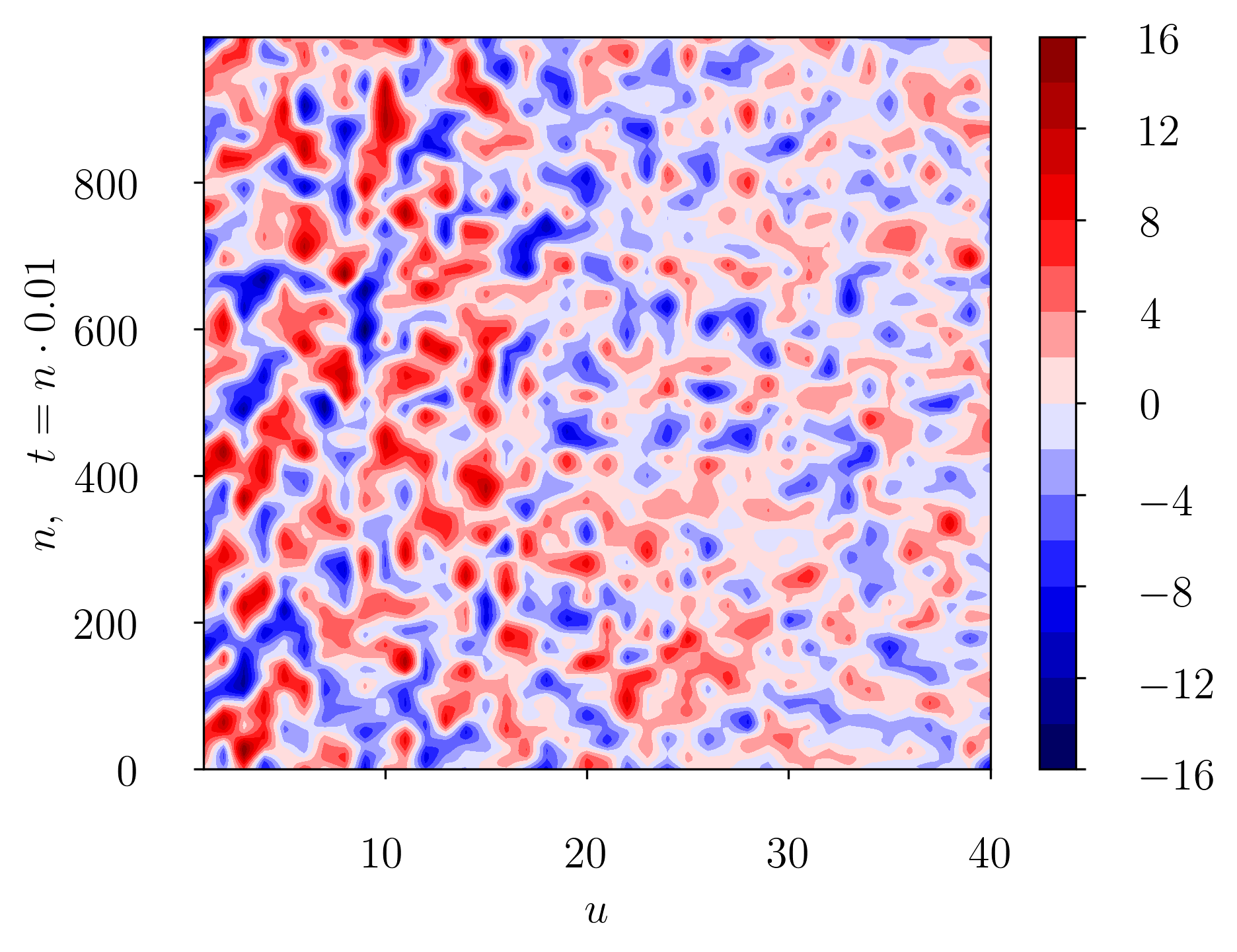}
\caption{SVD mode evolution $F=10$}
\label{fig:L96Energy:Plot_UR_F10}
\end{subfigure}
\caption{
Energy spectrum of Lorenz-96
}
\label{fig:L96Energy}
\end{figure*}

In the Lorenz-96 system, we construct a reduced order observable with $d_o =35$ modes of the system.
The cumulative energy distribution along with a contour plot of the state and the mode evolution is illustrated in \Cref{fig:L96Energy}.

%% file: appendix/appendix-3-lyapunov-spectrum.tex
\clearpage
\section{Calculation of Lyapunov Spectrum}
\label{app:sec:lamethod}

The true Lyapunov exponents of the KS equation are computed as in~\citep{Pathak2018a} by solving the KS equations in the Fourier space with a fourth order time-stepping method called ETDRK4~\citep{Kassam2005} and utilizing a QR decomposition approach.
The trained RNN model with GRU cell is used as a surrogate to compute the full Lyapunov spectrum of the Kuramoto-Sivashinsky system.
Recall that the RNN dynamics are given by
\begin{equation}
\begin{aligned}
\h_{t}&=f_h^h(\bmo_t, \h_{t-1}) \\
\bmo_{t+1}&=f_h^o(\h_{t}),
\end{aligned}
\label{eq:app:lernn}
\end{equation}
where $f_h^h$ is the hidden-to-hidden and $f_h^o$ is the hidden-to-output mapping, $\bmo \in \R^{d_o}$ is an observable of the state, and  $\h_t \in \R^{d_h}$ is the hidden state of the RNN.
All models utilized in this work share this common architecture.
They only differ in the forms of $f_h^o$ and $f_h^h$.
More importantly, the output mapping is linear, i.e.,
\begin{equation}
\bmo_{t+1}=f_h^o(\h_{t})= \W_{o} \, \h_t.
\end{equation}
The LEs are calculated based on the Jacobian $J=\frac{ d h_{t} }{d h_{t-1}}$ of the hidden state dynamics along the trajectory.
In the following we compute the Jacobian using \Cref{eq:app:lernn}.
By writing down the equations for two consecutive time-steps, we get
\begin{align}
\text{Timestep $t-1$ :  }& h_{t-1} = f_h^h(o_{t-1}, h_{t-2}) \\
&o_{t} = f_h^o(h_{t-1}) = W_o h_{t-1} \\
\text{Timestep $t$ :  }&h_{t} = f_h^h(o_{t}, h_{t-1}).
\end{align}
The partial Jacobians needed to compute the total Jacobian are:
\begin{gather}
\frac{\partial f_h^h}{\partial o} = J^{hh}_o \in \mathbb{R}^{d_h \times d_o} \\
\frac{\partial f_h^h}{\partial h} = J^{hh}_h \in \mathbb{R}^{d_h \times d_h} \\
\frac{\partial f_h^o}{\partial h} = J^{oh}_h \in \mathbb{R}^{d_o \times d_h}.
\end{gather}
In total we can write:
\begin{gather}
\frac{ d h_{t} }{d h_{t-1}} =
\frac{d 
f_h^h(o_{t}, h_{t-1})
}{d h_{t-1}} =
\frac{
\partial f_h^h(o_{t}, h_{t-1})
}{
\partial o_{t}
}
\frac{
\partial o_{t}
}{
\partial h_{t-1}
}
+
\frac{
\partial f_h^h(o_{t}, h_{t-1})
}{
\partial h_{t-1}
} \implies \\
\frac{ d h_{t} }{d h_{t-1}} = \frac{
\partial f_h^h(o_{t}, h_{t-1})
}{
\partial o_{t}
}
\frac{
\partial f_h^o(h_{t-1})
}{
\partial h_{t-1}
}
+
\frac{
\partial f_h^h(o_{t}, h_{t-1})
}{
\partial h_{t-1}
}  \implies \\
\frac{ d h_{t} }{d h_{t-1}} =
\underbrace{
J^{hh}_o \bigg\rvert_{(o_{t}, h_{t-1})}
}_{\text{evaluated at t}}
\cdot 
\underbrace{
J^{oh}_h\bigg\rvert_{h_{t-1}} + 
}_{\text{evaluated at t-1}}
\underbrace{
J^{hh}_h \bigg\rvert_{(o_{t}, h_{t-1})}
}_{\text{evaluated at t}}
\end{gather}
A product of this Jacobian along the orbit $\delta$ is developed and iteratively orthonormalized every $T_n$ steps using the Gram-Schmidt method to avoid numerical divergence and keep the columns of the matrix $R$ independent.
We check the convergence criterion by tracking the estimated LE values every $T_{c}$ time-steps.
The input provided to the algorithm is a short time-series of length $T_w$ to initialize the RNN and warm-up the hidden state $\bm{\widetilde{o}}_{1:T_{w}+1}$ (where the tilde denotes experimental or simulation data), the length of this warm-up time-series $T_w$, the number of the LE to calculate $N$, the maximum time to unroll the RNN $T$, a normalization time $T_n$ and an additional threshold $\epsilon$ used as an additional termination criterion.
The function $\mathrm{ColumnSum}(\cdot)$ computes the sum of each column of a matrix, i.e., $\mathrm{sum}(\cdot, \mathrm{axis}=1)$.
This method can be applied directly to RNNs with one hidden state like RC or GRUs.
An adaptation to the LSTM is left for future research.
The pseudocode of the algorithm to calculate the Lyapunov exponents of the RNN is given in \Cref{lecalcrnn}.

\begin{algorithm}
\caption{Algorithm to calculate Lyapunov Exponents of a trained surrogate RNN model}\label{lecalcrnn}
\begin{algorithmic}
\Procedure{LE\_RNN}{$\bm{\widetilde{o}}_{1:T_{w}+1}, T_{w}, N, T, T_{n}, \epsilon$}
\State Initialize $\h_0\leftarrow0$.
\For{$t=1:T_{w}$}
\Comment{Warming-up the hidden state of the RNN based on true data}
\State $\h_{t} \leftarrow f_h^h(\bm{\widetilde{o}}_t, \h_{t-1})$
\EndFor
\State $\h_0 \leftarrow \h_{T_{w}}$
\State $\bmo_1 \leftarrow \bm{\widetilde{o}}_{T_{w}+1}$
\State Pick a random orthonormal matrix $\delta \in \R^{d_h \times N_{LE}}$.
\Comment{Initializing $N_{LE}$ deviation vectors}
\State $\widetilde{T} \leftarrow T/T_{n}$
\State Initialize $\widetilde{R}\leftarrow\bm{0} \in \R^{N \times \widetilde{T}}$.
\State $l_{prev}, l \leftarrow \bm{0} \in \R^{N}$
\Comment{Initializing the $N$ LE to zero.}
\State $J_0 \leftarrow \nabla_{\bm{h}} f_h^o(\h_{0})$.
\For{$t=1:T$}
\Comment{Evolve the RNN dynamics}
\State $\h_{t} \leftarrow f_h^h(\bmo_t, \h_{t-1})$
\State $\bmo_{t+1} \leftarrow f_h^o(\h_{t})$
\State $J_1 \leftarrow \nabla_{\bm{h}} f_h^h(\bmo_{t+1}, \h_{t})$.
\Comment{Calculating the partial Jacobians}
\State $J_2 \leftarrow \nabla_{\bm{o}} f_h^h(\bmo_{t+1}, \h_{t})$.
\State $J \leftarrow J_1 + J_2 \cdot J_0$.
\Comment{Calculating the total Jacobian}
\State $\delta \leftarrow J \cdot \delta$
\Comment{Evolving the deviation vectors $\delta$}
\If{$\mod(t,T_{norm})=0$}
\Comment{Re-orthonormalizing with $QR$-decomposition}
\State $Q,R \leftarrow QR(\delta)$
\State $\delta \leftarrow Q[:,:N]$
\Comment{Replacing the deviation vectors with the columns of $Q$}
\State $\widetilde{R}[:,t/T_{norm}] \leftarrow \log(\mathrm{diag}(R[:N, :N]))$
\If{$\mod(t,T_{c})=0$}
\Comment{Checking the convergence criterion}
\State $l\leftarrow \mathrm{Real}(\mathrm{ColumnSum}(\widetilde{R}))/(t* \delta t)$
\Comment{Divide with the total timespan}
\State $l \leftarrow \mathrm{sort}(l)$
\State $d \leftarrow |l-l_{prev}|_2$
\If{$d<\epsilon$}
\State \textbf{break}
\EndIf
\EndIf
\EndIf
\State $J_0 \leftarrow \nabla_{\bm{h}} f_h^o(\h_{t})$.
\EndFor
\State \textbf{return} $l$
\Comment{Returning the estimated Lyapunov Exponents}
\EndProcedure
\end{algorithmic}
\end{algorithm}

%% file: appendix/appendix-4-hyperparameters.tex
\clearpage
\section{Model Hyperparameters}
\label{app:sec:hyperparameters}

For the Lorenz-96 system space with $d_o \in \{35,40 \}$ (in the PCA mode), we used the hyperparameters reported on \Cref{app:tab:hyprclorenz} for RC and \Cref{app:tab:hypgrulstmlorenz} for GRU/LSTM models.
For the parallel architectures in the state space of Lorenz-96 the hyperparameters are reported on \Cref{app:tab:hyprcparallellorenz} and \Cref{app:tab:hypgrulstmparallellorenz} for the parallel RC and GRU/LSTM models respectively.
For the parallel architectures in the state space of the Kuramoto-Sivashinsky architecture the hyperparameters are reported on \Cref{app:tab:hyprcparallelks} and \Cref{app:tab:hypgrulstmparallelks} for the parallel RC and GRU/LSTM models respectively.
We note here that in all RNN methods, the optimizer used to update the network can also be optimized.
To alleviate the computational burden we stick to Adam.


\begin{table}
\caption{Hyperparameters of \textbf{RC} for Lorenz-96}
\label{app:tab:hyprclorenz}
\centering
\begin{tabular}{ |c|c|c| } 
\hline
\text{Hyperparameter} & 
\text{Explanation} &
\text{Values} \\  \hline \hline
$D_r$ &
reservoir size &
$\{ 6000, \, 9000, \, 12000, \, 18000 \}$  \\ 
$N$ &
training data samples&
$10^{5}$  \\
 &
Solver &
Pseudoinverse/LSQR/Gradient descent  \\
$d$ &
degree of $W_{h,h}$&
$\{ 3, \, 8 \}$  \\
$\rho$ &
radius of $W_{h,h}$& 
$\{ 0.4, \, 0.8, \, 0.9, \, 0.99 \}$  \\
$\omega$ &
input scaling &
$ \{ 0.1,\,  0.5 ,\, 1.0 ,\, 1.5 ,\, 2.0 \} $  \\
$ \eta $&
regularization & 
$\{ 10^{-3},\,10^{-4},\,  10^{-5},\,  10^{-6} \}$  \\
 $ d_o $&
observed state dimension & 
$\{ 35, \, 40 \}$  \\
 $ n_{w} $&
warm-up steps (testing) & 
$2000$  \\
$\kappa_{n} $&
noise level in data & 
$\{0,\, 0.5\%,\, 1\% \}$  \\
\hline
\end{tabular}
\end{table}

\begin{table}
\caption{Hyperparameters of \textbf{GRU/LSTM} for Lorenz-96}
\label{app:tab:hypgrulstmlorenz}
\centering
\begin{tabular}{ |c|c|c| } 
\hline
\text{Hyperparameter} & 
\text{Explanation} &
\text{Values} \\  \hline \hline
$d_h$ &
hidden state size &
$\{1,\, 2, \, 3 \}$ layers of $\{  500, \, 1000, \, 1500\}$  \\ 
$N$ &
training data samples&
$10^{5}$  \\
$B$ &
batch-size&
$32$  \\
$\kappa_1$ &
BPTT forward time steps &
$\{1, \, 8 \}$  \\
$\kappa_2$ &
BPTT truncated backprop. length&
$\{8, \, 16 \}$  \\
$\kappa_3$ &
BPTT skip gradient parameter &
$=\kappa_2+\kappa_1-1$ \\
$ \eta $&
initial learning rate & 
$10^{-3}$  \\
$ p $&
zoneout probability & 
$\{0.99, \, 0.995 \, 1.0\}$  \\
$ d_o $&
observed state dimension & 
$\{ 35,\, 40 \}$  \\
 $ n_{w} $&
warm-up steps (testing) & 
$2000$  \\
$\kappa_{n} $&
noise level in data & 
$\{0,\, 0.2\% \}$  \\
\hline
\end{tabular}
\end{table}

\begin{table}
\caption{Hyperparameters of \textbf{Unitary Evolution} networks for Lorenz-96}
\label{app:tab:hypunitarylorenz}
\centering
\begin{tabular}{ |c|c|c| } 
\hline
\text{Hyperparameter} & 
\text{Explanation} &
\text{Values} \\  \hline \hline
$d_h$ &
hidden state size &
$\{1,\, 2, \, 3 \}$ layers of $\{  500, \, 1000, \, 1500\}$  \\ 
$N$ &
training data samples&
$10^{5}$  \\
$B$ &
batch-size&
$32$  \\
$\kappa_1$ &
BPTT forward time steps &
$\{1, \, 8 \}$  \\
$\kappa_2$ &
BPTT truncated backprop. length&
$\{8, \, 16 \}$  \\
$\kappa_3$ &
BPTT skip gradient parameter &
$=\kappa_2+\kappa_1-1$ \\
$ \eta $&
initial learning rate & 
$10^{-3}$  \\
$ p $&
zoneout probability & 
$ 1.0 $  \\
$ d_o $&
observed state dimension & 
$\{ 35,\, 40 \}$  \\
 $ n_{w} $&
warm-up steps (testing) & 
$2000$  \\
$\kappa_{n} $&
noise level in data & 
$\{0,\, 0.2\% \}$  \\
\hline
\end{tabular}
\end{table}

\begin{table}
\caption{Hyperparameters of \textbf{Parallel RC} for Lorenz-96}
\label{app:tab:hyprcparallellorenz}
\centering
\begin{tabular}{ |c|c|c| } 
\hline
\text{Hyperparameter} & 
\text{Explanation} &
\text{Values} \\  \hline \hline
$D_r$ &
reservoir size &
$\{ 1000, \, 3000, \, 6000, \, 12000\}$  \\ 
$N_g$ &
number of groups &
$20$  \\ 
$G$ &
group size &
$2$  \\ 
$I$ &
interaction length &
$4$  \\
$N$ &
training data samples&
$10^{5}$  \\
 &
Solver &
Pseudoinverse/LSQR/Gradient descent  \\
$d$ &
degree of $W_{h,h}$&
$10$  \\
$\rho$ &
radius of $W_{h,h}$& 
$ 0.6$  \\
$\omega$ &
input scaling &
$ 0.5 $  \\
$ \eta $&
regularization & 
$10^{-6}$  \\
 $ d_o $&
observed state dimension & 
$40$  \\
 $ n_{w} $&
warm-up steps (testing) & 
$2000$  \\
 \hline
\end{tabular}
\end{table}

\begin{table}
\caption{Hyperparameters of \textbf{Parallel GRU/LSTM} for Lorenz-96}
\label{app:tab:hypgrulstmparallellorenz}
\centering
\begin{tabular}{ |c|c|c| } 
\hline
\text{Hyperparameter} & 
\text{Explanation} &
\text{Values} \\  \hline \hline
$d_h$ &
hidden state size &
$\{  100, \, 250, \, 500 \}$  \\ 
$N_g$ &
number of groups &
$20$  \\ 
$G$ &
group size &
$2$  \\ 
$I$ &
interaction length &
$4$  \\ 
$N$ &
training data samples&
$10^{5}$  \\
$B$ &
batch-size&
$32$  \\
$\kappa_1$ &
BPTT forward time steps &
$4$  \\
$\kappa_2$ &
BPTT truncated backprop. length&
$4$  \\
$\kappa_3$ &
BPTT skip gradient parameter &
$4$  \\
$ \eta $&
initial learning rate & 
$10^{-3}$  \\
$ p $&
zoneout probability & 
$\{0.998, \, 1.0\}$  \\
$ d_o $&
observed state dimension & 
$40$  \\
 $ n_{w} $&
warm-up steps (testing) & 
$2000$  \\
 \hline
\end{tabular}
\end{table}

\begin{table}
\caption{Hyperparameters of \textbf{Parallel RC} for Kuramoto-Sivashinsky}
\label{app:tab:hyprcparallelks}
\centering
\begin{tabular}{ |c|c|c| } 
\hline
\text{Hyperparameter} & 
\text{Explanation} &
\text{Values} \\  \hline \hline
$D_r$ &
reservoir size &
$\{ 500, \, 1000, \, 3000, 6000, \, 12000 \}$  \\ 
$N_g$ &
number of groups &
$64$  \\ 
$G$ &
group size &
$8$  \\ 
$I$ &
interaction length &
$8$  \\
$N$ &
training data samples&
$10^{5}$  \\
 &
Solver &
Pseudoinverse  \\
$d$ &
degree of $W_{h,h}$&
$10$  \\
$\rho$ &
radius of $W_{h,h}$& 
$ 0.6$  \\
$\omega$ &
input scaling &
$ 1.0 $  \\
$ \eta $&
regularization & 
$10^{-5}$  \\
 $ d_o $&
observed state dimension & 
$512$  \\
 $ n_{w} $&
warm-up steps (testing) & 
$2000$  \\
 \hline
\end{tabular}
\end{table}

\begin{table}
\caption{Hyperparameters of \textbf{Parallel GRU/LSTM} for Kuramoto-Sivashinsky}
\label{app:tab:hypgrulstmparallelks}
\centering
\begin{tabular}{ |c|c|c| } 
\hline
\text{Hyperparameter} & 
\text{Explanation} &
\text{Values} \\  \hline \hline
$d_h$ &
hidden state size &
$\{  80, \, 100, \, 120 \}$  \\ 
$N_g$ &
number of groups &
$64$  \\ 
$G$ &
group size &
$8$  \\ 
$I$ &
interaction length &
$8$  \\ 
$N$ &
training data samples&
$10^{5}$  \\
$B$ &
batch-size&
$32$  \\
$\kappa_1$ &
BPTT forward time steps &
$4$  \\
$\kappa_2$ &
BPTT truncated backprop. length&
$4$  \\
$\kappa_3$ &
BPTT skip gradient parameter &
$4$  \\
$ \eta $&
initial learning rate & 
$10^{-3}$  \\
$ p $&
zoneout probability & 
$\{0.998, \, 1.0\}$  \\
$ d_o $&
observed state dimension & 
$512$  \\
 $ n_{w} $&
warm-up steps (testing) & 
$2000$  \\
 \hline
\end{tabular}
\end{table}

\begin{table}
\caption{Hyperparameters of \textbf{Parallel Unitary Evolution} networks for Kuramoto-Sivashinsky}
\label{app:tab:hypunitaryparallelks}
\centering
\begin{tabular}{ |c|c|c| } 
\hline
\text{Hyperparameter} & 
\text{Explanation} &
\text{Values} \\  \hline \hline
$d_h$ &
hidden state size &
$\{  100, \,200, \, 400 \}$  \\ 
$N_g$ &
number of groups &
$64$  \\ 
$G$ &
group size &
$8$  \\ 
$I$ &
interaction length &
$8$  \\ 
$N$ &
training data samples&
$10^{5}$  \\
$B$ &
batch-size&
$32$  \\
$\kappa_1$ &
BPTT forward time steps &
$4$  \\
$\kappa_2$ &
BPTT truncated backprop. length&
$4$  \\
$\kappa_3$ &
BPTT skip gradient parameter &
$4$  \\
$ \eta $&
initial learning rate & 
$10^{-2}$  \\
$ p $&
zoneout probability & 
$1.0$  \\
$ d_o $&
observed state dimension & 
$512$  \\
 $ n_{w} $&
warm-up steps (testing) & 
$2000$  \\
 \hline
\end{tabular}
\end{table}

%% file: appendix/appendix-5-lorenz96-f8-additional.tex
\clearpage
\section{Additional Results - Lorenz-96 - Divergence of Unitary and RC RNNs}

\begin{figure*}
\centering
\begin{subfigure}[t]{0.4\textwidth}
\centering
\includegraphics[height=3.5cm]{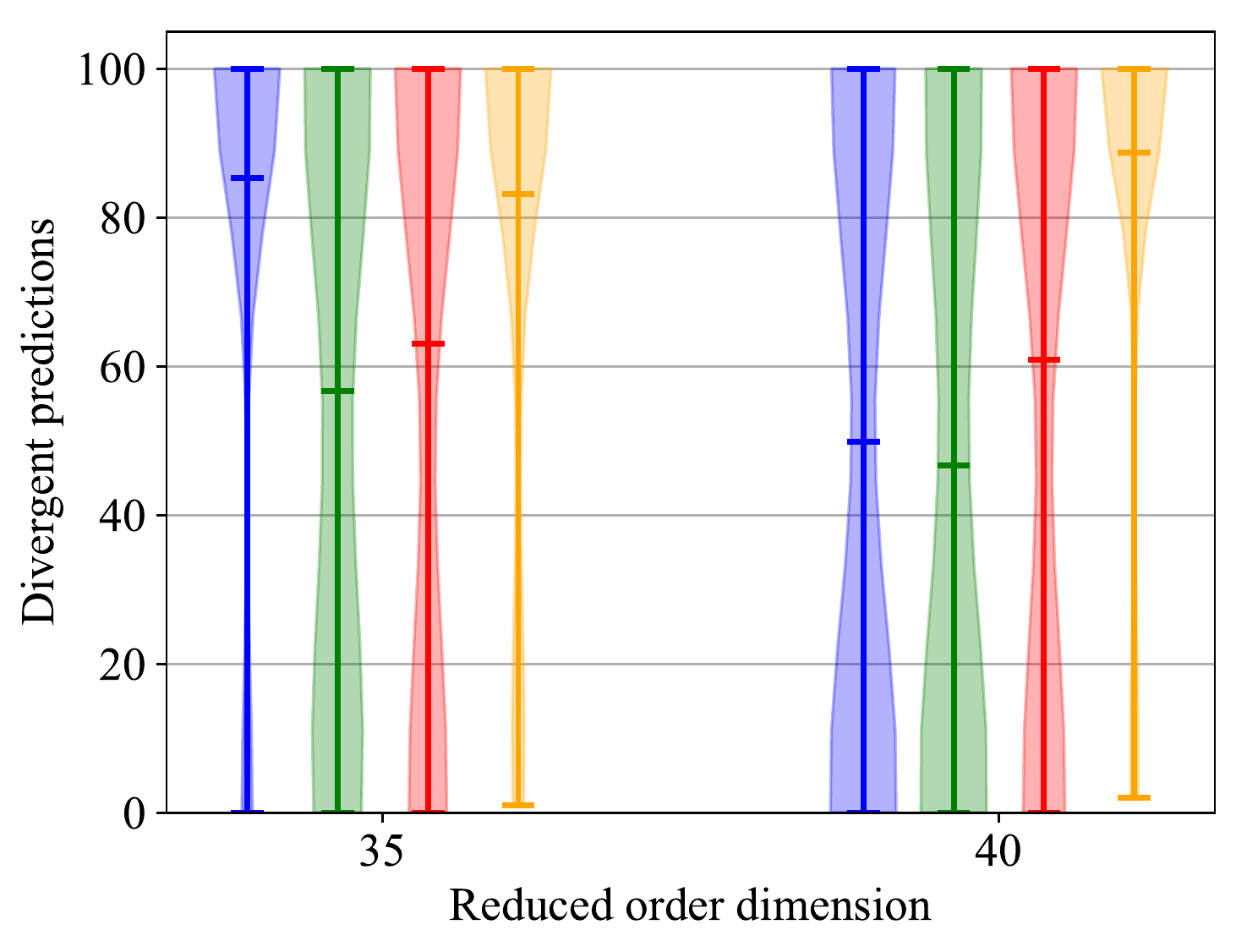}
\caption{$F=8$}
\label{fig:L96F8GP40R40:DIVERGPRED_BBO_NAP_2_RDIM}
\end{subfigure}
\begin{subfigure}[t]{0.4\textwidth}
\centering
\includegraphics[height=3.5cm]{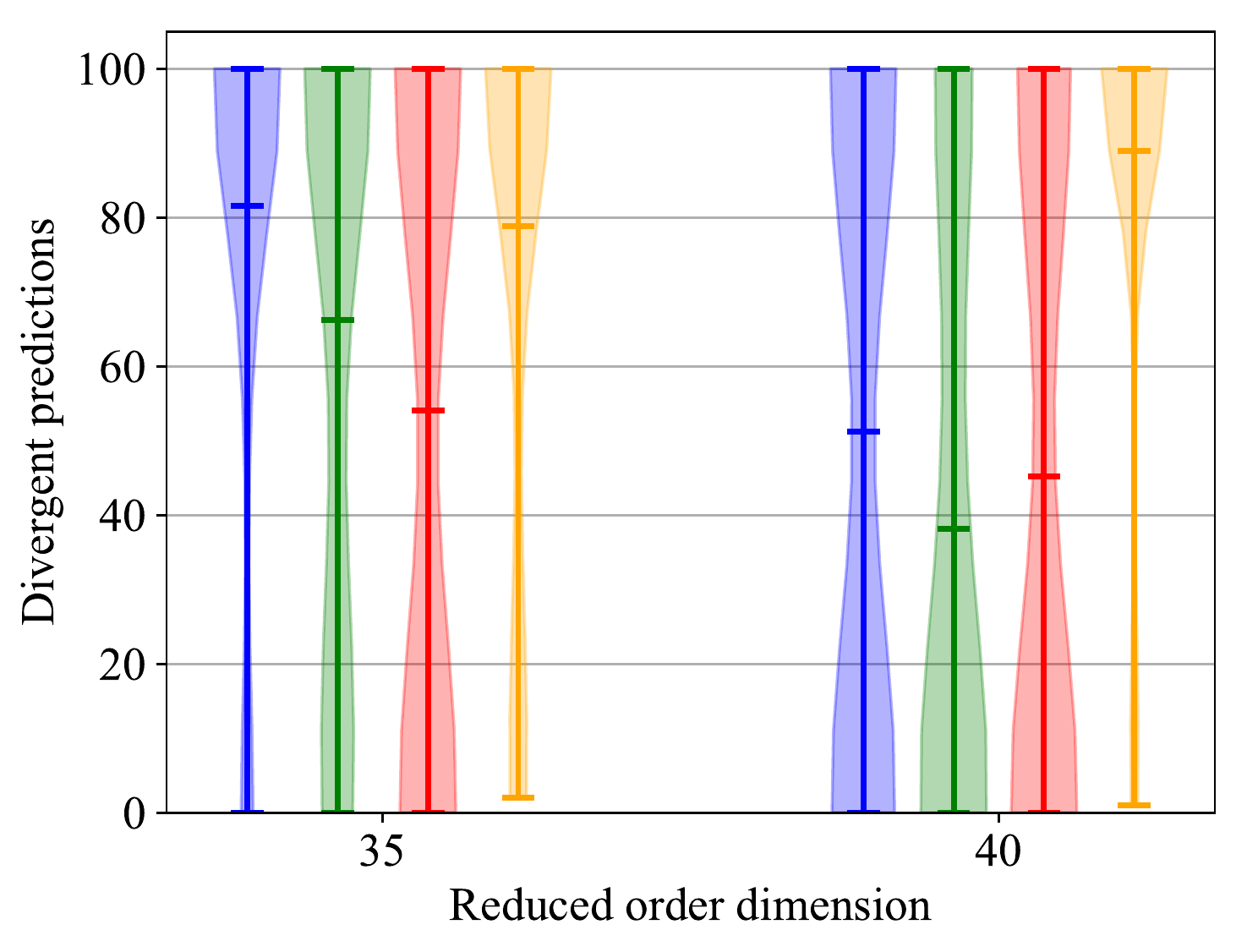}
\caption{$F=10$}
\label{fig:L96F10GP40R40:DIVERGPRED_BBO_NAP_2_RDIM}
\end{subfigure}
\caption{
Violin plots with kernel density estimates of the number of divergent predictions over the 100 initial conditions from the test data, over all hyperparameter sets for RC, GRU, LSTM, and Unitary networks, for reduced order state $d_o=35$ and full order state $d_o=40$ in two forcing regimes \textbf{(a)} $F=8$ and \textbf{(b)} $F=10$ in the Lorenz-96 system.
Most hyperparameter sets of Unitary networks, lead to models that diverge in iterative forecasting in both reduced order and full order scenario for both $F\in \{8,10\}$.
Although the divergence effect is a non-issue in RC in the full state scenario $d_o=40$, indicated by the wide part in the lower end of the density plot, the effect is more prominent in the reduced order scenario compared to GRU and LSTM.
Identification of hyperparameters for LSTM and GRU networks that show stable iterative forecasting behavior in the reduced order space is significantly easier compared to RC and Unitary networks, as indicated by the wide/thin lines in the lower part of the density plots of the first/latter.
\\
RC \protect \violinBlueRectangle;
GRU \protect \violinGreenRectangle;
LSTM \protect \violinRedRectangle;
Unit \protect \violinOrangeRectangle;
}
\label{fig:L96DivergentPredictions}
\end{figure*}

\begin{figure*}
\centering
\begin{subfigure}[t]{0.8\textwidth}
\centering
\includegraphics[width=.98\textwidth]{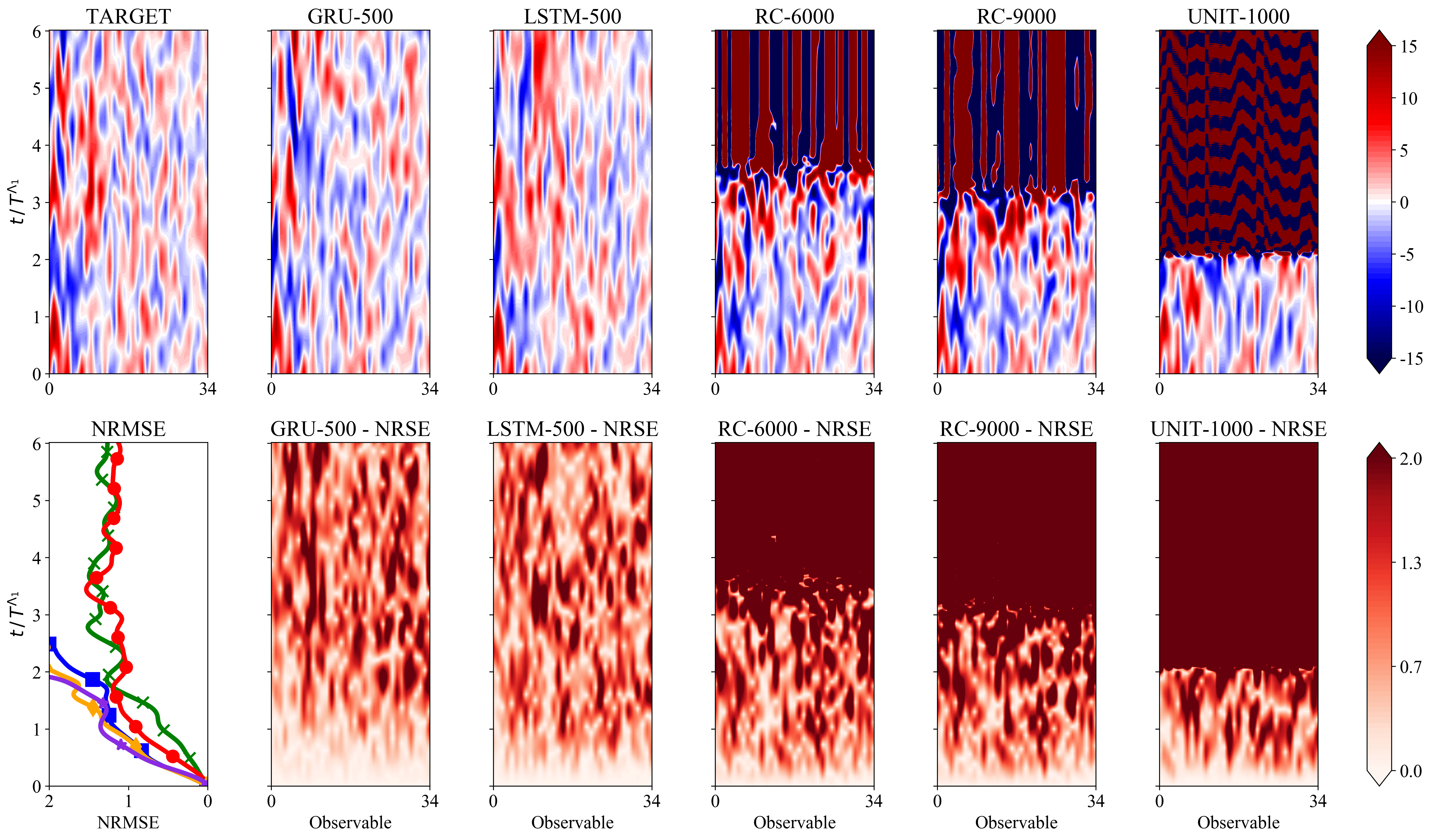}
\caption{Reduced order state ( $\bm{d_o=35}$), $F=8$}
\label{fig:L96F8GP40R40:CONTOUR_BBO_NAP_2_M_RDIM_35_IC15}
\end{subfigure}
\hfill
\begin{subfigure}[t]{0.8\textwidth}
\centering
\includegraphics[width=.98\textwidth]{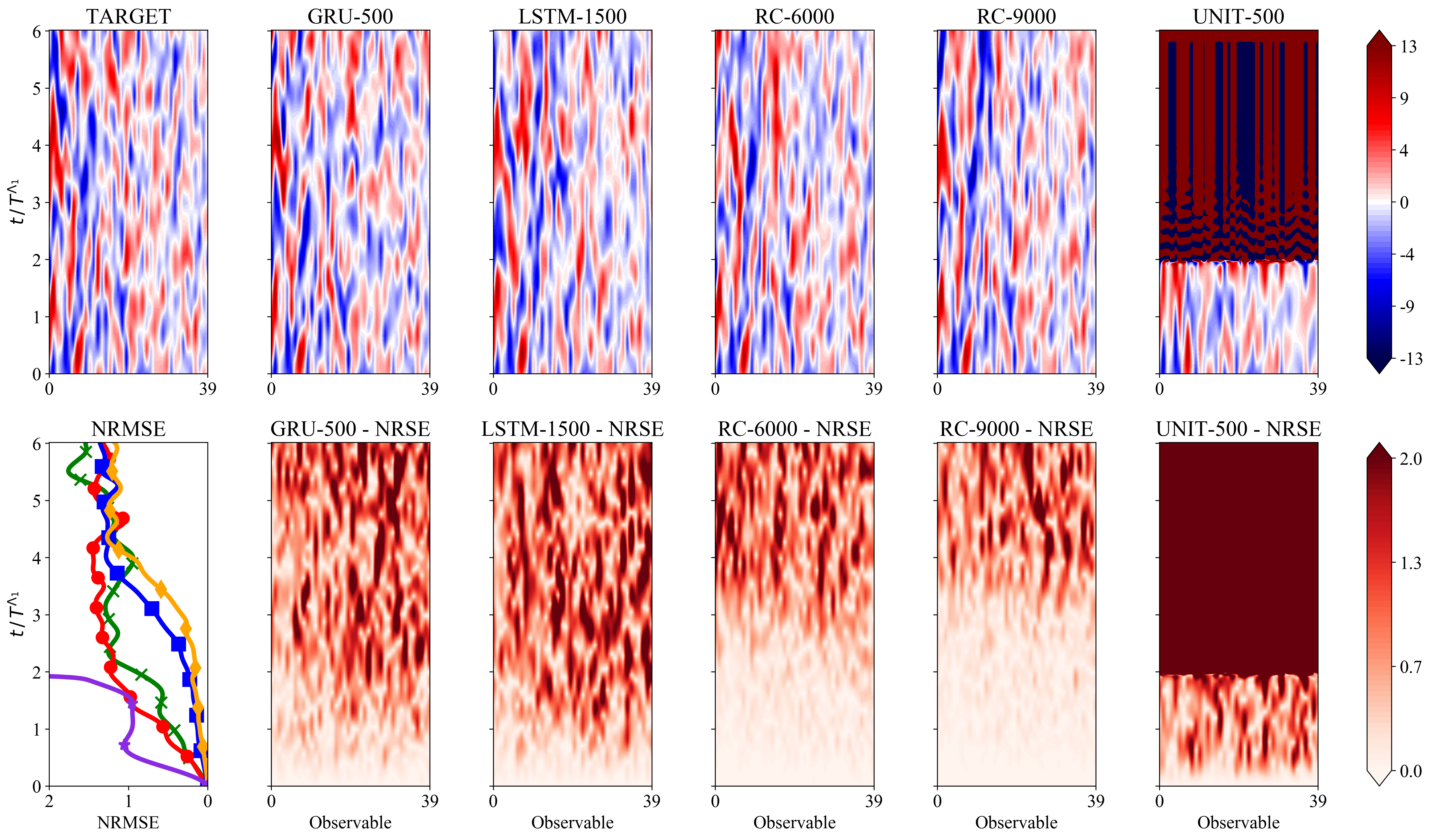}
\caption{Full order state ( $\bm{d_o=40}$), $F=8$}
\label{fig:L96F8GP40R40:CONTOUR_BBO_FREQERROR_TEST_2_M_RDIM_40_IC35.png}
\end{subfigure}
\caption{
Contour plots of a spatio-temporal forecast on the SVD modes of the Lorenz-96 system with $F=8$ in the testing dataset with GRU, LSTM, RC and a Unitary network along with the true (target) evolution and the associated NRSE contours for the reduced order observable \textbf{(a)} $d_o=35$ and the full state \textbf{(b)} $d_o=40$.
The evolution of the component average NRSE (NMRSE) is plotted to facilitate comparison.
Unitary networks suffer from propagation of forecasting error and eventually their forecasts diverge from the attractor.
Forecasts in the case of an observable dimension $d_o=40$ diverge slower as the dynamics are deterministic.
In contrast, forecasting the observable with $d_o=35$ is challenging due to both \textbf{(1)} sensitivity to initial condition and \textbf{(2)} incomplete state information that requires the capturing of temporal dependencies.
In the full-state setting, RC models achieve superior forecasting accuracy compared to all other models.
In the challenging reduced order scenario, LSTM and GRU networks demonstrate a stable behavior in iterative prediction and reproduce the long-term statistics of the attractor.
In contrast, in the reduced order scenario RC suffer from frequent divergence.
The divergence effect is illustrated in this chosen initial condition. 
\\
GRU \protect \greenlineX;
LSTM \protect \redlineCircle;
RC-6000 \protect \bluelineRectangle;
RC-9000 \protect \orangelineDiamond; 
Unit \protect \bluevioletlineStar;
}\label{fig:L96F8GP40R40:CONTOUR_BBO_NAP_2_M_DIVERGING}
\end{figure*}

In this section, we try to quantify the divergence effect due to the accumulation of the forecasting error in the iterative prediction.
In \Cref{fig:L96DivergentPredictions} we present violin plots with fitted kernel density estimates for the number of divergent predictions of each hyperparameter set, computed based on all tested hyperparameter sets for forcing regimes $F\in \{8,10\}$ and observable dimensions $d_o \in \{35, 40\}$.
The annotated lines denote the minimum, mean and maximum number of divergent predictions over the 100 initial conditions of all hyperparameter sets.
In the fully observable systems $d_o=40$, in both forcing regimes $F\in \{8,10\}$, there are many models (hyperparameter sets) with zero divergent predictions for RC, GRU and LSTM, as illustrated by the wide lower part of the violin plot.
In contrast, most hyperparameter sets in Unitary networks lead to models whose iterative predictions diverge from the attractor, as illustrated by the wide upper part in the violin plot. 
In the reduced order scenario, this divergence effect seems to be more prominent in RC and Unitary networks, as indicated by the very thin lower part of their violin plots, for both forcing regimes.
In contrast, many hyperparameter sets of GRU and LSTM models lead to stable iterative prediction.
This indicates that hyperparameter tuning in RC and Unitary networks when the system state is not fully observed, is cumbersome compared to LSTM and GRU networks.
One example of this divergence effect in an initial condition from the \textbf{test} dataset is illustrated in \Cref{fig:L96F8GP40R40:CONTOUR_BBO_NAP_2_M_DIVERGING}.
The RC and the Unitary networks diverge in the reduced order state predictions after approximately two Lyapunov times.

%% file: appendix/appendix-6-lorenz96-f10-additional.tex
\clearpage
\section{Additional Results - Lorenz-96 - $F=10$}

\begin{figure*}
\begin{subfigure}[t]{0.32\textwidth}
\includegraphics[width=.95\textwidth]{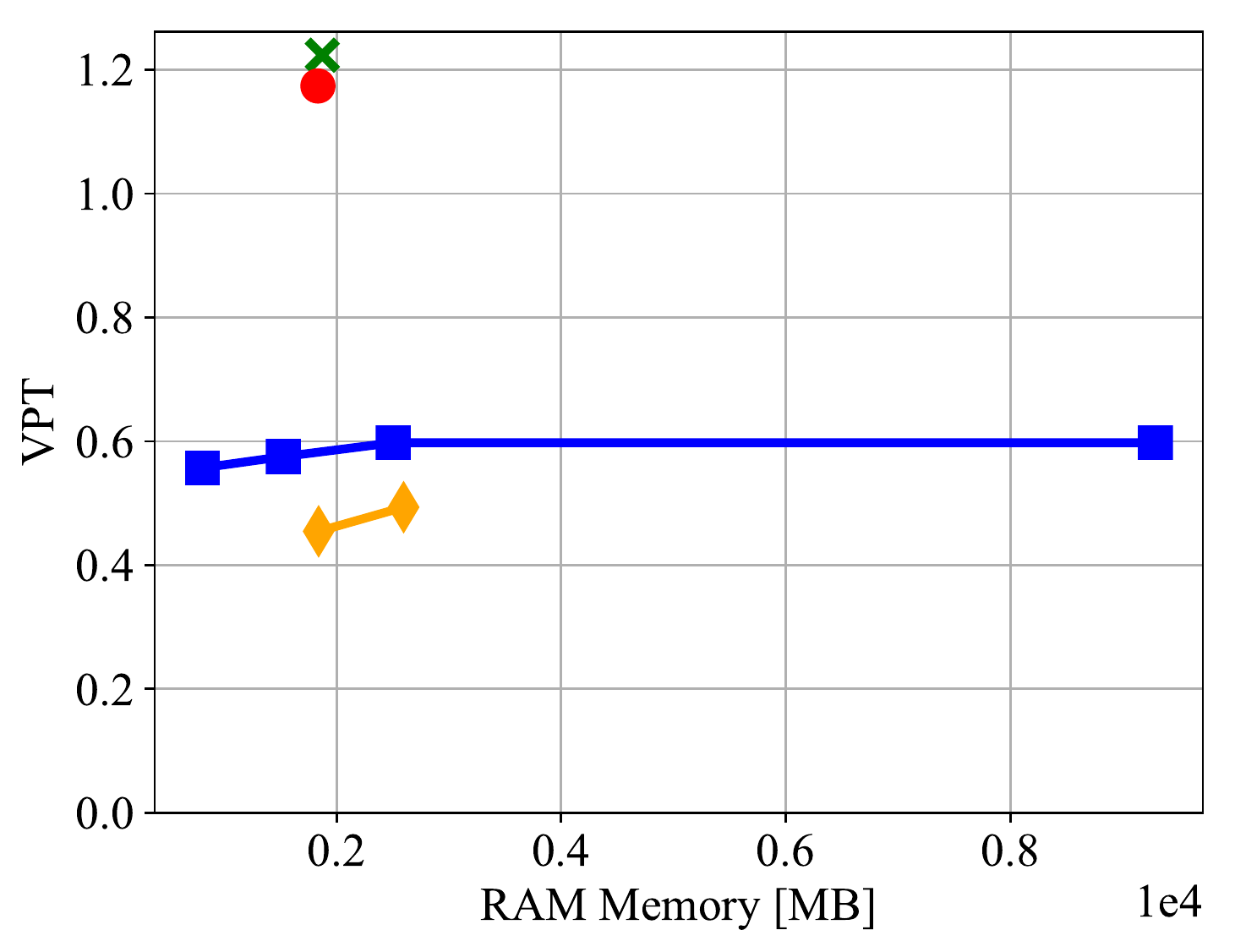}
\caption{
VPT w.r.t. RAM memory for $d_o=35$.
}
\label{fig:L96F10GP40R40:NAP_2_RAM_RDIM35}
\end{subfigure}
\hfill
\begin{subfigure}[t]{0.32\textwidth}
\includegraphics[width=.95\textwidth]{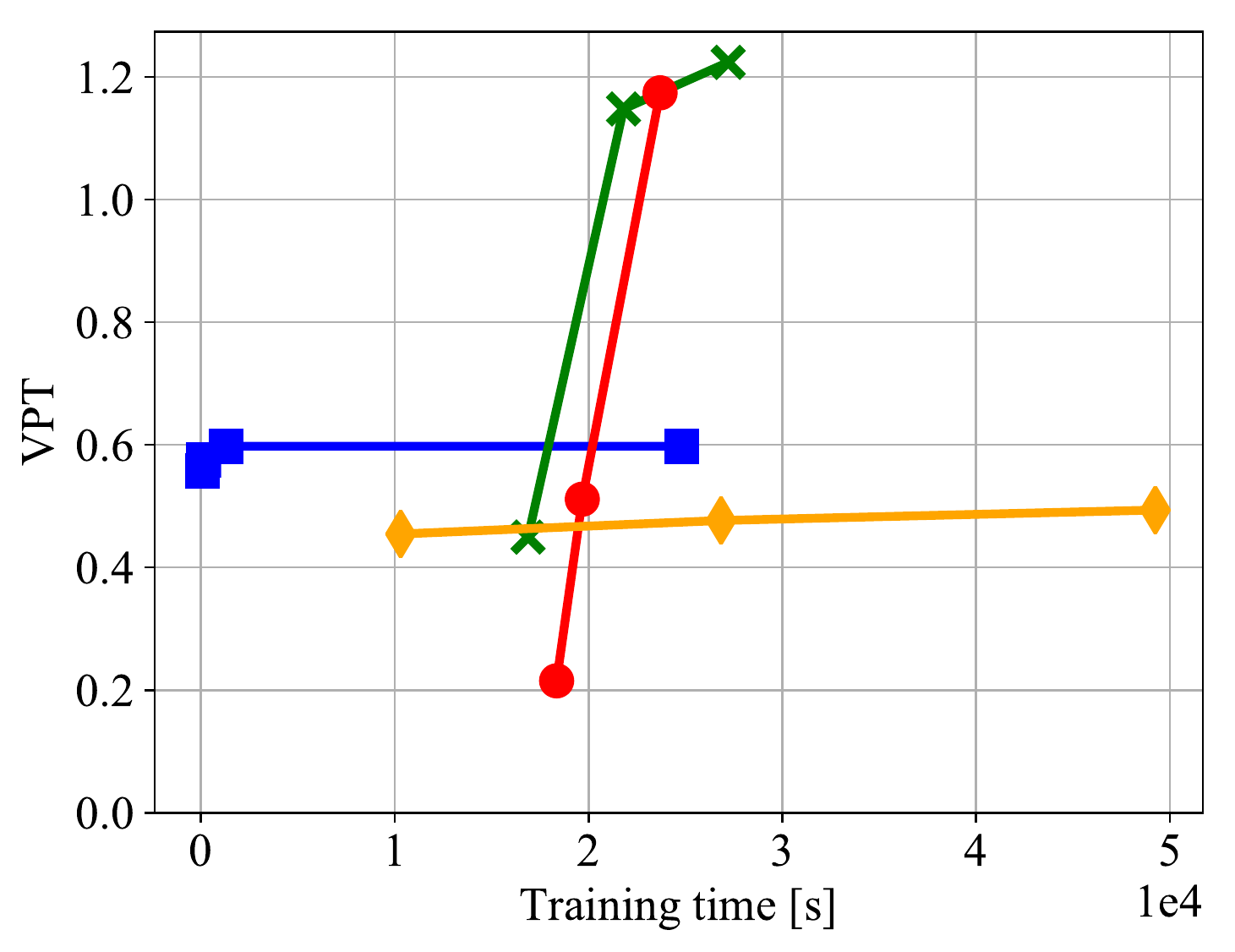}
\caption{
VPT w.r.t. the training time for $d_o=35$.
}
\label{fig:L96F10GP40R40:NAP_2_TRAINTIME_RDIM35}
\end{subfigure}
\hfill
\begin{subfigure}[t]{0.32\textwidth}
\includegraphics[width=.95\textwidth]{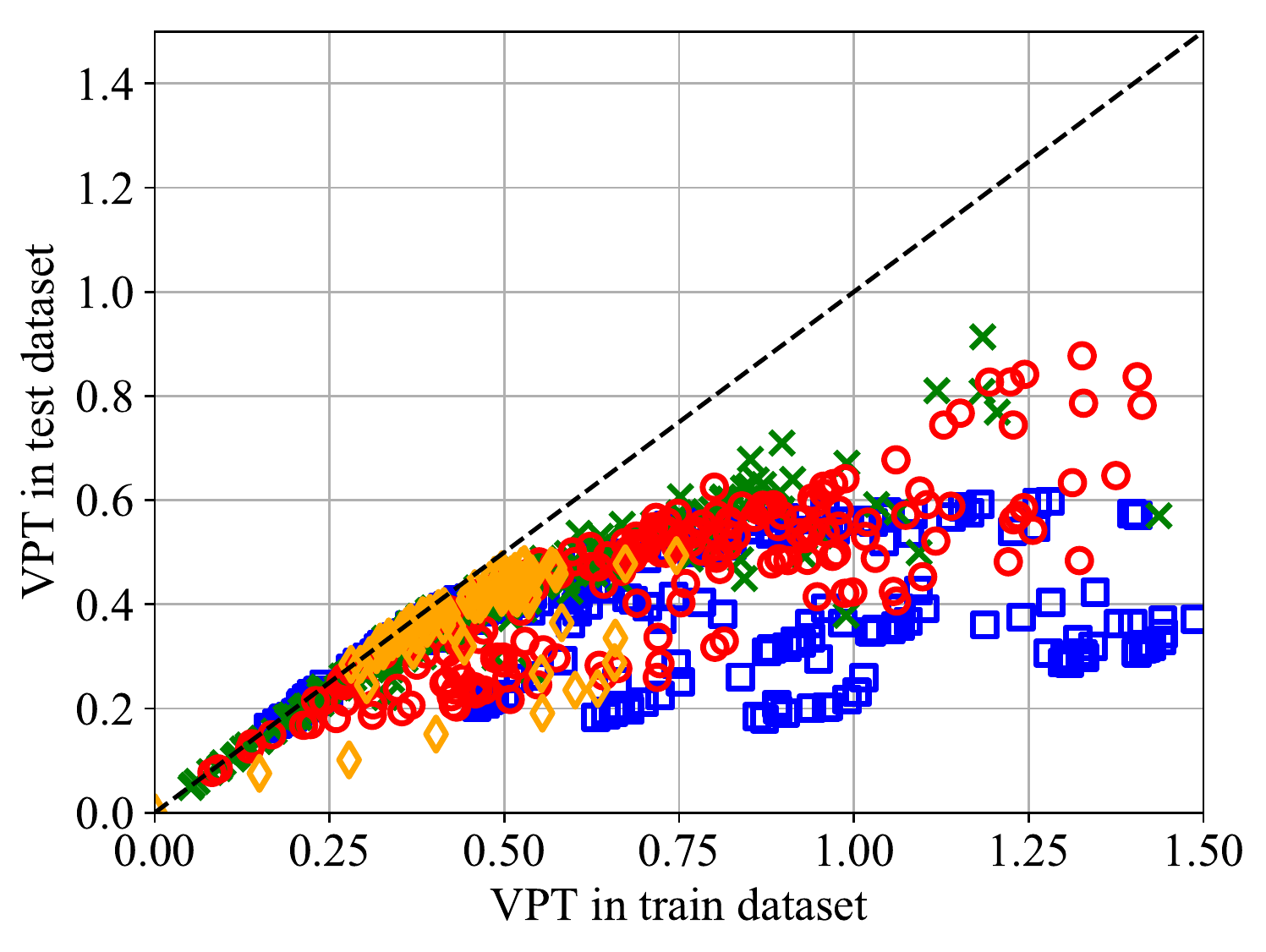}
\caption{
VPT in \textbf{test} data w.r.t. VPT in the \textbf{training} data for $d_o=35$.
}
\label{fig:L96F10GP40R40:OFSP_BBO_NAP_RDIM_35}
\end{subfigure}
\hfill
\begin{subfigure}[t]{0.32\textwidth}
\includegraphics[width=.95\textwidth]{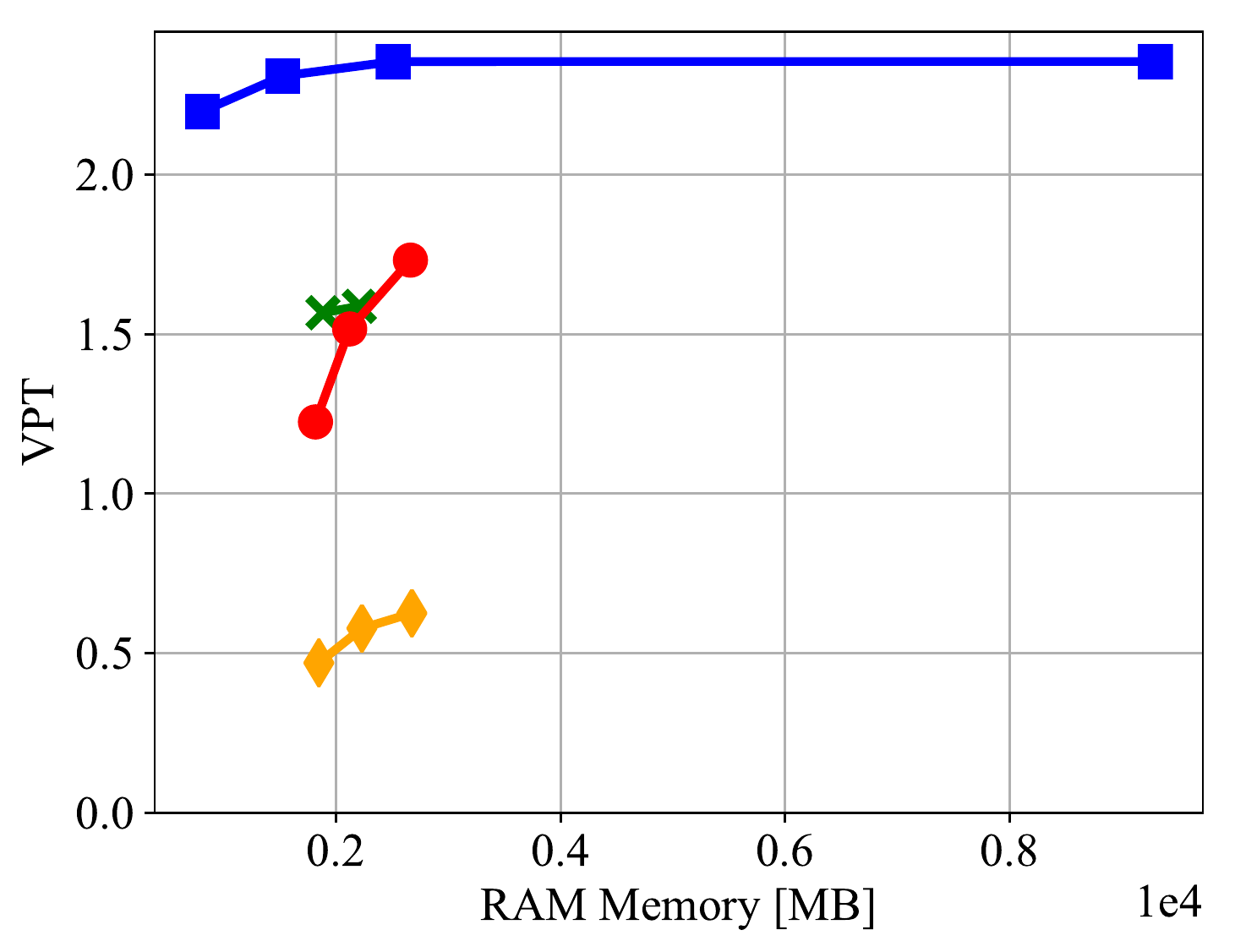}
\caption{
VPT w.r.t. RAM memory for $d_o=40$.
}
\label{fig:L96F10GP40R40:NAP_2_RAM_RDIM40}
\end{subfigure}
\hfill
\begin{subfigure}[t]{0.32\textwidth}
\includegraphics[width=.95\textwidth]{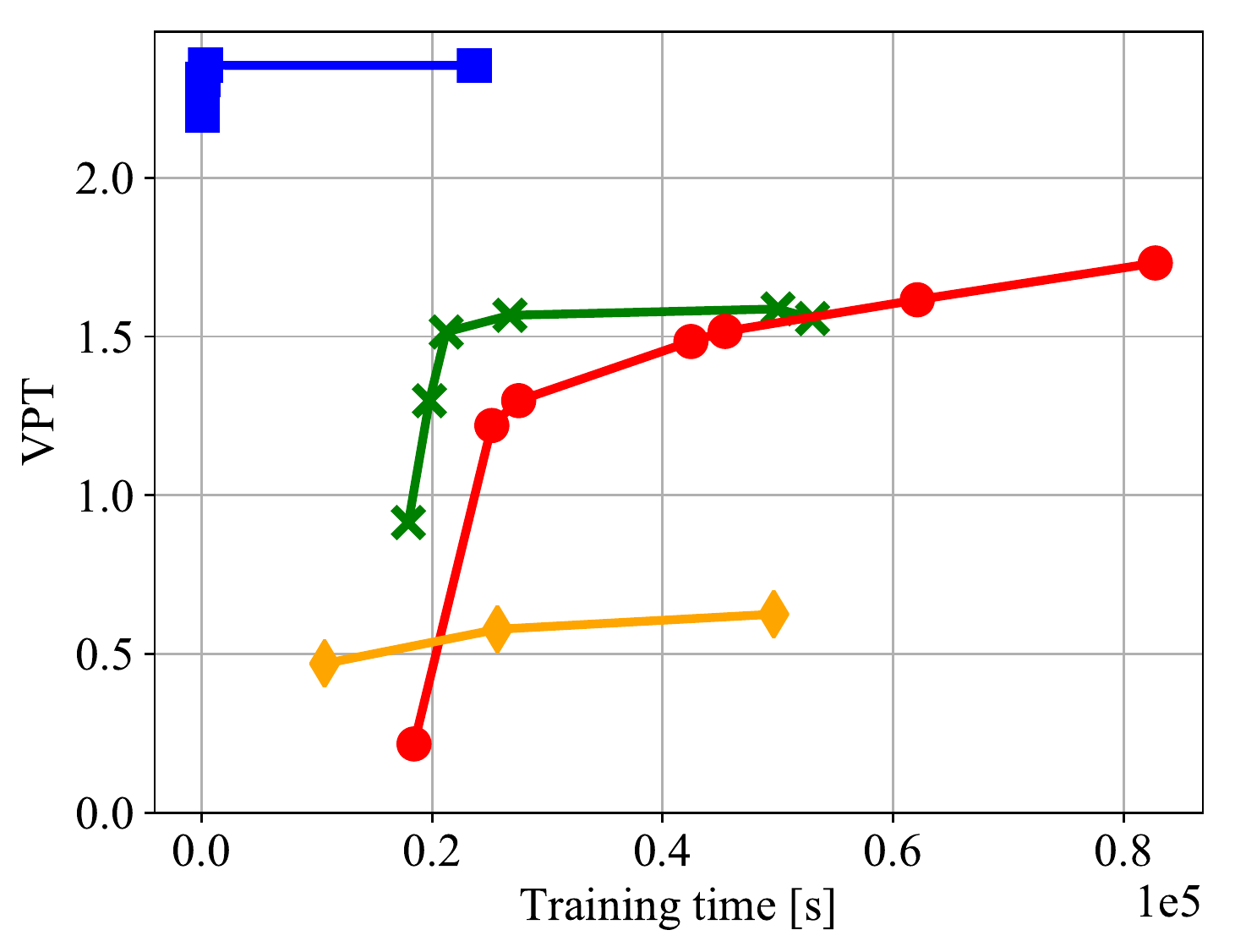}
\caption{
VPT w.r.t. the training time for $d_o=40$.
}
\label{fig:L96F10GP40R40:NAP_2_TRAINTIME_RDIM40}
\end{subfigure}
\hfill
\begin{subfigure}[t]{0.32\textwidth}
\includegraphics[width=.95\textwidth]{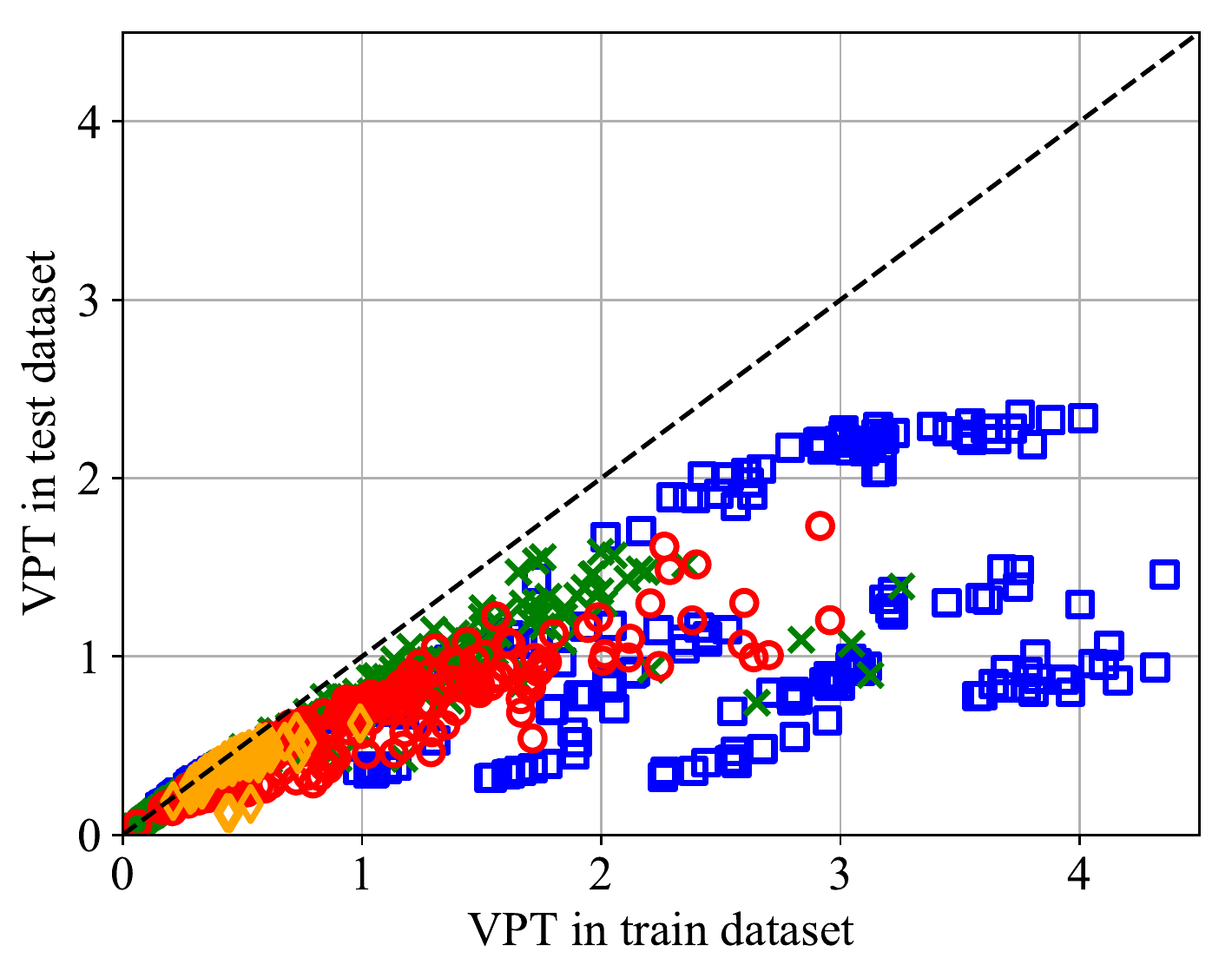}
\caption{
VPT in \textbf{test} data w.r.t. VPT in the \textbf{training} data for $d_o=40$.
}
\label{fig:L96F10GP40R40:OFSP_BBO_NAP_RDIM_40}
\end{subfigure}
\caption{
Forecasting results on the dynamics of an observable consisting of the SVD modes of the Lorenz-96 system with $F=10$ and state dimension $40$.
The observable consists of the $d_o \in \{35, 40 \}$ most energetic modes.
(a), (d) Valid prediction time (VPT) plotted w.r.t. the required RAM memory for dimension $d_o\in \{ 35, 40\}$.
(b), (e) VPT plotted w.r.t. training time for dimension $d_o\in \{ 35, 40\}$.
(c), (f) VPT measured from $100$ initial conditions sampled from the test data plotted w.r.t. VPT from $100$ initial conditions sampled from the training data for each model for $d_o\in \{ 35, 40\}$.
RCs tend to overfit easier compared to GRUs/LSTMs that utilize validation-based early stopping. \\
RC \protect \bluelineRectangle;
GRU \protect \greenlineX;
LSTM \protect \redlineCircle;
Unit \protect \orangelineDiamond;
Ideal \protect \blacklineDashed;
}
\label{fig:L96F10GP40R40VPTparametersplot}
\end{figure*}

\begin{figure*}
\centering
\begin{subfigure}[t]{0.8\textwidth}
\centering
\includegraphics[width=.98\textwidth]{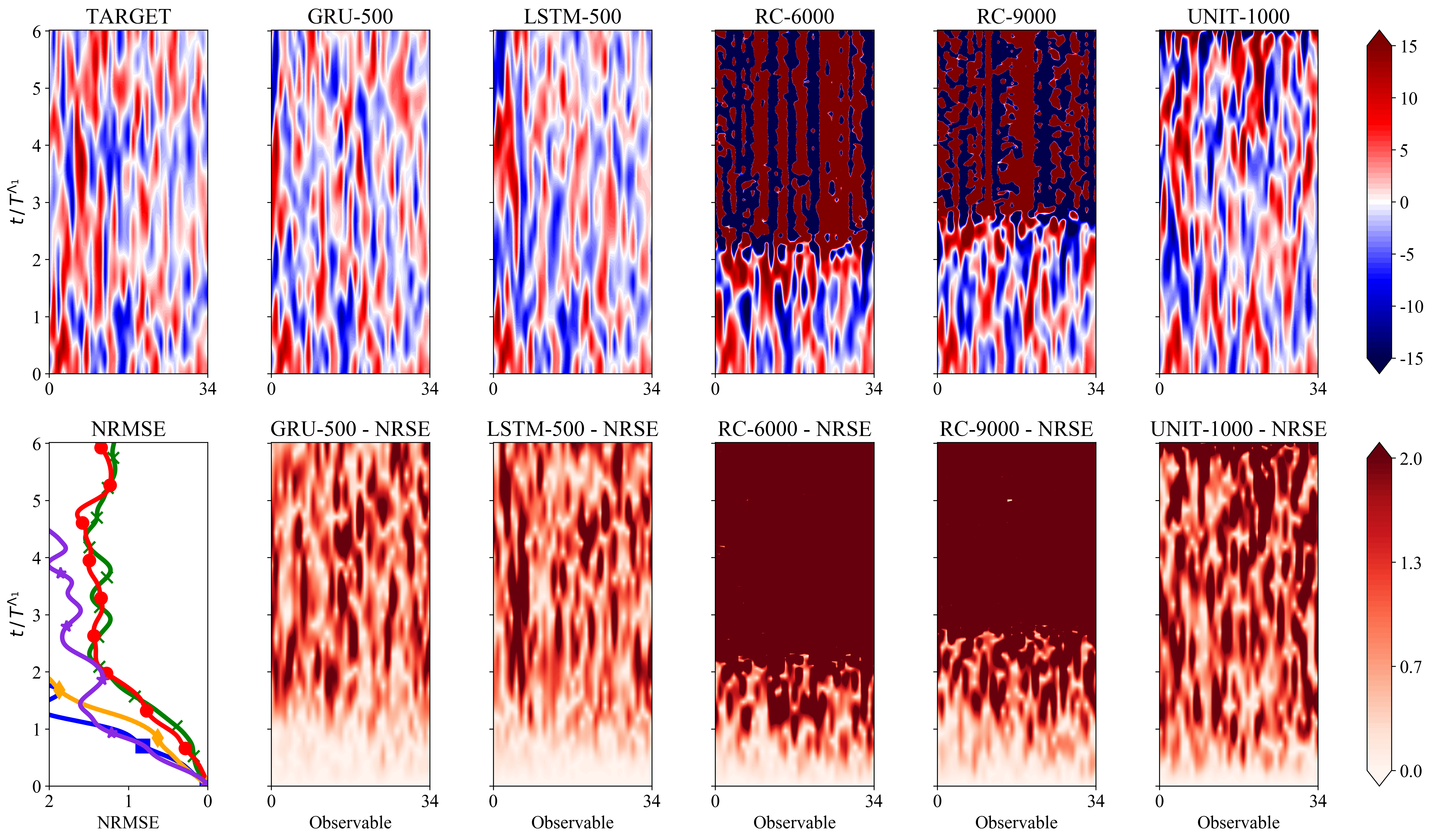}
\caption{Reduced order state ($\bm{d_o=35}$), $F=10$}
\label{fig:L96F10GP40R40:CONTOUR_BBO_NAP_2_M_RDIM_35_IC20}
\end{subfigure}
\hfill
\begin{subfigure}[t]{0.8\textwidth}
\centering
\includegraphics[width=.98\textwidth]{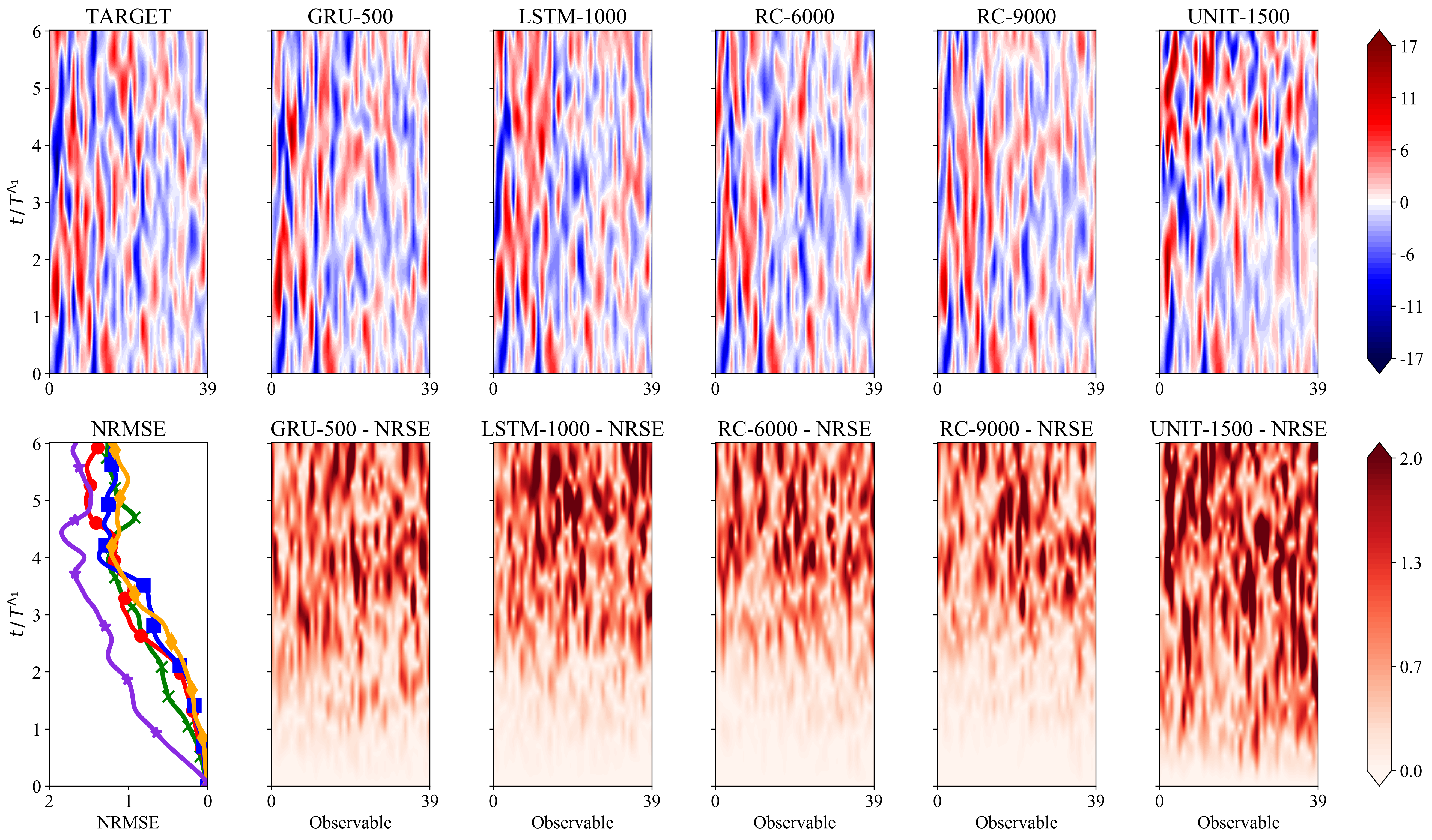}
\caption{Full state ($\bm{d_o=40}$), $F=10$}
\label{fig:L96F10GP40R40:CONTOUR_BBO_NAP_2_M_RDIM_40_IC10}
\end{subfigure}
\caption{
Contour plots of a spatio-temporal forecast on the SVD modes of the Lorenz-96 system with $F=10$ in the testing dataset with GRU, LSTM, RC and a Unitary network along with the true (target) evolution and the associated NRSE contours for the reduced order observable \textbf{(a)} $d_o=35$ and the full state \textbf{(b)} $d_o=40$.
The evolution of the component average NRSE (NMRSE) is plotted to facilitate comparison.
Forecasts in the case of an observable dimension $d_o=40$ diverge slower as the dynamics are deterministic.
In contrast, forecasting the observable with $d_o=35$ is challenging due to both \textbf{(1)} sensitivity to initial condition and \textbf{(2)} incomplete state information that requires the capturing of temporal dependencies.
Even in this challenging scenario, LSTM and GRU networks demonstrate a stable behavior in iterative prediction and reproduce the long-term statistics of the attractor.
Accurate short-term predictions can be achieved with very large RC networks ($d_h=9000$) at the cost of high memory requirements.
However, even in this case, RC models may diverge from the attractor and do not reproduce the attractor climate. \\
GRU \protect \greenlineX;
LSTM \protect \redlineCircle;
RC-6000 \protect \bluelineRectangle;
RC-9000 \protect \orangelineDiamond; 
Unit \protect \bluevioletlineStar;
}\label{fig:L96F10GP40R40:CONTOUR_BBO_NAP_2_MODELS}
\end{figure*}

In \Cref{fig:L96F10GP40R40VPTparametersplot}, we provide additional results for the forcing regime $F=10$ that are in agreement with the main conclusions drawn in the main manuscript for the forcing regime $F=8$.
An example of a single forecast of the models starting from an initial condition in the \textbf{test} dataset is given in \Cref{fig:L96F10GP40R40:CONTOUR_BBO_NAP_2_MODELS}

%% file: appendix/appendix-7-BBTT.tex
\clearpage
\section{Temporal Dependencies and Backpropagation}

In our study, in order to train the GRU and LSTM models with back-propagation through time (BPTT), we need to tune the parameters $\kappa_1$ and $\kappa_2$.
The first one denotes the truncated back-propagation length (also referred to as sequence length) and the second the number of future time-steps used to compute the loss and backpropagate the gradient during training at each batch.
In the hyperparameter study considered in this work, we varied $\kappa_1 \in \{8, 16 \}$ and $\kappa_2 \in \{1, 8\}$.

\begin{figure*}
\centering
\begin{subfigure}[t]{0.48\textwidth}
\centering
\includegraphics[height=3.75cm]{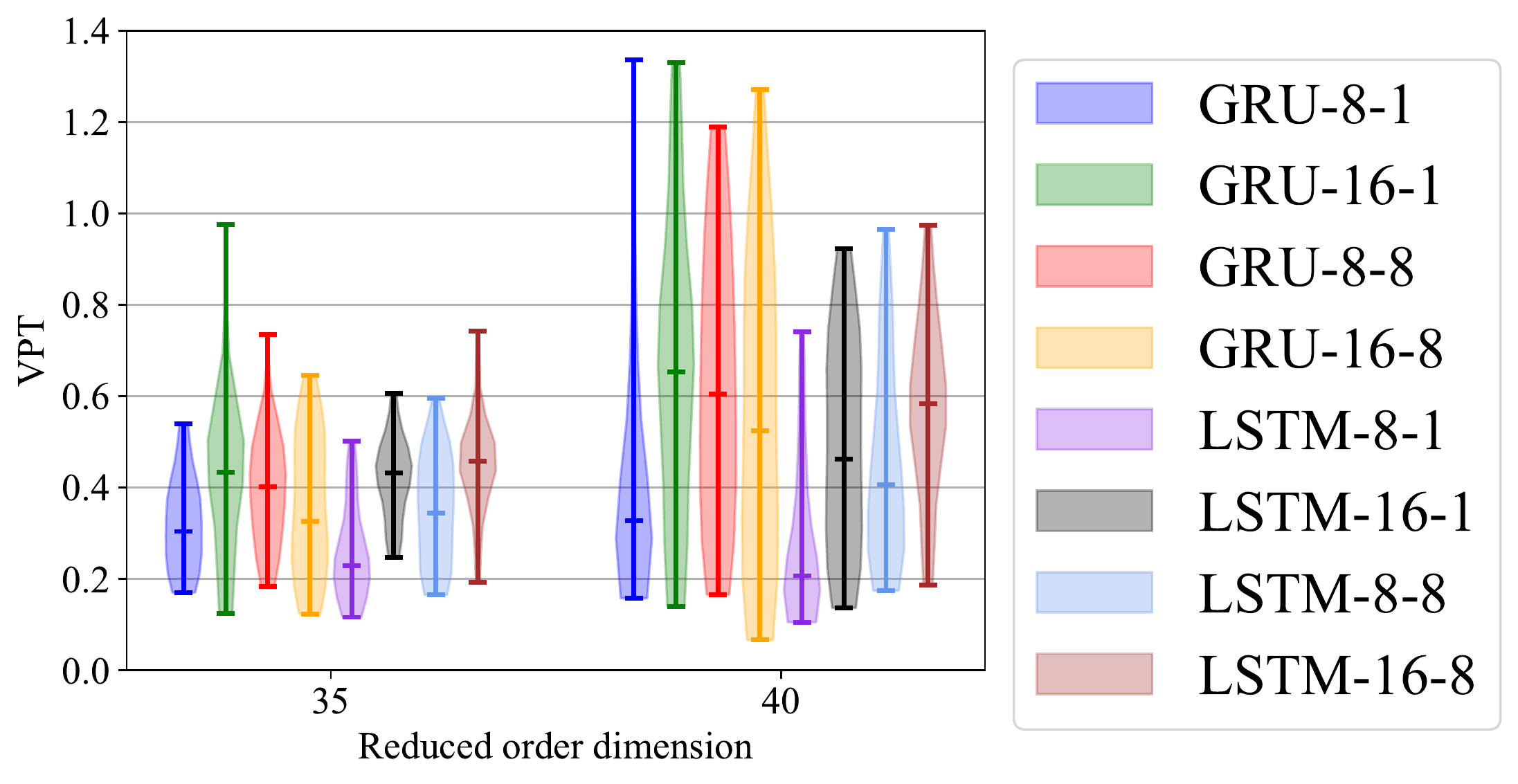}
\caption{$F=8$}
\label{fig:L96F8GP40R40:BBTT_STUDY_VPT_NAP_2_RDIM_LEGEND}
\end{subfigure}
\begin{subfigure}[t]{0.48\textwidth}
\centering
\includegraphics[height=3.75cm]{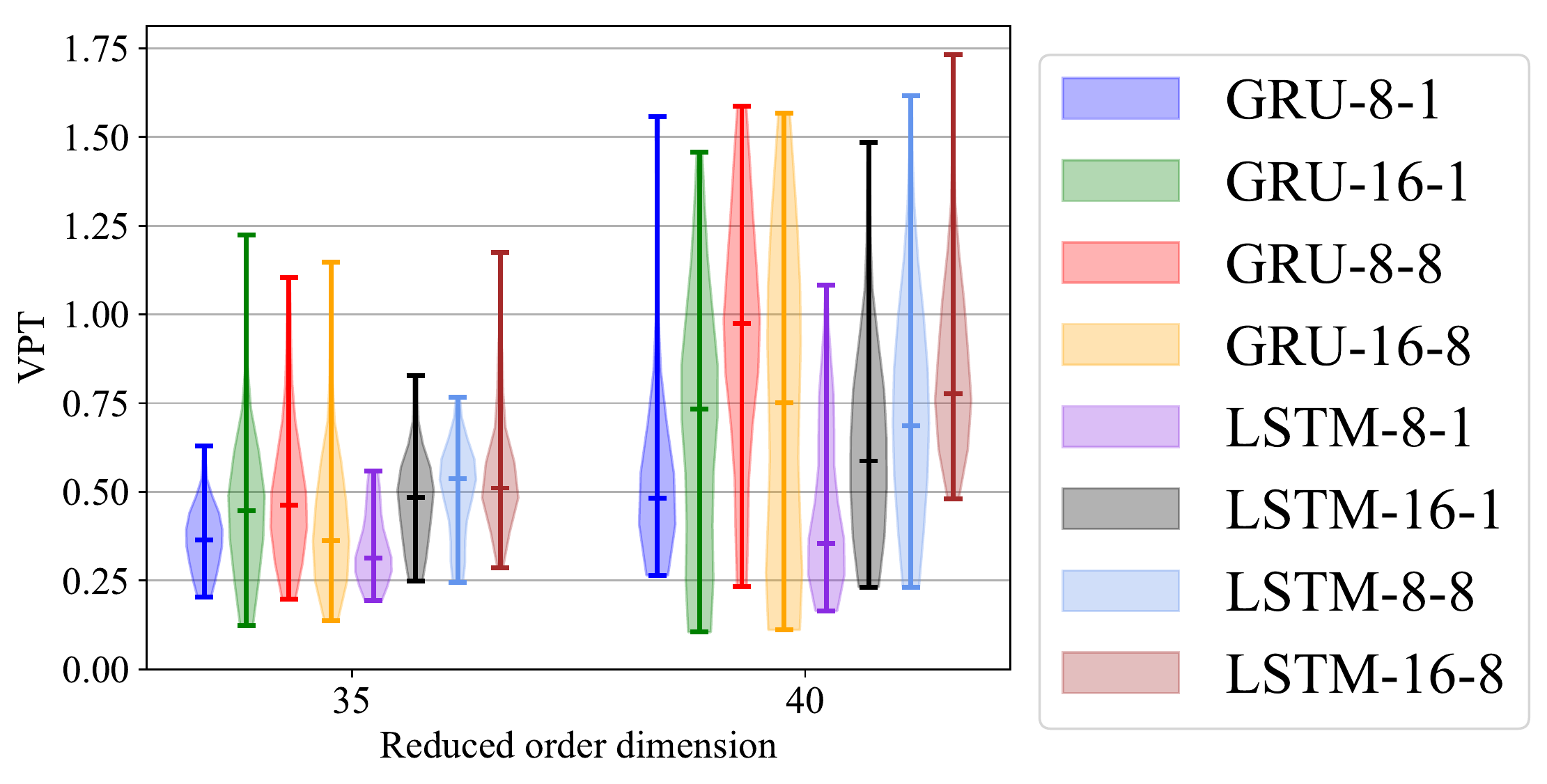}
\caption{$F=10$}
\label{fig:L96F10GP40R40:BBTT_STUDY_VPT_NAP_2_RDIM_LEGEND}
\end{subfigure}
\caption{
VPT in the testing data for stateful LSTM and GRU models trained with different truncated Backpropagation through time parameters $\kappa_1$ and $\kappa_2$ in the (reduced) SVD mode observable of the Lorenz-96 system.
The legend of each plot reports the models along with their $\kappa_1-\kappa_2$ parameters used to train them.
Each marker reports the mean VPT, while the errorbars report the minimum and maximum VPT.
We observe that especially in the reduced order observable scenario ($d_0=35$), having a large truncated back-propagation parameter $\kappa_1$ (also referred to as sequence length) is vital to capture the temporal dependencies in the data and achieve high forecasting efficiency.
In contrast in the full-state scenario ($d_0=40$) a model with a small back-propagation horizon suffices.
}
\label{fig:BBTT_STUDY_VPT}
\end{figure*}

In \Cref{fig:BBTT_STUDY_VPT} we plot the forecasting efficiency of LSTM and GRU models trained with different parameters in terms of the Valid Prediction Time (VPT) in the test dataset (averaged over 100 initial conditions) on the Lorenz-96 system for $F\in \{8, 10\}$.
In the reduced order scenario case, we observe that models with a large sequence length $\kappa_1$ and a large prediction length $\kappa_2$ are pivotal in order to achieve a high forecasting efficiency.
This implies that there are temporal correlations in the data that cannot be easily captured by other models with smaller horizons.
In contrast, in the full order scenario, models with smaller $\kappa_1$ perform reasonably well, demonstrating that the need of capturing temporal correlations in the data in order to forecast the evolution is less prominent, since the full information of the state of the system is available.

%% file: appendix/appendix-8-parallel-overfitting.tex
\clearpage
\section{Over-fitting of Parallel Models}

\begin{figure*}
\centering
\begin{subfigure}[t]{0.48\textwidth}
\centering
\includegraphics[height=4cm]{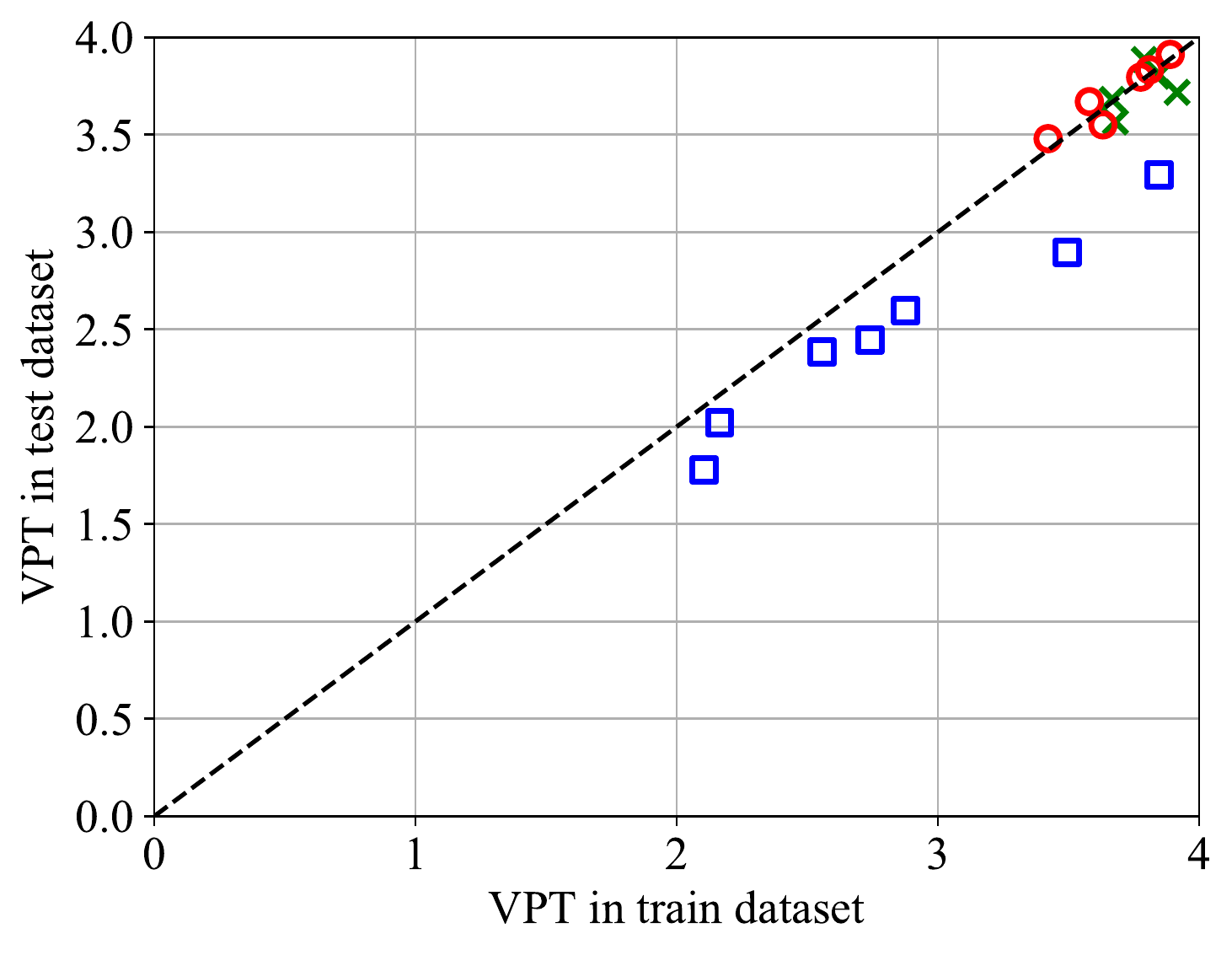}
\caption{VPT in test w.r.t. VPT in train}
\label{fig:L96F8GP40:P_OFSP_BBO_NAP_RDIM_512}
\end{subfigure}
\hfill
\begin{subfigure}[t]{0.48\textwidth}
\centering
\includegraphics[height=4cm]{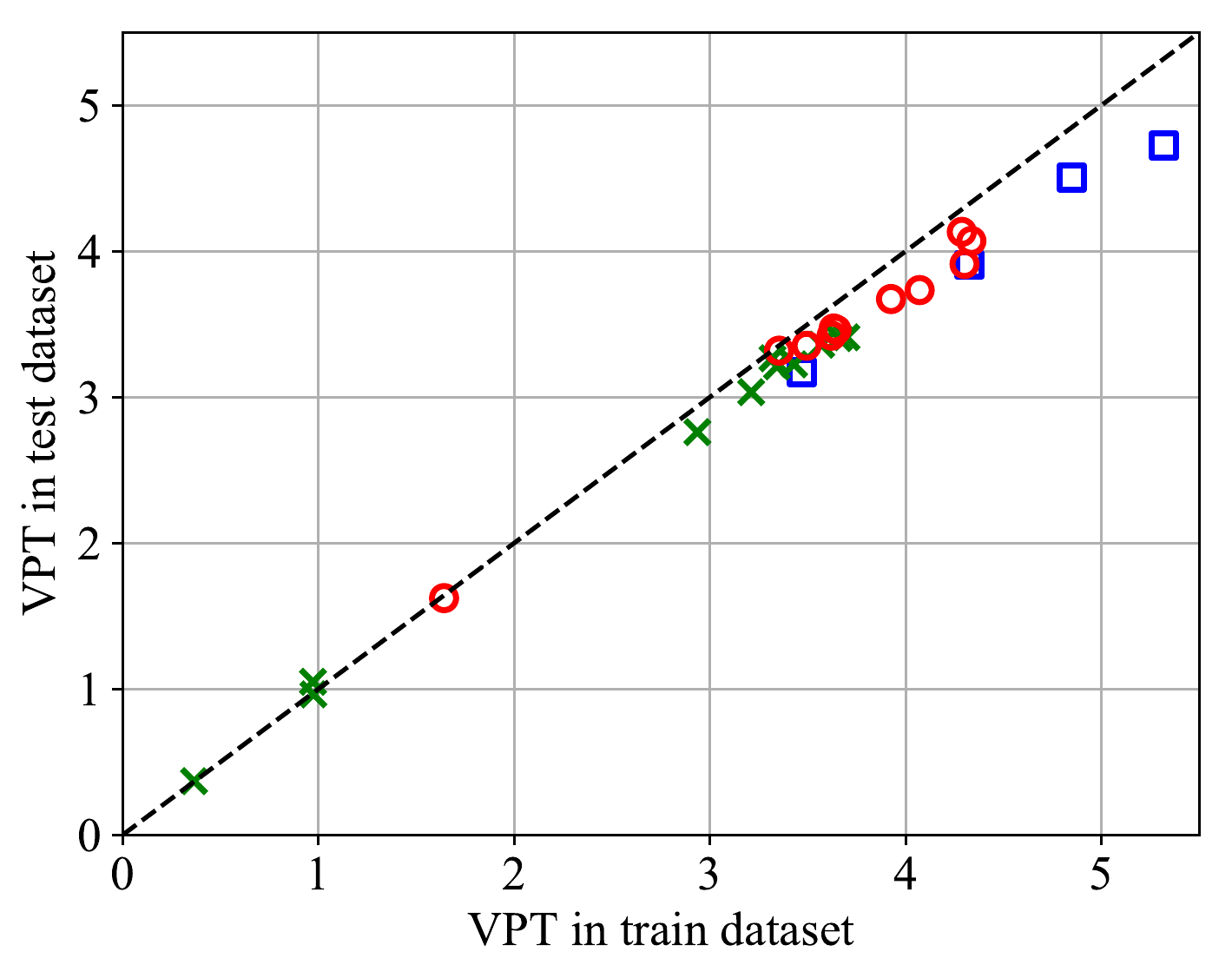}
\caption{VPT in test w.r.t. VPT in train}
\label{fig:KSGP512:P_OFSP_BBO_NAP_RDIM_512}
\end{subfigure}
\caption{
The average VPT measured from $100$ initial conditions sampled from the \textbf{test} dataset is plotted against the average VPT measured from $100$ initial conditions sampled from the \textbf{training} dataset for parallel models forecasting the dynamics of (a) the Lorenz-96 system with state dimension $d_o=40$ and (b) the Kuramoto Sivashinsky equation with state dimension $d_o=512$. 
\\
RC \protect \blueRectangle ;
GRU \protect \greenCross ;
LSTM \protect \redCircle ;
}\label{fig:P_OFSP_BBO_NAP_RDIM_512}
\end{figure*}

In this section, we provide results on the overfitting of the models in the parallel setting in the Lorenz-96 model in \Cref{fig:L96F8GP40:P_OFSP_BBO_NAP_RDIM_512}
and the Kuramoto-Sivashinsky equation in \Cref{fig:KSGP512:P_OFSP_BBO_NAP_RDIM_512}.
In both cases we do not observe overfitting issues as the predictive performance in terms of the VPT of the models in the test dataset is very close to that in the training dataset, emphasizing that the generalization ability of the models is good.

